\documentclass[a4paper,11pt]{article}
\pdfoutput=1
\usepackage{jheppub}
\usepackage{amsmath}
\usepackage{amssymb}
\usepackage{color}
\usepackage{graphicx}
\usepackage{hyperref}
\usepackage{xspace,slashed}
\usepackage[utf8]{inputenc}
\usepackage{multirow}
\usepackage{algorithm}
\usepackage{algorithmic}
\usepackage{stackengine}
\usepackage{enumitem}
\usepackage{etoolbox}
\usepackage{multirow}
\usepackage{array}

\newtoggle{draft}

\togglefalse{draft}

\iftoggle{draft} {
  \usepackage{changes}
}{%
  \usepackage[final]{changes}
}

\definechangesauthor[name=pn, color=blue]{pn}
\definechangesauthor[name=km, color=magenta]{km}

\usepackage{booktabs}

\makeatletter
\g@addto@macro\bfseries{\boldmath}
\makeatother

\newcommand{\oS}{\overline S}

\newcommand{\xa}{\mathfrak{m}}
\newcommand{\yb}{\mathfrak{n}}
\newcommand\numberthis{\addtocounter{equation}{1}\tag{\theequation}}

\newcommand{\EulerGamma}{\gamma_{\mathrm{E}}}

\newcommand{\ep}{\epsilon}

\iftoggle{draft} {
  \newcommand{\tobedeleted}[1]{\textcolor{azure}{#1}}
  
}{%
  \newcommand{\tobedeleted}[1]{}
  
}

\newcommand{\be}{\begin{equation}}
\newcommand{\ee}{\end{equation}}

\usepackage{soul}
\allowdisplaybreaks

\definecolor{azure}{rgb}{0.0, 0.9, 1.0}

\newcommand{\TTPaff}{Institute for Theoretical Particle Physics,
  KIT, 76128 Karlsruhe, Germany}
\newcommand{\IAPaff}{Institute for Astroparticle Physics, KIT, 76344 Eggenstein-Leopoldshafen, Germany}

\preprint{
  \begin {flushright}
    TTP24-004, P3H-24-014 
  \end{flushright}
}

\title{
$N$-jettiness soft function at next-to-next-to-leading order in perturbative QCD
}
\author[a]{Prem Agarwal,}
\author[a]{Kirill Melnikov,}
\author[a,b]{and Ivan Pedron}

\affiliation[a]{\TTPaff}
\affiliation[b]{\IAPaff}

\abstract{ 
We derive a compact representation of the \emph{renormalized} $N$-jettiness soft function that is free of infrared and collinear divergences through next-to-next-to-leading order in perturbative QCD. The number of hard partons $N$ is a parameter in the formula for the finite remainder. Cancellation of all infrared and collinear singularities between the bare soft function and its renormalization matrix in color space is demonstrated analytically. 
}

\begin{document}

\maketitle 

\section{Introduction}
The need for precise description of high-multiplicity final states produced in hard processes at the Large Hadron Collider (LHC)  led to many studies of the so-called subtraction and slicing schemes for perturbative QCD computations \cite{Caola:2017dug, Frixione:2004is, Gehrmann-DeRidder:2005btv, Currie:2013vh, Somogyi:2005xz, Somogyi:2006db, DelDuca:2016csb, DelDuca:2016ily, Czakon:2010td, Czakon:2011ve, Czakon:2014oma, Anastasiou:2003gr, Catani:2007vq, Boughezal:2011jf, Grazzini:2017mhc, Jouttenus:2011wh, Gaunt:2015pea, Boughezal:2015eha, Sborlini:2016hat, Herzog:2018ily, Magnea:2018hab, Capatti:2019ypt, TorresBobadilla:2020ekr, Bertolotti:2022aih, Moult:2016fqy, Boughezal:2016zws,Moult:2017jsg,Boughezal:2018mvf,Ebert:2018lzn,Boughezal:2019ggi}. The role of these schemes is to enable cancellation of infrared and collinear divergences between virtual and real contributions to cross sections without compromising the fully-differential nature of theoretical  predictions.  

In spite of the fact that slicing and subtraction schemes aim at achieving identical goals, their discussions in the literature  proceed  quite differently.  Indeed, a typical pathway to the construction of a subtraction scheme involves studies  of how  factorization of  matrix elements 
in the soft and collinear limits, and factorization of  phase spaces for final-state particles 
can be combined to  make the subtraction terms both observable-independent and  
analytically (or numerically) integrable in $d$-dimensions.\footnote{We use dimensional regularization 
throughout this paper and assume that 
$d=4-2\ep$.}  On the contrary, 
advanced slicing schemes start from the choice of an observable for which physical cross sections factorize 
into universal quantities, in the limit where the selected 
observable is small. 

In spite of a somewhat different philosophy, 
it is  clear that a very tight connection  between  slicing and  subtraction approaches must exist since  variables employed  for  slicing calculations are infrared safe.  It follows that a differential cross section for the small value of the selected slicing parameter -- which can be considered as the major  quantity to understand in order to enable the construction of a slicing scheme  -- 
should be computable with subtractions.

We believe that  existing computations in the context 
of existing slicing schemes do not make this connection sufficiently clear and do not exploit it. To see this, consider available  computations of 
the $N$-jettiness soft function at next-to-leading (NLO) \cite{Jouttenus:2011wh} and at next-to-next-to-leading order (NNLO) \cite{Kelley:2011ng, Monni:2011gb, Hornig:2011iu, Boughezal:2015eha,Gaunt:2015pea,Campbell:2017hsw, Bell:2018mkk, Jin:2019dho, Bell:2023yso} in perturbative QCD.
These computations are performed by mapping the available phase space of soft-gluon emissions onto 
hemispheres along the different hard directions, computing many of the required  integrals numerically, and studying  cancellation of infrared and collinear divergences against the soft-function renormalization matrix  
\textit{a posteriori}.  Furthermore, the soft-function renormalization matrix  is constructed from the \emph{consistency requirement} that soft and collinear divergences cancel when 
different elements of the factorization theorem (the hard function, the beam function, the jet function and the soft function) are combined to produce a physical cross section.
While this computational method works fine in practice, it is in stark  contrast with the current developments in 
understanding subtraction schemes where one attempts to find  suitable subtraction terms for a  generic  $N$-jet problem,
integrate them over unresolved phase spaces, 
show analytically the cancellation of all the  $1/\ep$ poles 
and derive a compact representation for the finite remainder
\cite{Bertolotti:2022aih,Devoto:2023rpv}. 

The goal of this paper is to show that  borrowing  certain  ideas from the recent developments in constructing generic NNLO QCD subtraction schemes \cite{Devoto:2023rpv} is beneficial for computing essential ingredients for modern slicing computations. Our prime example is the $N$-jettiness soft function \cite{Jouttenus:2011wh} at NNLO QCD that we study for a \emph{generic value of} $N$.
We analytically demonstrate the cancellation of all 
soft and collinear $1/\ep$ divergences   
that appear in the calculation of the $N$-jettiness soft function against the corresponding renormalization constant, 
and derive a simple formula for the finite remainder that is valid in  four space-time dimensions. We note that very recently a calculation  of $N$-jettiness soft function for arbitrary $N$ was reported  in  Ref.~\cite{Bell:2023yso}. The computational  method used in that reference is, however, very different from ours since it relies  on the extension of 
numerical methods developed in Refs.~\cite{Bell:2018vaa, Bell:2018mkk, Bell:2018oqa, Bell:2020yzz}.

The remainder of the paper is organized as follows. In the 
next Section we define the $N$-jettiness variable \cite{Stewart:2010tn} and the soft function that we study in this paper. In Section~\ref{sect:3}
we discuss the renormalization of the soft function. We continue with the calculation 
of the $N$-jettiness soft 
function at NLO QCD in Section~\ref{sect:4}. We then  proceed to the 
analysis  of the NNLO case. 
We start by discussing in 
Section~\ref{sect:5}
how the 
NNLO QCD computation of the $N$-jettiness soft function can be organized. We  first focus on
the two-gluon  final state.  We discuss the 
 contribution of the uncorrelated emissions of two soft gluons in Section~\ref{sect:6}. 
 We analyze the triple-color correlated terms in $N$-jettiness soft function in  Section~\ref{sect:7}.   Correlated emissions of two gluons 
 are discussed in Sections~\ref{sect:8}, \ref{sect:9}, \ref{sect:10} and \ref{sect:11}. We study the contribution of the soft $q \bar q$ 
 pair to  $N$-jettiness soft function   in Section~\ref{sect:12}. 
 The final result for the soft 
 function  can be found in Section~\ref{sect:13}. 
Numerical implementation and comparison with results available in the literature is described 
in Section~\ref{sec:num}.
 We conclude in Section~\ref{sect:14}. Some useful formulas are collected in 
 appendices. 

\section{Definitions of basic quantities}

We consider  partonic processes with an arbitrary number of hard color-charged partons 
that we denote by $N$
\be
\begin{split}
& 0 \to h_1 + h_2 + h_3 + h_4 + \cdots h_N,
\\
& h_1  \to h_2+h_3  + h_4 + \cdots h_N,
\\
&  h_1 + h_2 \to h_3 + h_4 + \cdots h_N. 
\label{eq2.1}
\end{split}
\ee
The first process refers to $e^+e^-$ annihilation, the second to deep-inelastic scattering and the third to  
hadron collisions. 
Each parton in 
Eq.~(\ref{eq2.1}) carries a four-momentum $p_i$ and a color 
charge ${\bf T_i}$. All partons are assumed to be massless.

The cross sections of processes shown in Eq.~(\ref{eq2.1})  are affected by the  radiation of soft  quarks and gluons, both real and virtual. To quantify their impact,
an  observable  that controls 
the softness of the radiation is needed. 
In this paper, we will use 
  the  $N$-jettiness ${\cal T}$ \cite{Stewart:2010tn} for this purpose. 

To define ${\cal T}$, we proceed as follows. For a generic  process,  we split all final-state partons that we have
to consider in a particular 
order of perturbation theory 
into resolved ${\cal R}$ and unresolved ${\cal U}$ ones. We define a list
\be
L_x = \left \{ \frac{2 p_x p_{h_1}}{P_{h_1} }, \frac{2 p_x p_{h_2}}{P_{h_2} }, \frac{2 p_x p_{h_3}}{P_{h_3}} ...
\right \},
\label{eq2.2}
\ee
where $x \in {\cal U}$ and $h_i \in {\cal R} $. Quantities 
$P_{h_i}$ are arbitrary normalization constants. 
We then define the $N$-jettiness variable
as \cite{Stewart:2010tn}
\be
   {\cal T}({\cal R}, {\cal U}) = \sum \limits_{x \in {\cal U}} {\rm min} \left [ L_x \right ].
   \label{eq2.3}
\ee

To compute  radiative corrections to  processes in  Eq.~(\ref{eq2.1}), we need to  consider virtual and real-emission
corrections. Since we work through next-to-next-to-leading order, 
we need to include  virtual corrections up to two loops, one-loop corrections with and without one soft parton and the real emission processes with up to two soft partons. 
For virtual corrections to processes without additional soft radiation, 
${\cal U}$ is an empty set and so ${\cal T}$ is zero.  The same is true 
for  leading order processes shown in Eq.~(\ref{eq2.1}).  

We note that virtual corrections do not need to be computed explicitly. Indeed, since infrared and collinear singularities of a generic two-loop amplitude are 
known \cite{Catani:1998bh,Becher:2009qa,Becher:2013vva}, it is customary  to separate  them from the short-distance 
contribution to the two-loop amplitude known as the ``hard function'' and include them into the renormalization constant 
of the soft function, that we discuss in the next section.
Hence, to compute  the soft function at NNLO we need to  study contributions with one or two soft partons to cross sections 
of processes in Eq.~(\ref{eq2.1}), subject to $N$-jettiness 
constraint. 
\\

The perturbative expansion of the \emph{bare} soft function  can be written as 
\be
S_\tau(\tau) = \delta(\tau) + 
a_s S_{\tau,1}(\tau) + 
a_s^2 S_{\tau,2}(\tau)+
\cdots, 
\ee
where $\tau$ is the value of $N$-jettiness and  $a_s = \alpha_s(\mu)/(2\pi)$ is the  renormalized QCD coupling constant. The quantities $S_{\tau,i} (\tau)$, $i = 1,2$,
can be computed starting  from  universal  functions that describe the behaviour of QCD matrix elements 
when some of the 
radiated partons are  soft. 

Before proceeding, it is useful to  point out that it is very  convenient to  work with  the Laplace transform 
of the soft function \cite{Bell:2023yso}.  The Laplace transform is defined  as follows  
\be
S(u) = 
\int \limits_{0}^{\infty}
{\rm d} \tau \; S_\tau(\tau) e^{-u \tau}.
\ee
 As we will  see from the explicit computations described below, the dependence of functions $S_{\tau,i}$, $i=1,2$,  on $\tau$ is very particular and is given by  
simple powers  $\tau^{-1-n\ep}$,
where $n=2,4$ etc.   For these functions, the Laplace transform evaluates to 
\be
\int \limits_{0}^{\infty}
{\rm d} \tau \; \tau^{-1-n\ep} e^{-u \tau} = -u^{n\ep} \frac{ \Gamma(1-n \ep)}{n \ep} = 
- \frac{ {\bar u}^{n\ep} e^{-\ep n \EulerGamma} \Gamma(1-n \ep) }{n \ep}.
\label{eq2.6}
\ee
Note that in the last step we introduced a variable $\bar u =
u e^{\gamma_E }$ to simplify 
the expansion of the corresponding $\Gamma$-functions in $\ep$.
In what follows we will use Eq.~(\ref{eq2.6})  to compute  the 
Laplace transform of the perturbative expansion of the  bare soft function.

\section{Renormalization of the soft function}
\label{sect:3}

Perturbative computations of the soft function lead to expressions that contain soft and collinear divergences 
that manifest themselves through $1/\ep$ poles. 
These divergences are removed by a dedicated  renormalization \cite{Jouttenus:2011wh}.

The renormalization of the soft function is a multiplicative (matrix) renormalization in the Laplace space \cite{Bell:2023yso}. We denote the Laplace transforms of the bare and renormalized 
soft functions by   
$S$ and $\tilde S$, respectively.  The two functions 
are related by the following equation
\be
S = Z \tilde S Z^+,
\label{eq3.1}
\ee
where $Z$ is a matrix in color space, see Appendix~\ref{app:a}.
While it is entirely possible to compute  the unrenormalized soft function $S$ and then remove the divergences 
by using Eq.~(\ref{eq3.1}), it is 
beneficial to study suitable  combinations of $Z$ and 
$S$ that are actually needed for 
computing $\tilde S$.

To find these combinations,   
we write expansions of  $Z$, $S$ and $\tilde S$ in powers of $\alpha_s$
\be
\begin{split}
& Z = 1 + Z_1 + Z_2,\\
& S = 1 +S_1 + S_2,
\\
& 
\tilde S  = 1 + \tilde S_1
+ \tilde S_2, 
\end{split}
\ee
substitute them into Eq.~(\ref{eq3.1}) and solve for $
{\tilde S}_{1,2}$.
We find 
\be
\begin{split}
& \tilde S_1 = S_1 - Z_1 - Z_1^+,\\
& \tilde S_2 = S_2 - Z_2 - Z_2^+ + Z_1 Z_1 + Z_1^+ Z_1^+ - Z_1 S_1 - S_1 Z_1^+ + Z_1 Z_1^+.
\end{split}
\label{eq3.3}
\ee

It was recently pointed out 
in Ref.~\cite{Devoto:2023rpv}, in the context of the application of the nested soft-collinear subtraction scheme to arbitrary multi-jet processes, that it is beneficial to 
separate iterations of 
${\cal O}(\alpha_s)$ soft, soft-collinear and virtual contributions to cross sections from the rest of NNLO 
contributions. Following the same logic, we write 
\be
{S}_2 = \frac{1}{2} {S}_1 {S}_1
+{S}_{2,r}.
\label{eq3.4}
\ee
It is useful to do the same for the NNLO contribution to the renormalization matrix
\be
Z_2 = \frac{1}{2} Z_1 Z_1 
+ Z_{2,r}.
\ee
It is then easy 
to see that  
$\tilde S_2$
can be written 
as 
\be
\tilde S_2 
= \frac{1}{2}
\tilde S_1 \tilde S_1 + \frac{1}{2}[Z_1,Z_1^+]
+ \frac{1}{2}
\left [S_1, Z_1 - Z_1^+
\right ] 
+ S_{2,r} - Z_{2,r}-Z_{2,r}^+.
\label{eq3.6}
\ee
We will show below that 
this representation is 
beneficial for computing the renormalized $N$-jettiness 
soft function. Indeed,  when 
representation in Eq.~\eqref{eq3.6} 
is used, the  required  effort is minimized because  cancellations of $1/\ep$ poles between the different quantities are identified relatively early in the course of the computation.  However, 
before discussing how the NNLO computation is performed, 
we will illustrate our approach by calculating the NLO 
contribution to the $N$-jettiness soft function. 

\section{\texorpdfstring{$N$}{N}-jettiness soft function at NLO}
\label{sect:4}

We will explain main ideas  of our  approach 
by considering   $N$-jettiness soft function at NLO. 
Since one-loop virtual corrections to $N$-jettiness soft function do not need to be considered, we  focus on the single real-emission contribution below.
 We use label $\xa$ to denote a soft  gluon  and
write its momentum as $p_\xa$.   The gluon $\xa$ is the only unresolved parton, therefore ${\cal U} = \{\xa\}$.

 We choose $P_{h_i} = E_i$, where 
 $E_i$ is the energy  of the parton $h_i$ in the chosen 
 reference frame, 
 and obtain (c.f. Eqs.~(\ref{eq2.2},\ref{eq2.3}))
\begin{equation}
    \mathcal{T}(\xa)=  E_\xa \psi_\xa.
\label{eq4.1}
\end{equation}
In Eq.~(\ref{eq4.1})  $E_\xa$ is the energy of the unresolved gluon $\xa$ and  $\psi_\xa$  is  a function defined as 
\be
\psi_\xa = {\rm min}\{\rho_{1 \xa},\rho_{2\xa},\rho_{3\xa},...,\rho_{N \xa}
\},
\ee
where $\rho_{i j} = 1 - \vec n_i \cdot \vec n_j$ and 
$\vec n_{i}$
is a unit vector which points in the direction of the three-momentum of parton $i$.
\\

The real-emission  single-gluon  soft contribution at fixed $N$-jettiness $\tau$ reads
\be
S_{\tau,1}(\tau) =  -\sum \limits_{(ij) }  {\bf T}_i \cdot {\bf T}_j \;  I_{S,ij}^{(1)}(\tau),
\label{eq4.3}
\ee
where we use the notation 
$(ij)$ to indicate that the sum runs over all $i,j \in {\cal R}$ with the constraint $i \ne j$.
Furthermore, 
 ${\bf T}_i$ is  the color charge operator of parton $i$, 
\be
I_{S,ij}^{(1)} = g_s^2 \int [{\rm d} p_\xa ] \; \delta( \tau - E_{\xa} \psi_{\xa} )  \; S_{ij}(\xa),
\label{eq4.4}
\ee
 $g_s$ is the bare QCD coupling constant
 and 
\be
S_{ij}(\xa) = \frac{p_i p_j}{(p_i p_\xa) ( p_j p_\xa)} = \frac{1}{ E_\xa^2} \; \frac{\rho_{ij} }{\rho_{i \xa} \rho_{j \xa} },
\label{eq4.5}
\ee
is the soft eikonal function.
It is convenient to write the 
phase space element of the gluon 
$\xa$ as follows
\be
[ {\rm d} p_m]  = \frac{{\rm d} \Omega_\xa^{(d-1)}}{2 (2 \pi)^{d-1}} \; \frac{{\rm d} E_\xa}{E_{\xa}^{1+2 \ep}} \; E_{\xa}^2 
 = \frac{\Omega^{(d-2)}}{2 (2 \pi)^{d-1}} \;  [{\rm d} \Omega^{(d-1)}_{\xa}]
  \; \frac{{\rm d} E_\xa}{E_{\xa}^{1+2 \ep}} \; E_{\xa}^2.
  \label{eq4.6}
\ee
We then  combine the 
coupling constant squared 
with the remnant of phase space 
normalization factor 
\be
g_s^2 \frac{\Omega^{(d-2)}}{2 (2 \pi)^{d-1}} = [\alpha_s],
\ee
and note that 
\be
[\alpha_s] = a_s(\mu) \;\mu^{2 \ep}\frac{ e^{\gamma_E \epsilon}}{\Gamma(1-\ep)}
\left ( 1 + {\cal O}(a_s) \right ),
\ee
where $a_s(\mu) = \alpha_s(\mu)/(2\pi)$ and 
$\alpha_s(\mu)$ is the QCD coupling constant renormalized in the $\overline {\rm MS}$ scheme. 

We work to first order in 
$a_s$ and  use Eq.~(\ref{eq4.6}) to 
 integrate over energy of the gluon $E_\xa$ in Eq.~(\ref{eq4.4}). We find 
\be
I_{S,ij}^{(1)}(\tau) =  \frac{   [\alpha_s] }{\tau^{1+2\ep}} \; 
\int  [{\rm d} \Omega^{(d-1)}_\xa ] \; \psi_\xa^{2\ep} \;
\frac{\rho_{ij} }{\rho_{i \xa} \rho_{j \xa} }
= 
\frac{[\alpha_s] }{\tau^{1+2\ep}} \; 
\left \langle 
\psi_\xa^{2\ep} \;
\frac{\rho_{ij} }{\rho_{i \xa} \rho_{j \xa} }
\right \rangle_\xa \; ,
\label{eq4.9}
\ee
where  $\langle .. \rangle_\xa$ 
indicates integration over directions  of the vector 
$\vec n_\xa$.
\\

To proceed further, we compute the
Laplace transform 
of the soft function and obtain 
\be
S_{1} = \frac{
{\bar u}^{2\ep} e^{-2\ep\EulerGamma} \Gamma(1-2\ep)}{2\ep} \; [\alpha_s]
\sum \limits_{(i  j)  }  {\bf T}_i \cdot {\bf T}_j \left \langle \psi_\xa^{2\ep} \frac{\rho_{ij}}{\rho_{i \xa} \rho_{j \xa}} \right \rangle_\xa .
\label{eq4.10}
\ee

Integration over directions of the vector $\vec n_\xa$ induces collinear singularities when 
$\vec n_\xa || \vec n_i$ and $\vec n_\xa || \vec n_j$. To isolate them,  we make use 
of the fact that collinear 
singularities of the 
integrand are  fully controlled by the  remnants of the eikonal function in Eq.~(\ref{eq4.10}) 
and that the collinear limits of the 
function $\psi_\xa$ can be predicted  
for an \emph{arbitrary  hard process}. 
\\

Indeed, consider the collinear limit
$\xa || i$ as an example. 
In that limit,  $\rho_{i \xa}$ is the smallest $\rho$-value so that
\be
\lim_{\xa || i} \psi_{\xa}  = \rho_{i \xa}.
\ee
It turns out useful to rewrite the integrand in 
Eq.~(\ref{eq4.9}) as follows
\be
\psi_\xa^{2\ep} \;
\frac{\rho_{ij} }{\rho_{i \xa} \rho_{j \xa} }
=
\left ( \frac{ \psi_\xa \rho_{ij} }{ \rho_{i \xa} \rho_{j \xa} }  \right )^{2\ep} \;
\frac{\rho^{1-2\ep}_{ij} }{\rho^{1-2\ep}_{i \xa} \rho^{1-2\ep}_{j \xa} }.
\ee
We then define 
\be
\left ( \frac{ \psi_\xa \rho_{ij} }{ \rho_{i \xa} \rho_{j \xa} }  \right )^{2\ep}
 = 1 + 2\ep g^{(2)}_{ij,\xa}, 
\ee
where the function $g^{(\alpha)}_{ij,\xa}$  contains  higher powers of $\ep$ in addition to the first power   shown explicitly. 
Hence, 
we write
\be
\left \langle \psi_\xa^{2\ep} \frac{\rho_{ij}}{\rho_{i \xa} \rho_{j \xa}} \right \rangle_\xa
=
\left \langle \left (
1 + 2 \ep g^{(2)}_{ij, \xa } 
\right )
\frac{\rho^{1-2\ep}_{ij} }{\rho^{1-2\ep}_{i \xa} \rho^{1-2\ep}_{j \xa} }
\right \rangle_\xa.
\label{eq4.14}
  \ee
   Since the function $\ep g^{(2)}_{ij,m}$ vanishes in all collinear limits and is ${\cal O}(\ep)$, it only provides
  a finite contribution to the $N$-jettiness soft function that will have to be computed numerically for a given configuration of hard partons.  
  On the contrary,
  the  first integral in the right-hand side of  Eq.~(\ref{eq4.14}) 
  can be 
  computed explicitly, see Eq.~(\ref{eqb.5}). 
  We write 
  \be
\left \langle 
\frac{\rho^{1-2\ep}_{ij} }{\rho^{1-2\ep}_{i \xa} \rho^{1-2\ep}_{j \xa} }
\right \rangle_\xa
= \frac{2 \eta_{ij}^{\ep}} {\ep} K_{ij}^{(2)} ,
\label{eq4.15}
\ee
  where $\eta_{ij} =\rho_{ij}/2$
  and 
  \be
K_{ij}^{(2)} = \frac{\Gamma(1+\ep)^2}{\Gamma(1+2\ep)} \;
  {}_2 F_{1} \left (\ep,\ep,1-\ep,1-\eta_{ij} \right ). 
  \ee
Putting everything together, we find
\be
S_{1}
= a_s \; (\mu \bar u )^{2\ep} \;
\frac{\Gamma(1-2\ep)}{
\Gamma(1-\ep) e^{\ep \gamma_E}}
\sum \limits_{(ij)}
{\bf T}_i \cdot {\bf T}_j 
 \; 
\left [  \frac{ \eta_{ij}^{\ep} }{\ep^2} K_{ij}^{(2)} + 
\left \langle  \; g^{(2)}_{ij, \xa } \; 
\frac{\rho^{1-2\ep}_{ij} }{ \rho^{1-2\ep}_{i \xa} \rho^{1-2\ep}_{j \xa} } \right \rangle_{\xa} \;
\right ],
\label{eq4.17a}
 \ee
 It follows from the above equation that  all   $1/\ep$ poles have been separated 
from the finite remainder 
and that the remnant of the complicated $N$-jettiness function appears in the finite remainder \emph{only}.
\\

It remains  to combine 
$S_1$ with the 
renormalization matrices $Z_{1}$ and
$Z_1^+$, c.f. 
Eq.~(\ref{eq3.3}).
 The expression 
for $Z_1$ can be found in Appendix~\ref{app:a};  similar to the  NLO soft function
shown in Eq.~(\ref{eq4.17a}), 
it is given by the  sum of  products of the color-charge operators 
with some coefficients.  Combining the different contributions 
and discarding all terms beyond ${\cal O}(\ep^0)$, we find  
\be
{\tilde S}_1
= a_s \sum \limits_{(ij)}
{\bf T}_i \cdot {\bf T}_j
\left [ 
2 L_{ij}^2
+ {\rm Li}_{2}(1-\eta_{ij}) + \frac{\pi^2}{12}
+ 
\left \langle
\ln \left ( 
\frac{\psi_\xa \rho_{ij}}{\rho_{i \xa} \rho_{j \xa}}
\right )
\; 
\frac{\rho_{ij} }{ \rho_{i \xa} \rho_{j \xa} } \right \rangle_{\xa}
+{\cal O}(\ep) \right ],
\ee
where 
\be
L_{ij} =  \ln \left ( \mu \bar u \sqrt{\eta_{ij}} \right ).
\label{eq4.19}
\ee
This result can be easily evaluated numerically for an arbitrary number of hard 
partons $N$. The logarithm $\ln (\psi_\xa \rho_{ij}/(\rho_{i\xa} \rho_{j \xa}))$ provides an infrared regulator for an integral that  would have exhibited 
 collinear divergencies without it.

\section{\texorpdfstring{$N$}{N}-jettiness soft function at NNLO}
\label{sect:5}

We will now explain how to extend  the above 
approach  to compute the NNLO contribution to the $N$-jettiness soft function. It is clear that the NNLO computation is significantly more complex than the NLO one. However, we will show that it is possible to split the calculation  into several  independent parts  making the entire problem simpler.  
\\

The  NNLO contribution to the bare  soft function  reads 
\be
S_2  
 = S_{2,RR} 
 + S_{2,RV}
 - a_s \;  \frac{\beta_0}{\ep}  S_1,
 \ee
where $S_{2,RR}$
is the double real-emission contribution, 
$S_{2,RV}$ is the real-virtual contribution and the 
last term arises because of the renormalization of the strong coupling constant in the NLO soft function. The leading order QCD $\beta$-function  is defined as
\be
\beta_0 = \frac{11}{6}C_A - \frac{2}{3}n_F T_R,
\ee
where $C_A = N_c$ is the Casimir  
of the group $SU(N_c)$, $T_R=1/2$ 
and $n_f$ is the number of massless  quarks.

The double-real emission contribution 
is obtained by integrating the  double-soft
eikonal current  \cite{Catani:1999ss}
over the phase space of two soft gluons subject to the $N$-jettiness constraint. 
Following Ref.~\cite{Catani:1999ss},
we  further split the  double-real 
contribution  into 
correlated and uncorrelated pieces 
\be
S_{2,RR,\tau}
= S_{2,RR,\tau,T^4}
+S_{2,RR,\tau,T^2}.
\ee
We note that the subscript $\tau$ indicates
that the Laplace transform of the 
corresponding quantities has not yet been computed.

The uncorrelated contribution reads 
\be
S_{2,RR,\tau,T^4}
= 
\frac{1}{2} \; \sum \limits_{(ij),(k,l)} \{ {\bf T}_i \cdot {\bf T}_j, {\bf T}_k \cdot {\bf T}_l \}
\;I_{T^4,ij,kl},
\label{eq5.4}
\ee
where 
\be
I_{T^4,ij,kl} = \frac{ g_s^4}{2} 
\int [{\rm d} p_\xa ] \;   [{\rm d} p_\yb ]
\delta( \tau - E_{\xa} \psi_{\xa} -E_{\yb} \psi_{\yb} ) \; S_{ij}(\xa) \; S_{kl}(\yb),
 \ee
 and the eikonal functions  $S_{ij}(\xa)$ and 
 $S_{kl}(\yb)$ are defined in 
 Eq.~(\ref{eq4.5}).
The  correlated  contribution $S_{2,RR,T^2}$
reads
\be
S_{2,RR,\tau,T^2}
=-\frac{C_A}{2} \sum \limits_{(ij)}
\;{\bf T}_i \cdot {\bf T}_j \;I_{T^2,ij},
\label{eq5.6}
\ee
where 
\be
I_{T^2,ij} = 
\frac{g_s^4}{2} \; \int [{\rm d} p_\xa ] [{\rm d} p_\yb] \delta(\tau - E_\xa
\psi_\xa - E_\yb \psi_{\yb} ) \tilde S_{ij}^{gg}(\xa,\yb).
\ee
The eikonal function 
${\tilde S}_{ij}^{gg}(\xa, \yb)$
can be found in Appendix~\ref{app:c}.
\\

The real-virtual 
contribution reads \cite{Catani:1999ss}
\be
\begin{split} 
S_{2,RV,\tau}
& = \frac{[\alpha_s] \; 2^{-\ep}  }{\ep^2} C_A A_K(\ep)
\sum \limits_{(ij)}
\;{\bf T}_i \cdot {\bf T}_j
\;I_{RV,ij}  +
[\alpha_s]
\frac{4\pi N_\ep}{\ep}
\sum \limits_{(kij)}
\kappa_{ij} 
F^{kij} I_{kij}.
\end{split}
\label{eq5.8}
\ee
where
\be
\kappa_{ij}
= \lambda_{ij} 
- \lambda_{i
\xa} - \lambda_{j \xa},
\label{eq5.9}
\ee
and $\lambda_{ij} = 1$ if 
both $i$ and $j$ refer to incoming or outgoing partons and  zero otherwise. 
Furthermore, the normalization 
factors in Eq.~(\ref{eq5.8}) 
read 
\be
A_K(\ep) 
= \frac{\Gamma^3(1+\ep) \Gamma^5(1-\ep)}{
\Gamma(1+2\ep) 
\Gamma^2(1-2\ep)
},
\;\;\;
N_\ep = \frac{\Gamma(1+\ep)
\Gamma^3(1-\ep)}
{ \Gamma(1-2\ep)},
\ee
and we have defined 
$F^{kij} =  f_{abc} T^a_k T^b_i T^c_j $
to describe   
anti-symmetrized  product of  three color charge operators. Finally, 
the functions $I_{RV,ij}$ 
and $I_{kij}$ read
\be
I_{RV,ij}
= g_s^2 \int 
[{\rm d} p_\xa]
\delta(\tau - E_\xa \psi_\xa)
(S_{ij}^{gg}(\xa)) ^{1+\ep},
\ee
and
\be
I_{kij}
= g_s^2 \int
[{\rm d} p_\xa]
\delta( \tau - E_\xa \psi_\xa )
S_{ki}(\xa) (S_{ij}^{gg}(\xa ) )^{\ep}.
\ee
\\

In the next section we will show that 
\be
S_{2,RR,T^4} = \frac{1}{2}
S_1 S_1.
\ee
Therefore, the quantity $S_{2,r}$ 
introduced in Eq.~(\ref{eq3.4}) reads 
\be
S_{2,r} = S_{2,RR,T^2} 
+ S_{2,RV} - \frac{a_s \beta_0}{\ep} S_1.
\ee

\section{Uncorrelated emissions of two soft gluons }
\label{sect:6}

We have argued that  $S_2$ contains an iterated contribution of the next-to-leading order  soft function
$S_1$ and that it is beneficial to separate it from the 
rest of $S_2$,  c.f. Eq.~(\ref{eq3.4}). We will now show 
that this iteration 
can be  identified 
with the Laplace transform of the 
term $S_{2,RR,\tau, T^4}$
in Eq.~(\ref{eq5.4}).
 To this end, we write 
\be
I_{T^4,ij,kl} = \frac{ [\alpha_s]^2}{2} 
\left \langle 
\int \limits_{0}^{\infty} \frac{{\rm d} E_\xa}{E_\xa^{1+2\ep}}
\frac{{\rm d} E_\yb}{E_\yb^{1+2\ep}}
\;
\delta( \tau - E_{\xa} \psi_{\xa} -E_{\yb} \psi_{\yb} ) 
\frac{\rho_{ij}}{\rho_{i \xa} \rho_{j \xa} } \frac{\rho_{k l}}{\rho_{k \yb} \rho_{l \yb} }
\right \rangle_{\xa \yb}.
 \ee

It is straightforward to 
integrate over energies of  two gluons $E_{\xa, \yb}$. Indeed, the first integration
(say over $E_\yb$) is elementary since it just removes the $\delta$-function. The second integration  over 
$E_\xa$ extends from  $0$ to 
$E_\xa^{\rm max}
 = \tau/\psi_{\xa}
$, i.e. the energy of the gluon $\xa$ which corresponds to  $E_{\yb} = 0$.
The integral  turns out to  be of a Beta-function type, and we obtain 
\be
\int \limits_{0}^{\infty} \frac{{\rm d} E_\xa}{E_\xa^{1+2\ep}}
\frac{{\rm d} E_\yb}{E_\yb^{1+2\ep}}
\;
\delta( \tau - E_{\xa} \psi_{\xa} -E_{\yb} \psi_{\yb} )
= \frac{\tau^{-1-4\ep}}{\Gamma(-4\ep)}
\frac{\psi_\xa^{2\ep} \Gamma(1-2\ep)}{2\ep}
\frac{\psi_\yb^{2 \ep} \Gamma(1-2\ep)}{2\ep}.
\ee
Using this result, 
and performing the 
Laplace transform, 
we derive 
\be
\begin{split}
S_{2,RR,T^4}
& = \frac{[\alpha_s]^2
}{4}
\sum \limits_{(ij),(kl)}
\{ {\bf T}_i \cdot {\bf T}_j, {\bf T}_k \cdot {\bf T}_l \}
\left (\frac{u^{2\ep} \Gamma^(1-2\ep)}{2\ep} 
\right )^2
\\
& \times 
\left \langle 
\psi_\xa^{2 \ep} \frac{ \rho_{ij}}{\rho_{i \xa} \rho_{j \xa} }
\right  \rangle_{\xa}
\;
\left \langle 
\psi_\yb^{2 \ep} \frac{ \rho_{kl}}{\rho_{k \yb} \rho_{l \yb} }
\right \rangle_{\yb}
= \frac{1}{2}\;
S_{1} S_{1},
\end{split}
\ee
where the last step follows from 
the symmetry between 
$(ij)$ and $(kl)$
summation indices, 
and from the comparison with the  results for the  NLO $N$-jettiness soft function derived in Section~\ref{sect:4}.
As we already explained
in Section~\ref{sect:3}, it is straightforward to combine the above result with the iterated contribution of the 
renormalization constants to arrive at the relevant contribution to the renormalized soft function. 

\section{Terms with three color charges }
\label{sect:7}

We will next discuss the computation of the contribution that depends on triple products of color charges.  It originates from the 
following  terms 
in the expression for ${\tilde S}_2$
\be
{\tilde S}_{2}^{tc} = 
\frac{1}{2}
\left [Z_1, Z_1^+
\right ]
+\frac{1}{2}
\left [S_1,Z_1-Z_1^+ \right ]
+ S_{RV,tc},
\ee
where the 
real-virtual  triple-color correlated 
contribution is given  by the Laplace transform 
of the last term 
in Eq.~(\ref{eq5.8}).

The commutators 
are easy to compute using a general formula in 
Ref.~\cite{Devoto:2023rpv}.\footnote{See Eqs.~(4.142, 4.143).} We find 
\be
\frac{1}{2} [Z_1,Z_1^+]
=
- \frac{2 \pi a_s^2}{\ep^2}
\sum \limits_{(kij)}
\lambda_{kj}
 L_{ij} F^{kij}
 = 
 - \frac{ \pi a_s^2}{\ep^2}
\sum \limits_{(kij)}
\lambda_{kj}
 \ln \eta_{ij}  \;F^{kij},
 \label{eq7.2}
 \ee
where the last step 
follows from the definition of $L_{ij}$ in 
Eq.~(\ref{eq4.19}) 
and the (anti)-symmetry  of $F^{kij}$
with respect to  permutations of its indices. 

 The second commutator reads 
 \be
 \begin{split}
\frac{1}{2}
[S_1,Z_1-Z_1^+]
& = \frac{a^2_s  \pi ( \mu u )^{2\ep}}{\ep^2}
\frac{e^{\gamma_E \ep} \Gamma(1-2\ep)}{\Gamma(1-\ep)}\;
\sum \limits_{(kij)}
\lambda_{kj} \;
\left \langle 
\psi^{2\ep}_\xa 
\frac{\rho_{ij}}{
\rho_{i \xa} \rho_{j \xa}
}
\right \rangle_\xa
F^{kij}
\\
& 
= -\frac{a^2_s  \pi ( \mu u )^{2\ep}}{\ep^2}
\frac{e^{\gamma_E \ep} \Gamma(1-2\ep)}{\Gamma(1-\ep)}\;
\sum \limits_{(kij)}
\lambda_{kj} \;
\left \langle 
\psi^{2\ep}_\xa 
\frac{\rho_{ki}}{
\rho_{k \xa} \rho_{i \xa}
}
\right \rangle_\xa
F^{kij},
\label{eq7.3}
\end{split}
 \ee
where in the last step  we replaced 
$k \leftrightarrow j$
and used the symmetry  of $\lambda_{kj}$ and the anti-symmetry of $F^{kij}$. Furthermore, it is convenient to 
replace $\lambda_{kj}$ 
in the above expression 
with 
$\kappa_{kj}$ 
defined in  Eq.~(\ref{eq5.9}). 
This is possible
because 
\be
\sum \limits_{(kij)}
\lambda_{j \xa} \;
\left \langle 
\psi^{2\ep}_\xa 
\frac{\rho_{ki}}{
\rho_{k \xa} \rho_{i \xa}
}
\right \rangle_\xa
F^{kij} 
=
\sum \limits_{(kij)}
\lambda_{k \xa} \;
\left \langle 
\psi^{2\ep}_\xa 
\frac{\rho_{ki}}{
\rho_{k \xa} \rho_{i \xa}
}
\right \rangle_\xa
F^{kij} 
=0.
\ee
The first of these  equation holds because 
the summation over 
$i$ and $k$ involves a symmetric and an anti-symmetric tensor.  The 
second is fulfilled 
because 
one can sum over $j$ using  conservation of color charges and use the antisymmetry of $F^{kij}$
to set the resulting terms to zero. 
Hence, we can write 
\be
 \begin{split}
\frac{1}{2}
[S_1,Z_1-Z_1^+]
= -\frac{a^2_s  \pi ( \mu u )^{2\ep}}{\ep^2}
\frac{e^{\gamma_E \ep} \Gamma(1-2\ep)}{\Gamma(1-\ep)}\;
\sum \limits_{(kij)}
\kappa_{kj} \;
\left \langle 
\psi^{2\ep}_\xa 
\frac{\rho_{ki}}{
\rho_{k \xa} \rho_{i \xa}
}
\right \rangle_\xa
F^{kij} .
\label{eq7.5}
\end{split}
\ee
The real-virtual 
triple-color correlated 
contribution evaluates to 
\be
S_{RV,tc}
=
\frac{a_s^2 \pi
( \mu\; \bar u)^{4\ep} N_\ep 2^{-\ep}}{2\ep^2 }
\frac{\Gamma(1-4\ep)}{
\Gamma^2(1-\ep)
e^{2 \gamma_E \ep}}
\sum \limits_{(kij)}
\kappa_{k j} 
\left \langle 
\psi_\xa^{4\ep}
\frac{\rho_{ki}}{\rho_{k \xa} \rho_{i \xa}}
\left ( 
\frac{\rho_{kj}}{\rho_{k \xa} \rho_{j \xa}}
\right )^{\ep}
\right \rangle_\xa
\; F^{kij} \;.
\ee

We need to 
combine the three
contributions to 
${\tilde S}_2^{tc}$
shown above, demonstrate that 
all  $1/\ep$ poles cancel and derive a representation for the finite remainder.  We will do this in steps starting from 
contributions that exhibit  dependence on the $N$-jettiness function. 
There are two 
such contributions -- the commutator in Eq.~(\ref{eq7.5})
and $S_{RV,tc}$. We define
\be
\Sigma_{\text{tcc}}^{\psi} 
=\frac{1}{2}
[S_1,Z_1-Z_1^+]
+ S_{RV,tc} .
\label{eq7.7a}
\ee
Both terms in Eq.~(\ref{eq7.7a})  involve 
integrals of various 
combinations of $\rho_{ij}$'s and $\psi_\xa$ over the 
directions of the 
momentum of the gluon $\xa$.  Following the  discussion of the NLO contribution 
to the soft function, 
 we write
\be
\begin{split} 
& \left \langle
\psi_\xa^{2\ep}
\frac{\rho_{ki}}{\rho_{k \xa} \rho_{i \xa}}
\right \rangle_\xa
=\left \langle 
(1 + 2\ep g^{(2)}_{ki,\xa} )\frac{\rho_{ki}^{1-2\ep}}{\rho_{k \xa}^{1-2\ep} \rho_{i \xa }^{1-2\ep}}
\right \rangle_\xa, 
\\
& \left \langle 
\psi_\xa^{4\ep}
\frac{\rho_{ki}}{\rho_{k \xa} \rho_{i \xa}}
\left ( 
\frac{\rho_{kj}}{\rho_{k \xa} \rho_{j \xa}}
\right )^{\ep}
\right \rangle_\xa
=
\left \langle 
\left ( 
1 + 4 \ep g^{(4)}_{ki,\xa}
\right )
\frac{\rho_{ki}^{1-4\ep}}{\rho_{k \xa}^{1-4\ep}  \rho_{i \xa}^{1-4\ep}}
\left ( 
\frac{\rho_{kj}}{\rho_{k \xa} \rho_{j \xa}}
\right )^{\ep}
\right \rangle_\xa.
\end{split}
\label{eq7.7}
\ee
Using Eq.~(\ref{eq7.7}),  and focusing  on the \emph{divergent}
contributions that contain $g^{(2)}$ and $
g^{(4)}$,
we find 
\be
\Sigma_{\text{tcc}}^{\psi}
 \to -\frac{2 a_s^2 \pi}{\ep} 
\sum \limits_{(kij)} 
\kappa_{kj}
\left \langle 
\frac{\rho_{ik}}{\rho_{i \xa} \rho_{k \xa}} 
\left ( 
 g_{ki,\xa}^{(2)} 
-
g_{ki,\xa}^{(4)} 
\right )
\right \rangle_\xa 
 F^{kij} = {\cal O}(\ep^0),
\ee
where the 
last step follows from the equality  $g^{(2)} = g^{(4)}$ through  order ${\cal O}(\ep^0)$. 

Extending the above computation in 
a straightforward way, 
we  derive  the finite $N$-jettiness-dependent triple-color correlated remainder. We obtain
\be
\Sigma_{\text{tcc}}^{\psi}
 \to 2 a_s^2 \pi
\sum \limits_{(kij)} 
\kappa_{kj}
\left \langle 
\frac{\rho_{ki}}{\rho_{i \xa} \rho_{k \xa}} 
\ln \left ( 
\frac{\psi_\xa \rho_{ki}}{\rho_{k \xa} \rho_{i \xa}}
\right ) 
\ln \left ( 
\frac{
({\bar u}\mu)^2
\psi_\xa \rho_{i \xa } \rho_{k j} 
}{2 \rho_{j \xa} \rho_{k i}}
\right ) 
\right \rangle_\xa 
 F^{kij}.
\label{eq7.9}
\ee
\\

Next, we  discuss triple-color correlated 
contributions that  do not involve jettiness functions.  The 
first one is 
given in Eq.~(\ref{eq7.2}); 
it contains explicit $1/\ep^2$ divergence and 
does not require further discussion. 
The second contribution is 
extracted from Eq.~(\ref{eq7.5}) 
using Eq.~(\ref{eq7.7}).  The required angular integral 
is straightforward to compute and we do not discuss it further. 

It is slightly more  challenging to calculate 
a similar contribution 
to  $S_{RV,tc}$. 
However, we can 
cast it into a  form that 
allows us to expand
it 
in powers of $\ep$. The corresponding 
expression reads 
\be
\begin{split}
& \left \langle 
\frac{\rho_{ki}^{1-4\ep}}{\rho_{k \xa}^{1-4\ep}  \rho_{i \xa}^{1-4\ep}}
\left ( 
\frac{\rho_{kj}}{\rho_{k \xa} \rho_{j \xa}}
\right )^{\ep}
\right 
\rangle_\xa 
=
\left \langle 
\frac{\rho_{ki}^{1-4\ep}}{\rho_{k \xa}^{1-3\ep}  \rho_{i \xa}^{1-4\ep}}
\right 
\rangle_\xa 
+ 
\left \langle 
\frac{\rho_{ki}^{-\ep}}{ \rho_{i \xa}^{1-4\ep}}
\left ( 
\left ( 
\frac{\rho_{kj}}{\rho_{i j} }
\right )^{\ep}
- 1 \right )
\right 
\rangle_\xa
\\
& + 
\left \langle 
\frac{\rho_{ki}^{1-4\ep}}{\rho_{k \xa}^{1-3\ep}  \rho_{i \xa}^{1-4\ep}}
\left ( 
\left ( 
\frac{\rho_{kj}}{\rho_{j \xa}}
\right )^{\ep}
- 1 \right )
- 
\frac{\rho_{ki}^{-\ep}}{ \rho_{i \xa}^{1-4\ep}}
\left ( 
\left ( 
\frac{\rho_{kj}}{\rho_{i j} }
\right )^{\ep}
- 1 \right )
\right 
\rangle_\xa.
\end{split}
\label{eq7.12}
\ee
The first two integrals on the right-hand side are straightforward to compute but the last one is challenging. Since this integral  is  ${\cal O}(\ep)$, it contributes to the divergence of the triple-color correlated term so that we need to analytically compute it. 

The simplest way to do this is to use the results
of  Ref.~\cite{Devoto:2023rpv} where the 
analytic expression for the following integral 
\be
\left \langle 
\frac{\rho_{ki}}{\rho_{k \xa}  \rho_{i \xa}}
\left ( 
\frac{\rho_{kj}}{\rho_{k \xa} \rho_{j \xa}}
\right )^{\ep}
\right 
\rangle_\xa, 
\label{eq7.13}
\ee
was derived.  The idea is to rewrite the integral in Eq.~(\ref{eq7.13}) 
in the same way as  the integral 
in Eq.~(\ref{eq7.12}) and then 
take the difference of the two 
results. Upon doing that, 
one finds that 
the 
complicated finite integral 
appears at order ${\cal O}(\ep^2)$ only; this integral
can be computed  \emph{numerically}
alongside with 
$N$-jettiness dependent contributions. 

Putting everything together, we find that all divergences in triple-color correlated contributions cancel, and a  finite result 
is obtained. It reads 
\be
\begin{split} 
 \frac{1}{2}[Z_1, Z_1^+]
+
\frac{1}{2}
[S_1,Z_1-Z_1^+]
+ S_{RV,tc}
= a_s^2  \pi \sum \limits_{(kij)}
F^{kij} \; \kappa_{kj} G_{kij}^{\text{triple}},
\end{split} 
\ee
where 
\be
\begin{split} 
& G_{kij}^{\text{triple}} = 
 \Bigg[ \frac{8}{3} L_{ki}^3 + 4 L_{ki} \left({\rm Li}_2(1-\eta_{ki})+\frac{\pi ^2}{12}
 + \left \langle 
\frac{\rho_{ki}}{\rho_{i \xa} \rho_{k \xa} } \ln \left ( 
\frac{\psi_\xa \rho_{ki}}{\rho_{k \xa} \rho_{i \xa}}
\right ) 
 \right \rangle_{\xa} 
 \right) 
\\
& 
+2 \text{Li}_3(1-\eta_{ki})-6 \text{Li}_3(\eta_{ki})
+{\rm Li}_2(1-\eta_{ki}) (2 \ln 2-8 \ln (\eta_{ki})) - \ln ^3(\eta_{ki}) 
\\
& -3 \ln (1-\eta_{ki}) \ln ^2(\eta_{ki})
 -\ln 2 \ln^2 (\eta_{ki}) + \frac{\pi^2}{6} \ln (\eta_{ki})
-{\bar G}_{r,\rm fin}^{ikj} - 2 W_{kij}
\\
& + 2 \left \langle 
\frac{\rho_{ki}}{\rho_{i \xa} \rho_{k \xa}} 
\ln \left ( 
\frac{\psi_\xa \rho_{ki}}{\rho_{k \xa} \rho_{i \xa}}
\right ) 
\ln \left ( 
\frac{
\psi_\xa \rho_{i \xa } \rho_{k j} 
}{ \rho_{j \xa} \rho_{k i}^2}
\right ) 
\right \rangle_\xa +{\cal O}(\ep)\Bigg].
\end{split}
\label{eq7.15}
\ee
We note that the function  $L_{ki} $ is given in Eq.~(\ref{eq4.19}), the function 
$W_{kij}$ reads 
\be
W_{kij} =  \left \langle 
\frac{\rho_{ki}}{\rho_{i \xa} \rho_{k \xa}}
 \ln \frac{\rho_{kj}}{\rho_{j \xa}} \ln \frac{\rho_{k i} }{\rho_{i \xa} \rho_{k \xa} } + \frac{1}{\rho_{i \xa}} \ln \rho_{i \xa} 
 \ln \frac{\rho_{kj}}{\rho_{ij}}
 \right \rangle_\xa,
  \ee
and the function ${\bar G}_{r,\rm fin}^{ikj}$ can be found in Eq.~(H.16) 
of Ref.~\cite{Devoto:2023rpv}.

\section{The remaining contributions} 
\label{sect:8}

We are left with the discussion 
of the contributions to the soft function that are not iterations of  next-to-leading order terms, and do not contain triple-color correlators. We write them as 
\be
S_{2,RR,T^2} 
+ S_{RV,T^2} - Z_{2,r} - Z_{2,r}^+
-\frac{a_s \beta_0}{\ep} S_1,
\ee
where $S_{2,RR,T^2}$ is defined 
in Eq.~(\ref{eq5.6}) and 
\be
S_{RV,T^2,\tau} = \frac{[\alpha_s] 2^{-\ep}}{\ep^2} C_A A_K(\ep) 
 \sum \limits_{(ij)} {\bf T}_i \cdot {\bf T}_j \; I_{RV,ij},
\ee
is the part of the real-virtual contribution that depends on   
the products of two 
color-charge operators. 

The calculation of $I_{RV,ij}$
is straightforward. We obtain
\be
I_{RV,ij} = 
\frac{ [\alpha_s]
}{\tau^{1+4\ep}}
\left \langle
\; \psi_\xa^{4\ep}
\left ( 
 \frac{\rho_{ij}}{\rho_{i \xa } \rho_{j \xa}}
 \right )^{1+\ep}
\right \rangle_{\xa}.
\ee
Next, we perform the Laplace  transform and find
\be
S_{RV,T^2} = - \frac{[\alpha_s]^2 ( \bar u)^{4\ep}  e^{-4\gamma_E \ep } \Gamma(1-4\ep)}{2^{2+\ep} \; \ep^3} C_A A_K(\ep) 
 \sum \limits_{(ij)} {\bf T}_i \cdot {\bf T}_j
\left \langle
\psi_\xa^{4\ep}
\left ( 
 \frac{\rho_{ij}}{\rho_{i \xa } \rho_{j \xa}}
 \right )^{1+\ep}
\right \rangle_{\xa}.
\label{eq8.4}
\ee
This  contribution contains both 
explicit $1/\ep$
terms and  collinear divergences that 
arise after  integration over directions 
of the gluon $\xa$.  The latter 
can be isolated 
by  following what has been done for computing the soft function at next-to-leading order.  We also note that  terms that involve the jettiness function $\psi_\xa$ start contributing  at 
order $1/\ep^2$. 
We will discuss how the jettiness-dependent divergent 
contributions cancel out  once  we isolate 
similar terms in $S_{2,RR,T^2}$ in the next Section.

\section{Correlated emissions of two soft gluons: the strongly ordered limit}
\label{sect:9}

We consider  the remaining contribution to $S_{2,RR,T^2}$.
It reads 
\be
S_{2,RR,T^2,\tau}=
-\frac{C_A}{2} \sum 
\limits_{(ij)}
{\bf T}_i \cdot 
{\bf T}_j \;
I_{ij,\tau},
\ee
where 
\be
  I_{ij,\tau}  =  \frac{g_s^4}{2} \int [{\rm d} p_\xa ] [{\rm d} p_\yb] \; 
  \; \delta \left (\tau - E_\xa \psi_{\xa} - E_\yb \psi_{\yb}
   \right ) \tilde S_{ij}^{gg}(\xa,\yb).
\ee
Since the function $\tilde S_{ij}^{gg}$ is symmetric under the permutation of $\xa$ and $\yb$,  we can  introduce the 
energy-ordering $E_\yb < E_\xa$ ($E_\yb = \omega E_\xa$) and use the fact that the  scaling of $\tilde S_{ij}^{gg}$ with
the common gluon energy  ``scale'' 
$E_{\xa}$ is uniform. We then write 
\be
\begin{split} 
  I_{ij,\tau} & =  g_s^4 \int [{\rm d} p_\xa ] [{\rm d} p_\yb] \; \theta ( E_\xa - E_\yb)
  \; \delta \left (\tau - E_\xa \psi_{\xa} - E_\yb \psi_{\yb}
   \right ) \tilde S_{ij}^{gg}(\xa,\yb)
\\
&   =   [\alpha_s]^2 \; 
\; \int \limits_{0}^{1} \; \frac{{\rm d} \omega}{\omega^{1+2\ep} } \left \langle
   \int \limits_{0}^{\infty}  \frac{ {\rm d}E_{\xa}}{ E_{\xa}^{1+4\ep}} 
    \delta ( \tau - E_{\xa} \psi_{\xa \yb} ) \;
    \left [ 
    \omega^2 \; \tilde S^{gg}_{ij}(\xa,\yb) 
    \right ]_{E_{\xa} \to 1, E_{\yb} \to \omega}
    \right \rangle_{\xa \yb} 
    \\
  &   =  \frac{ [\alpha_s]^2}{\tau^{1+4\ep}}
     \int \limits_{0}^{1} \; \frac{{\rm d} \omega}{\omega^{1+2\ep} }
   \left \langle  \psi_{\xa \yb}^{4\ep}  \; \left [ 
   \omega^2 {\tilde S}^{gg}(\xa,\yb) 
   \right ]
   \right \rangle_{\xa \yb}, 
     \end{split}
\ee
where $\psi_{\xa \yb} = \psi_\xa + \omega \psi_{\yb}$. We note that from now on, for simplicity of notation, we will always assume that 
in the evaluation of 
${\tilde S}^{gg}(\xa,\yb)$ the 
energy of the gluon $\xa$ is set  to $1$ and the energy 
of the gluon $\yb$ is set to $\omega$.
Finally, we 
perform the Laplace transform and find
\be
I_{ij} 
= \frac{N_u}{\ep} 
     \int \limits_{0}^{1} \; \frac{{\rm d} \omega}{\omega^{1+2\ep} }
   \left \langle  \psi_{\xa \yb}^{4\ep}  \;
   \left [ 
   \omega^2 \tilde S_{ij}^{gg}(\xa,\yb) 
   \right ]
   \right \rangle_{\xa \yb}, 
   \label{eq9.4}
\ee
where 
\be
N_u = -
\frac{ [\alpha_s]^2 {\bar u }^{4\ep}
e^{-4 \ep \gamma_E} \Gamma(1-4\ep)}{4}.
\label{eq9.5}
\ee
\\

The representation of function 
$I_{ij}$ in 
Eq.~(\ref{eq9.4}) is the starting point of our analysis 
of the correlated emissions of two gluons.  
To compute $I_{ij}$ we
identify and subtract various singular limits 
of the integrand in Eq.~(\ref{eq9.4}). We begin with the $\omega \to 0$
limit which we will refer to as 
the  ``strongly-ordered'' one, 
in the  sense of the energy ordering. 
The strongly-ordered limit of ${\tilde S}^{gg}_{ij}$ evaluates to  
\be
\lim_{\omega \to 0}  \; \left [ \omega^2 \tilde S_{ij}^{gg}(\xa,\yb)
\right ] = 
\left [ 
  \frac{2 \rho_{ij}}{\rho_{\xa \yb} \rho_{ i \xa } \rho_{j \yb}}
  +\frac{2 \rho_{ij}}{\rho_{\xa \yb}
  \rho_{i\yb} \rho_{j\xa }} -\frac{2 \rho_{ij}^2}{\rho_{i \xa} \rho_{i\yb} \rho_{j\xa} \rho_{j \yb}}
\right ] = F_{ij}(\xa ,\yb).
\ee
We then need to compute
\be
\begin{split} 
I_{ij}^{\rm so} = 
\frac{N_u}{\ep}
     \int \limits_{0}^{1} \;  \frac{{\rm d} \omega}{\omega^{1+2\ep} } \;  \left \langle
   \left ( \psi_\xa + \omega \psi_\yb \right )^{4\ep}  \;
    F_{ij}(\xa,\yb) \right \rangle_{\xa \yb}.
     \end{split}
\ee
To  integrate over $\omega$, 
we change the integration variable  $\omega \to 1/\omega$ 
and find
\be
I_{ij}^{\rm so} = 
\frac{N_u}{\ep}
     \; \int \limits_{1}^{\infty} \; \frac{{\rm d} \omega}{\omega^{1+2\ep} } \left \langle 
   \left ( \omega \psi_\xa +  \psi_\yb \right )^{4\ep}  \;
    F_{ij}(\xa,\yb) \right \rangle_{\xa \yb}.
\ee
We then rename $\xa \leftrightarrow \yb$ and obtain
\be
I_{ij}^{\rm so} = 
\frac{N_u}{\ep}
\;
     \int \limits_{1}^{\infty} \; \frac{{\rm d} \omega}{\omega^{1+2\ep} } \left \langle 
   \left ( \psi_\xa + \omega \psi_\yb    \right )^{4\ep}  \;
    F_{ij}(\yb,\xa) \right \rangle_{\xa \yb}.
    \ee
    Since $F_{ij}(\yb,\xa) = F_{ij}(\xa,\yb)$, we finally find
    \be
    \begin{split} 
I_{ij}^{\rm so} & = 
\frac{N_u}{2\ep} 
    \; \int \limits_{0}^{\infty} \; \frac{{\rm d} \omega}{\omega^{1+2\ep} }
  \left \langle    \left ( \psi_\xa +  \omega \psi_\yb \right )^{4\ep}  \;
        F_{ij}(\xa,\yb) \right \rangle_{\xa \yb}
   \\
   & = 
-\frac{N_u \Gamma[1-2\ep]^2}{
2\ep^2 \Gamma[1-4\ep]}
    \left \langle    \psi_\xa^{2\ep} \; \psi_\yb^{2\ep}   \;  F_{ij}(\xa,\yb) \right \rangle_{\xa \yb}
   \\
& =
-\frac{N_u \Gamma[1-2\ep]^2}{
\ep^2 \Gamma[1-4\ep]}
   \left \langle   \psi_\xa^{2\ep} \; \psi_\yb^{2\ep}
   \left [ 
     \frac{2 \rho_{ij}}{\rho_{ \xa \yb} \rho_{i \xa } \rho_{j \yb}} -\frac{ \rho_{ij}^2}{\rho_{i \xa } \rho_{i \yb} \rho_{j \xa } \rho_{j \yb}}
     \right ] \right \rangle_{\xa \yb}.
     \label{eq9.10}
      \end{split} 
   \ee
   The last term in Eq.~(\ref{eq9.10}) is fully  factorized and can be computed in the same way as the uncorrelated contribution.
   We only need to understand  how to deal with the first term in 
   Eq.~(\ref{eq9.10}).

      To this end, we note that the angular integral   
   \be
   \left \langle \psi_\xa^{2\ep} \; \psi_\yb^{2\ep}
   \frac{\rho_{ij}}{\rho_{ \xa \yb} \;  \rho_{i \xa } \; \rho_{j \yb}}
   \right \rangle_{\xa \yb},
   \label{eq9.11}
   \ee
 is singular in the three
collinear limits 
$\xa || i$, $\yb ||j$ and $\xa || \yb$.  
To regulate the first two, 
we define the following  functions
   \be
   \left ( \frac{\psi_\xa}{\rho_{i \xa} }  \right )^{2\ep} = 1 + \ep f_{i \xa },\;\;\;\;
  \left ( \frac{\psi_\yb}{\rho_{j \yb} }  \right )^{2\ep}  = 1 + \ep f_{j \yb}.
  \label{eq9.12}
   \ee
Then, we  split the integral
into contributions with and without functions $f_{i \xa}$
and $f_{j \yb}$. This splitting 
introduces unregulated singularities in some integrals;
to address this issue, we 
modify the 
integral in Eq.~(\ref{eq9.11}) by incorporating an additional (analytic) regulator, and write it as
  \be
   \left \langle \psi_\xa^{2\ep} \; \psi_\yb^{2\ep}
   \frac{\rho_{ij}}{\rho_{ \xa \yb} \;  \rho_{i \xa }^{1+\nu}  \; \rho_{j \yb}^{1+\nu}}
   \right \rangle_{\xa \yb}.
   \label{eq9.13}
   \ee
Using the functions $f_{i \xa},
f_{j \yb}$ defined in 
Eq.~(\ref{eq9.12}), we rewrite it as follows
   \be
   \begin{split} 
      & \left \langle \psi_\xa^{2\ep} \; \psi_\yb^{2\ep}
   \frac{\rho_{ij}}{\rho_{ \xa \yb} \;  \rho_{i \xa }^{1+\nu} \; \rho_{j \yb}^{1+\nu}}
   \right \rangle_{\xa \yb}
   \\
     & = \left \langle
   \left ( \ep^2 f_{i \xa } f_{j \yb }  + \left ( \frac{\psi_\xa}{\rho_{i \xa} }  \right )^{2 \ep}
   +  \left ( \frac{\psi_\yb}{\rho_{j \yb} }  \right )^{2 \ep} - 1
   \right ) \frac{\rho_{ij}}{\rho_{ \xa \yb} \;  \rho^{1+\nu -2\ep}_{i \xa } \; \rho^{1+\nu-2\ep}_{j \yb}}
      \right \rangle_{\xa \yb}.
      \end{split}
      \label{eq9.14}
   \ee
We need to compute the above integral 
and take the limit $\nu \to 0$ \emph{before}
the expansion around $\ep = 0$ is performed. 
   
   It is clear that in the second, the third and the forth  terms in the right-hand side 
   in Eq.~(\ref{eq9.14}),  
   we can integrate over directions of 
   $\yb$ or directions of $\xa$ or directions of both.  We begin 
   with the computation of the 
   last integral in  the right-hand side 
   of Eq.~(\ref{eq9.14}) 
   where we first 
   integrate over $\yb$. The integration is straightforward 
   \cite{Somogyi:2011ir}
and we obtain 
   \be
   \begin{split}
 \left \langle  \frac{\rho_{ij}}{\rho_{ \xa \yb} \;  \rho^{1+\nu -2\ep}_{i \xa } \; \rho^{1+\nu-2\ep}_{j \yb}}
      \right \rangle_{\xa \yb} &
      =
\frac{2^{-\ep} \Gamma(-\ep) \Gamma(
  \ep - \nu) }{\Gamma(-\nu)}
  \\
& \times
\left \langle 
{}_2 F_{1}
(-\ep, \ep - \nu, 1 - \ep, 1 - \eta_{j \xa }) 
\frac{ \rho_{ij} } {
\rho_{i \xa}^{1 - 2 \ep + \nu}
  \rho_{j \xa}^{1 - \ep + \nu}}
  \right \rangle_\xa.
  \end{split}
  \label{eq9.15}
  \ee
Since 
   $1/\Gamma(-\nu) = -\nu/\Gamma(1-\nu)$,
this integral appears 
to be proportional to the analytic regulator. Because we are supposed 
to take the $\nu \to 0$ limit 
at fixed $\ep$, it is tempting 
to conclude that the whole 
integral vanishes. However, 
this conclusion is wrong 
because \emph{the remaining 
integral over directions of 
$\vec n_\xa$ is ill-defined  
for $\nu = 0$}. 
   
    To extract the $1/\nu$ singularity from the remaining integration over $\vec n_{\xa}$
    in Eq.~(\ref{eq9.15}) we note that it
   originates from a kinematic configuration where $\xa || j$. Hence, to  extract it,  we  take the collinear limit in all  quantities 
   of the integrand except 
   $1/\rho_{j \xa}^{1-\ep+\nu}$, whose integration produces the  $1/\nu$ singularity.
  Taking the $\nu \to 0$ limit, we find 
   \be
   \begin{split}
&  \lim_{\nu \to 0} \left \langle  \frac{\rho_{ij}}{\rho_{ \xa \yb} \;  \rho^{1+\nu -2\ep}_{i \xa } \; \rho^{1+\nu-2\ep}_{j \yb}}
      \right \rangle_{\xa \yb}
      =
-\frac{\eta_{ij}^{2\ep} \Gamma(1+\ep) \Gamma^3(1-\ep)}{\ep^2\Gamma(1-2\ep)}.
\end{split} 
     \ee
It is easy to convince oneself that the second and the third integrals 
in Eq.~(\ref{eq9.14}) cannot be rescued using this mechanism and just vanish, and 
the first integral does not 
need the analytic regulator 
since all divergences are dimensionally regularized.

Hence, we conclude that the angular integral needed 
to describe the strongly-ordered 
soft limit is given by the following equation
   \be
    \left \langle \psi_\xa^{2\ep} \; \psi_\yb^{2\ep}
   \frac{\rho_{ij}}{\rho_{ \xa \yb} \;  \rho_{i \xa } \; \rho_{j \yb}}
   \right \rangle_{\xa \yb}
        = \ep^2 \left \langle
   f_{i \xa} f_{j \yb }  \frac{\rho_{ij}}{\rho_{ \xa \yb} \;  \rho^{1-2\ep}_{i \xa } \; \rho^{1-2\ep}_{j \yb}}
      \right \rangle_{\xa \yb}
       +\frac{\eta_{ij}^{2\ep} \Gamma(1+\ep) \Gamma^3(1-\ep)}{\ep^2\Gamma(1-2\ep)}.
      \ee
 The usefulness of this representation stems from the fact that in the remaining integral on the right-hand side $\xa || i$ and $\yb || j$ collinear singularities are regulated       since $f_{i \xa }$ and $f_{j \yb }$ vanish in these limits. The singular limit which remains is the      $\xa || \yb$ one. We therefore  subtract the 
$\xa || \yb$ singularity and write\footnote{As we will see later, sometimes it is useful  to take the collinear limit of the phase space as well. If this is the case, we  indicate this explicitly.} 
\be
\begin{split}
\left \langle
   f_{i \xa} f_{j \yb }  \frac{\rho_{ij}}{\rho_{ \xa \yb} \;  \rho^{1-2\ep}_{i \xa } \; \rho^{1-2\ep}_{j \yb}}
      \right \rangle_{\xa \yb}
        & = \left \langle
  {\bar C}_{\xa \yb}  f_{i \xa} f_{j \yb }  \frac{\rho_{ij}}{\rho_{ \xa \yb} \;  \rho^{1-2\ep}_{i \xa } \; \rho^{1-2\ep}_{j \yb}}
      \right \rangle_{\xa \yb}
      \\
      & +
     \left \langle
  C_{\xa \yb}  f_{i \xa} f_{j \yb }  \frac{\rho_{ij}}{\rho_{ \xa \yb} \;  \rho^{1-2\ep}_{i \xa } \; \rho^{1-2\ep}_{j \yb}}
      \right \rangle_{\xa \yb},
      \end{split}
      \ee
where $\bar C_{\xa \yb} = 
I - C_{\xa \yb}$.
We then simplify the last term.  We write
\be
\begin{split}
 \left \langle
  C_{\xa \yb}  f_{i \xa} f_{j \yb }  \frac{\rho_{ij}}{\rho_{ \xa \yb} \;  \rho^{1-2\ep}_{i \xa } \; \rho^{1-2\ep}_{j \yb}}
      \right \rangle_{\xa \yb} 
     &  = 
      \left \langle
    f_{i \xa} f_{j \xa }  \frac{\rho_{ij}}{\rho_{ \xa \yb} \;  \rho^{1-2\ep}_{i \xa } \; \rho^{1-2\ep}_{j \xa}}
      \right \rangle_{\xa \yb} 
\\
     &  =
      -\frac{2^{-2\ep}}{\ep}
      \frac{\Gamma^2(1-\ep)}{\Gamma(1-2\ep)}
     \left \langle
    f_{i \xa} f_{j \xa }  \frac{\rho_{ij}}{  \rho^{1-2\ep}_{i \xa } \; \rho^{1-2\ep}_{j \xa}}
      \right \rangle_{\xa}. 
\end{split}
\ee
The final result for the  
integral $I_{ij}^{\rm so}$ 
defined in Eq.~(\ref{eq9.10}) reads
\be
\begin{split} 
   I_{ij}^{\rm so}
     & =  
     - \frac{N_u \; \Gamma(1-2\ep)^2}{\ep^2 \Gamma(1-4\ep)} \;  \Bigg [
          \ep^2  \left \langle  {\bar C}_{\xa \yb} \;   f_{i \xa} f_{j \yb} \;  \frac{2  \rho_{ij}}{\rho_{ \xa \yb} \;  \rho^{1-2\ep}_{i \xa } \; \rho^{1-2\ep}_{j \yb}}
     \right  \rangle_{\xa \yb}     
\\
&
+\frac{2\eta_{ij}^{2\ep} \Gamma(1+\ep) \Gamma^3(1-\ep)}{\ep^2\Gamma(1-2\ep)} -\frac{2^{1-2\ep}}{\ep} \frac{\Gamma^2(1-\ep)}{\Gamma(1-2\ep)} \;  \left \langle 
    \ep^2 \;  f_{i \xa} f_{j \xa} \; \frac{\rho_{ij}}{   \rho^{1-2\ep}_{i \xa } \;
     \rho^{1-2\ep}_{j \xa}} \right \rangle_{\xa}
\\
&
   - \left \langle \psi_{\xa} ^{2\ep} \frac{\rho_{ij}}{\rho_{i \xa } \rho_{j \xa} } \right \rangle_{\xa}
   \left \langle \psi_{\yb} ^{2\ep} \frac{\rho_{ij}}{\rho_{i \yb} \rho_{j \yb} } \right \rangle_{\yb} 
\Bigg  ].
\end{split}   
\label{eq9.20}
\ee
We use this result to write 
the strongly-ordered 
contribution to the uncorrelated real-emission 
part of the $N$-jettiness soft function as follows 
\be
\begin{split}
 S^{\rm so}_{2,RR,T^2}
& = 
\frac{C_A  N_u \Gamma^2(1-2\ep)}
{2 \ep^2 \; \Gamma(1-4\ep) }
\sum \limits_{(ij)} {\bf T}_i \cdot {\bf T}_j 
 \;  \Bigg [
           \ep^2 \left \langle  {\bar C}_{\xa \yb}  \;  f_{i \xa} f_{j \yb} \;  \frac{2  \rho_{ij}}{\rho_{ \xa \yb} \;  \rho^{1-2\ep}_{i \xa } \; \rho^{1-2\ep}_{j \yb}}
     \right  \rangle_{\xa \yb}
\\
&
 +\frac{2 \eta_{ij}^{2\ep} \Gamma(1+\ep) \Gamma^3(1-\ep)}{\ep^2\Gamma(1-2\ep)}
-\frac{2^{1-2\ep}}{\ep} \frac{\Gamma^2(1-\ep)}{\Gamma(1-2\ep)} \;  \left \langle 
  \; \ep^2  \; f_{i \xa} f_{j \xa} \;  \frac{\rho_{ij}}{   \rho^{1-2\ep}_{i \xa } \;
     \rho^{1-2\ep}_{j \xa}} \right \rangle_{\xa}
     \\
  &  - \left \langle \psi_{\xa} ^{2\ep} \frac{\rho_{ij}}{\rho_{i \xa} \rho_{j \xa} } \right \rangle_{\xa}
   \left \langle \psi_{\yb} ^{2\ep} \frac{\rho_{ij}}{\rho_{i \yb} \rho_{j \yb} } \right \rangle_{\yb}
   \Bigg ].
   \end{split}
   \label{eq9.21}
\ee

We continue with the 
discussion of how to extract divergences from  $S^{\rm so}_{2,RR,T^2}$.
 We 
note that the first term in the sum 
on the r.h.s. in Eq.~(\ref{eq9.21}) is already finite 
because functions $f_{\xa i} \sim f_{\yb j} \sim {\cal O}(1)$
vanish in the collinear limits
$\xa || i$, $\yb || j$, respectively  and 
the singularity at $\xa || \yb$ is regulated by  ${\bar C}_{\xa \yb}$
operator.
Since the $\ep$-dependence  
of the second term 
on the right-hand side in Eq.~(\ref{eq9.21})
is  explicit, we need to 
focus on the remaining 
two terms.  We rewrite 
them to make their dependencies on the $N$-jettiness constraint explicit.  We find 
\be
\begin{split}
& \left \langle 
\ep^2 f_{i \xa} f_{j \xa} 
\frac{\rho_{ij}}{\rho_{i\xa}^{1-2\ep} \rho_{j \xa}^{1-2\ep}}
\right \rangle_\xa
= 
\left \langle 
 \frac{\rho^{1-4\ep}_{ij}}{\rho_{i\xa}^{1-4\ep} \rho_{j \xa}^{1-4\ep}}
 +
 \frac{\rho_{ij}}{\rho_{i\xa}^{1-2\ep} \rho_{j \xa}^{1-2\ep}}
-
 2 \frac{\rho_{ij}^{1-2\ep}}{\rho_{i\xa}^{1-2\ep} \rho_{j \xa}^{1-4\ep}}
\right \rangle_\xa
\\
& + 4\ep 
\left \langle 
g_{ij,\xa}^{(4)}
\frac{\rho_{ij}^{1-4\ep}}{\rho_{i\xa}^{1-4\ep} \rho_{j \xa}^{1-4\ep}}
-
g_{ij,\xa}^{(2)}
\frac{\rho_{ij}^{1-2\ep}}{\rho_{i\xa}^{1-2\ep} \rho_{j \xa}^{1-4\ep}}
\right \rangle_\xa, 
    \end{split}
\ee
and 
\be
\begin{split}
& \left \langle \psi_{\xa} ^{2\ep} \frac{\rho_{ij}}{\rho_{\xa i} \rho_{\xa j} } \right \rangle_{\xa}
   \left \langle \psi_{\yb} ^{2\ep} \frac{\rho_{ij}}{\rho_{\yb i} \rho_{\yb j} } \right \rangle_{\yb} 
   = 
   \left \langle \frac{\rho_{ij}^{1-2\ep}}{\rho^{1-2\ep}_{\xa i} \rho^{1-2\ep}_{\xa j} } \right \rangle_{\xa}^2
 \\
  &  +
  4\ep \left \langle 
  g_{ij,\xa}^{(2)}
  \frac{\rho^{1-2\ep}_{ij}}{\rho_{\xa i}^{1-2\ep} \rho_{\xa j}^{1-2\ep} } \right \rangle_{\xa}
   \left \langle  \frac{\rho^{1-2\ep}_{ij}}{\rho_{\yb i}^{1-2\ep} \rho^{1-2\ep}_{\yb j} } \right \rangle_{\yb} 
   +
   4\ep^2 \left \langle 
  g_{ij,\xa}^{(2)}
  \frac{\rho^{1-2\ep}_{ij}}{\rho_{\xa i}^{1-2\ep} \rho_{\xa j}^{1-2\ep} } \right \rangle_{\xa}^2.
  \end{split}
   \ee
\\
   When these equations are used in Eq.~(\ref{eq9.21}), higher order terms in the 
   expansion of functions $g^{(4)}$ and $g^{(2)}$
   in $\ep$ become necessary.   Fortunately, it is  easy to compute them 
    since 
   \be
g^{(k)}_{ij,\xa} = \frac{1}{k \epsilon} \left [ \left ( \frac{\psi_{\xa} \rho_{ij}}{\rho_{i \xa} \rho_{j \xa}} \right )^{k \ep} - 1 \right ].
\label{eq9.24}
   \ee
Furthermore, since these functions regulate all divergences 
in the angular  integrals where 
they appear, Eq.~(\ref{eq9.24}) 
can be expanded in 
powers of $\ep$ \emph{before  integration over directions 
of the momentum of gluon 
$\xa$  is performed}. Comparing the 
divergent 
contributions of \emph{all $\psi$-dependent terms}
in Eq.~(\ref{eq9.21}) with similar contributions 
to the 
real-virtual soft limit in 
Eq.~(\ref{eq8.4}), we find that they  cancel.
The remaining strongly-ordered contributions 
are accounted for when the final result for the 
soft function is assembled. 

\section{Collinear subtractions}
\label{sect:10}

We continue with the discussion 
of the uncorrelated contribution to $N$-jettiness soft function. 
Our starting point is Eq.~(\ref{eq9.4}). We have computed  the \emph{strongly-ordered}
contribution to that equation and we will 
consider the remaining contribution 
in this section. Hence, we write 
\be
{\bar S}_\omega [ I_{ij}  ] = \frac{N_u}{\ep}
\int \limits_{0}^{1} \frac{{\rm d} \omega }{ \omega^{1+2\ep}}
\left \langle \psi_{\xa \yb}^{4\ep} \;  
{\bar S}_\omega
\left [ \omega^2 {\tilde S}^{gg}_{ij
(\xa, \yb) } \right ]
\right \rangle_{\xa \yb},
\label{eq10.1}
 \ee
 where $N_u$ can be found 
 in Eq.~(\ref{eq9.5}), 
 ${\bar S}_\omega = I - S_\omega$
and  $S_\omega$ is the operator that 
 enforces the strongly-ordered limit.  Furthermore, 
 the 
 eikonal function in Eq.~(\ref{eq10.1}) should be evaluated assuming that the 
 four-momentum of the gluon 
 $\xa$ is $(1,\vec n_\xa)$ 
 and the four-momentum of the gluon 
 $\yb$ is $\omega (1,\vec n_\yb)$.
 
The integrand 
in Eq.~(\ref{eq10.1}) possesses 
a complicated singularity structure. We separate double-collinear 
and triple-collinear singularities 
introducing partition functions (see e.g. 
Ref.~\cite{Devoto:2023rpv})
and write
 \be
{\bar S}_\omega[ I_{ij} ]
= {\bar S}_\omega[ I_{ij}^{dc} ] + {\bar S}_\omega [ I_{ij}^{tc}] ,
 \ee
where the first term describes the  contribution of the double-collinear 
 and the second one -- of the triple-collinear partition.  We will discuss them in turn.  

 We begin with the double-collinear contribution. It reads 
 \be
 \begin{split}
  {\bar S}_\omega [  I_{ij}^{dc}
  ]=
   \frac{N_u}{\ep} \int \limits_{0}^{1}  \frac{{\rm d} \omega}{ \omega^{1+2\ep}}
  \left \langle \; \psi_{\xa \yb}^{4 \ep} 
\left ( w^{\xa i, \yb j} + 
w^{\yb i, \xa j} \right )
{\bar  S}_\omega
\left [ \omega^2 {\tilde S}^{gg}_{ij}(\xa, \yb) \right ]
\right \rangle_{\xa \yb} .
\label{eq10.3}
 \end{split}
 \ee
 By construction, the partitions ensure that the only  allowed singularities
 are $\xa || i$,  $\yb || j$, and $\xa || j$ and $\yb || i$, respectively. However,
 it is easy to see  that the integrand is not singular in these limits.
 Hence, we can expand $\psi_{\xa \yb}^{4\ep}$ in powers of  $\ep$
 in Eq.~(\ref{eq10.3}). The first term of this expansion 
  corresponds
 to the contribution of double-collinear partitions to the soft-subtracted integral of $\omega^2 {\tilde S}^{gg}_{ij}$ \emph{without
 the $N$-jettiness constraint}.  
 Since this term is multiplied 
 by $1/\ep$, we need to compute it analytically if we want to show the cancellation of all $\ep$-singular terms without relying on numerical calculations. 
 We will show 
 that it is possible to extract this 
 contribution from the result 
of Ref.~\cite{Caola:2018pxp},
so that its computation is not required. 
   The jettiness-dependent 
   term  that arises in the 
   expansion of 
   Eq.~(\ref{eq10.3}) in powers of $\ep$,
 contains a factor $4 \ep \ln \psi_{\xa \yb}$  and, thus, 
   only provides an $\ep$-finite
 contribution that we evaluate numerically.  
 \\
 
The term $I_{ij}^{tc}$ is a triple-collinear contribution. In this case  we write 
\be
{ \bar S}_\omega [ I_{ij}^{tc}] =
 \frac{N_u}{\ep}
\int \limits_{0}^{1} \frac{{\rm d} \omega }{ \omega^{1+2\ep}}
\left \langle 
\psi_{\xa \yb}^{4 \ep} \;
w^{tc} \;
{\bar S}_\omega \left [ \omega^2 {\tilde S}^{gg}_{ij}(\xa,\yb)\right ]
\right \rangle_{\xa \yb}, 
\label{eq10.4}
 \ee
where 
\be
w^{tc}
= w^{\xa i, \yb i} + w^{\xa j, \yb j}.
\ee

To facilitate the computation of ${\bar S}_\omega[I_{ij}^{tc}]$, 
we introduce the 
standard sectors to 
further partition angular 
integrations, see e.g. 
Refs.~\cite{Czakon:2010td,Czakon:2011ve,Caola:2017dug,Devoto:2023rpv}.  We also  assume that when we subtract $\xa || \yb$ singularity, we need to simplify the phase space but a similar simplification is not needed for the triple-collinear limit. 
Finally, we note that there are no double-collinear 
$\xa || i$, $\yb || i$ and similar singularities. Hence,  we only need to subtract triple-collinear and $\xa || \yb$ double-collinear singularities 
in the integrand in Eq.~(\ref{eq10.4}).
With this in mind, we write\footnote{As is customary, c.f. Refs.~\cite{Caola:2017dug,Devoto:2023rpv}, we use the convention that 
collinear operators act on everything that appears to the 
right of them.  The angular phase space appears to the right 
of $C_{\xa \yb}$ and to the left 
of $C_{x \xa \yb}$ indicating 
that the first operator acts on it and the second does not.}
\begin{align}
     & {\bar S}_{\omega}[I_{ij}^{tc}]
  = \frac{N_u}{\ep}
  \int \limits_{0}^{1}\frac{{\rm d} \omega}{\omega^{1+2\ep}} \left \langle C_{\xa \yb} \left [
    {\rm d} \Omega_{\xa \yb} \right ]  \theta^{b+d } w^{tc} \psi_{\xa \yb}^{4 \ep}
 {\bar S}_\omega \left [ w^2 \tilde S_{ij}^{gg}(\xa, \yb) \right ] \right \rangle_{\xa \yb}
 \nonumber 
\\
& 
 +\frac{N_u}{\ep}
 \sum \limits_{x \in \{i,j\}}
\int \limits_{0}^{1}\frac{{\rm d} \omega}{\omega^{1+2\ep}}  \left \langle (1 - \theta^{b+d}  C_{\xa \yb} )
     \left [ {\rm d} \Omega_{\xa \yb} \right ] C_{x \xa \yb}  w^{tc}
\psi_{\xa \yb}^{4 \ep}
{\bar S}_\omega\left [ w^2 \tilde S_{ij}^{gg}(\xa, \yb) \right ] \right \rangle_{\xa \yb}  \label{eq10.6}
  \\
& 
+\frac{N_u}{\ep}  \sum \limits_{x \in \{i,j\}}
  \int \limits_{0}^{1}\frac{{\rm d} \omega}{\omega^{1+2\ep}}  \left \langle (1 - \theta^{b+d}  C_{\xa \yb} )
       \left [ {\rm d} \Omega_{\xa \yb} \right ] {\bar  C}_{x \xa \yb} \; w^{\xa x, \yb x} \psi_{\xa \yb}^{4 \ep}
{\bar S}_\omega\left [ w^2 \tilde S_{ij}^{gg}(\xa, \yb) \right ] \right \rangle_{\xa \yb},
\nonumber 
\end{align} 
where ${\bar  C}_{x \xa \yb} = 
I - C_{x \xa \yb}$ and $[{\rm d} \Omega_{\xa \yb} ]
= [{\rm d} \Omega_{\xa } ] [{\rm d} \Omega_{ \yb} ]
$.
We expect that the first term 
in Eq.~(\ref{eq10.6}) contains an 
${\cal O}(1/\ep^3)$ singularity, the second term an 
${\cal O}(1/\ep^2)$ one and the last  term   is  
${\cal O}(1/\ep)$ because its collinear and soft singularities  are fully subtracted.
 \\

 Since the $1/\ep$ divergence 
 in the last term in Eq.~(\ref{eq10.6}) arises from a
prefactor, we can expand  the integrand in powers of $\ep$. 
In particular, we apply the expansion to the  function $\psi_{\xa \yb}^{4\ep}$.  The term 
with $\ln \psi_{\xa \yb}$ is ${\cal O}(\ep^0)$
and the term
without it provides an 
${\cal O}(1/\ep )$ divergent 
contribution. However, this contribution is 
the same as in the case when no $N$-jettiness constraint is imposed.  To make this more explicit, 
we combine double- and triple-collinear 
partitions, and write 
\begin{align*}
    & {\bar S}_{\omega}[I_{ij}]
  = \frac{N_u}{\ep}
  \int \limits_{0}^{1}\frac{{\rm d} \omega}{\omega^{1+2\ep}} \left \langle C_{\xa \yb} \left [
    {\rm d} \Omega_{\xa \yb} \right ]  \theta^{b+d } w^{tc} \psi_{\xa \yb}^{4 \ep}
 {\bar S}_\omega \left [ w^2 \tilde S_{ij}^{gg}(\xa, \yb) \right ] \right \rangle_{\xa \yb}
\\
& 
 +\frac{N_u}{\ep}
 \sum \limits_{x \in \{i,j\}}
\int \limits_{0}^{1}\frac{{\rm d} \omega}{\omega^{1+2\ep}}  \left \langle (1 - \theta^{b+d}  C_{\xa \yb} )
     \left [ {\rm d} \Omega_{\xa \yb} \right ] C_{x \xa \yb}  w^{tc}
\psi_{\xa \yb}^{4 \ep}
{\bar S}_\omega\left [ w^2 \tilde S_{ij}^{gg}(\xa, \yb) \right ] \right \rangle_{\xa \yb}
  \\
& 
+\frac{N_u}{\ep}  \sum \limits_{x \in \{i,j\}}
  \int \limits_{0}^{1}\frac{{\rm d} \omega}{\omega^{1+2\ep}}  \left \langle (1 - \theta^{b+d}  C_{\xa \yb} )
       \left [ {\rm d} \Omega_{\xa \yb} \right ] {\bar  C}_{x \xa \yb} 
       w^{x \xa, x \yb}\;
{\bar S}_\omega\left [ w^2 \tilde S_{ij}^{gg}(\xa, \yb) \right ] \right \rangle_{\xa \yb}
\\
& 
+\frac{N_u}{\ep} 
  \int \limits_{0}^{1}\frac{{\rm d} \omega}{\omega^{1+2\ep}}  \;
  \left \langle 
  \left ( 
       w^{\xa i, \yb j} 
       + 
    w^{\xa j, \yb i}  \right )    
       \;
{\bar S}_\omega\left [ w^2 \tilde S_{ij}^{gg}(\xa, \yb) \right ] \right \rangle_{\xa \yb} \numberthis \label{eq10.7}
\\
& +4 N_u \sum \limits_{x \in \{i,j\}}
  \int \limits_{0}^{1}\frac{{\rm d} \omega}{\omega^{1+2\ep}}  \left \langle (1 - \theta^{b+d}  C_{\xa \yb} )
       \left [ {\rm d} \Omega_{\xa \yb} \right ] {\bar  C}_{x \xa \yb}  \;
       w^{x \xa, x \yb} \;
       \ln \psi_{\xa \yb} \;
{\bar S}_\omega\left [ w^2 \tilde S_{ij}^{gg}(\xa, \yb) \right ] \right \rangle_{\xa \yb}
\\
& +4 N_u \sum \limits_{x \in \{i,j\}}
  \int \limits_{0}^{1}\frac{{\rm d} \omega}{\omega^{1+2\ep}}  \left \langle 
  ( 
w^{i \xa, j \yb}
+w^{i \yb, j \xa}
  )
  \ln \psi_{\xa \yb} \;
{\bar S}_\omega\left [ w^2 \tilde S_{ij}^{gg}(\xa, \yb) \right ] \right \rangle_{\xa \yb}.
\end{align*}
In the above formula, the first, the second,  
the third and the fourth terms are divergent. Our strategy 
will be to compute the first and the second terms explicitly, 
and extract the third and the fourth terms   from the result in 
Ref.~\cite{Caola:2018pxp} making 
 use of the fact that 
they do not 
depend on the $N$-jettiness 
function $\psi_{\xa \yb}$. 
 
To make this connection explicit, we write a formula that describes the extraction of singularities of 
the integral of the double-eikonal function in a situation when  \emph{no $N$-jettiness constraint is imposed}. This  quantity,
computed in Ref.~\cite{Caola:2018pxp}, can 
be cast into the following form 
\be
J_{ij} = \frac{N_E}{\ep}
\int \limits_{0}^{1}
\frac{{\rm d} \omega}{\omega^{1+2\ep}}
\left \langle \omega^2 {\tilde S}_{ij}(\xa, \yb ) 
\right \rangle_{\xa \yb},
\ee
where 
\be
N_E = -\frac{ E_{\rm max}^{-4\ep} [\alpha_s]^2}{4}.
\ee
We can now write an expression for $J_{ij}$, where the 
 various singularities are extracted. We obtain 
\be
\begin{split} 
  J_{ij}
  &= \frac{N_E}{\ep} 
  \int \limits_{0}^{1} 
  \frac{{\rm d} \omega}{\omega^{1+2\ep}}
  \left \langle S_\omega \left [ \omega^2 {\tilde S}_{ij}(\xa, \yb ) \right ]
\right \rangle_{\xa \yb}
\\
& +
 \frac{N_E}{\ep}
  \int \limits_{0}^{1}\frac{{\rm d} \omega}{\omega^{1+2\ep}} \left \langle C_{\xa \yb} \left [
    {\rm d} \Omega_{\xa \yb} \right ]  \theta^{b+d } w^{tc} 
 {\bar S}_\omega \left [ w^2 \tilde S_{ij}^{gg}(\xa, \yb) \right ] \right \rangle_{\xa \yb}
\\
& 
 +\frac{N_E}{\ep}
 \sum \limits_{x \in \{i,j\}}
\int \limits_{0}^{1}\frac{{\rm d} \omega}{\omega^{1+2\ep}}  \left \langle (1 - \theta^{b+d}  C_{\xa \yb} )
     \left [ {\rm d} \Omega_{\xa \yb} \right ] C_{x \xa \yb}  w^{tc}
{\bar S}_\omega\left [ w^2 \tilde S_{ij}^{gg}(\xa, \yb) \right ] \right \rangle_{\xa \yb}
  \\
& 
+\frac{N_E}{\ep}  \sum \limits_{x \in \{i,j\}}
  \int \limits_{0}^{1}\frac{{\rm d} \omega}{\omega^{1+2\ep}}  \left \langle (1 - \theta^{b+d}  C_{\xa \yb} )
       \left [ {\rm d} \Omega_{\xa \yb} \right ] {\bar  C}_{x \xa \yb} 
       \;  w^{\xa x, \yb x} 
{\bar S}_\omega\left [ w^2 \tilde S_{ij}^{gg}(\xa, \yb) \right ] \right \rangle_{\xa \yb}
\\
& 
+\frac{N_E}{\ep} 
  \int \limits_{0}^{1}\frac{{\rm d} \omega}{\omega^{1+2\ep}} 
  \; \left \langle 
  \left ( 
       w^{\xa i, \yb j} 
       + 
    w^{\xa j, \yb i}  \right )    
       \;
{\bar S}_\omega\left [ w^2 \tilde S_{ij}^{gg}(\xa, \yb) \right ] \right \rangle_{\xa \yb}.
\end{split}   
\label{eq10.10}
\ee
Since the left-hand side of the above equation 
was computed  through ${\cal O}(\ep^0)$ in Ref.~\cite{Caola:2018pxp}, 
we can use this result to determine the last two  terms that appear 
on  the right-hand side in 
Eq.~(\ref{eq10.10}) and 
\emph{also} enter Eq.~(\ref{eq10.7}). This is 
a sensible thing to do because their
 computation appears to be particularly challenging.
Clearly, to make this happen, we have to compute all the 
other terms  
that appear in 
Eqs.~(\ref{eq10.7},\ref{eq10.10}) through finite terms in  the expansion in $\ep$. We describe details of the required computations in what follows.

\subsection{The double-collinear 
\texorpdfstring{$\xa || \yb $}{m || n} contribution with 
\texorpdfstring{$N$}{N}-jettiness constraint}
\label{subsect:10.1}

To proceed, we need to compute the $\xa || \yb$ limit of the soft-subtracted eikonal function ${\tilde S}_{ij}$, 
given by the first term in Eq.~(\ref{eq10.7}).
In principle, this is straightforward, except for the fact that the $\xa || \yb$ singularity appears to be
stronger than logarithmic. Although this is not the case, 
it implies that the corresponding contribution needs to be carefully extracted. We find 
   \be
   \begin{split}
     \\
    C_{\xa \yb} \oS_\omega \; \left [ \omega^2 {\tilde S}^{gg}_{ij}(\xa,\yb) \right ] & = 
     \frac{4 \omega \rho_{ij} }{ (1 + \omega)^2 \rho_{\xa \yb} \rho_{i \xa} \rho_{j \xa} }
     \Bigg  \{ -2
\\
    & 
     +  \frac{(1-\ep ) \omega \rho_{i \xa} \rho_{j \xa} }{(1+\omega)^2  \rho_{ij}}
       \left ( \frac{  n_j \kappa_\yb }{\rho_{j \xa}} - \frac{ n_i \kappa_{\yb}}{\rho_{i \xa} } 
   \right )^2
   \Bigg  \},
   \end{split} 
   \ee
   where the three-vector $\kappa_\yb$ is defined through the following equation
   \be
\vec n_{\yb} = \cos \theta_{\xa \yb} \vec n_{\xa} + \sin  \theta_{\xa \yb} \; \vec \kappa_{\yb}, 
   \ee
   with the additional constraints 
   \be
\vec n_{\xa} \cdot \vec \kappa_{\yb}  = 0,\;\;\; {\vec \kappa_{\yb}}^2 = 1.
   \ee
   
 We also need the collinear limit of the phase space in $b$- 
 and $d$-sectors. The corresponding parametrization 
 can be found in 
Refs.~\cite{Caola:2017dug,Devoto:2023rpv}.
 We obtain 
   \be
   \begin{split} 
   & C_{\xa \yb} \;  [ {\rm d} \Omega_{\xa \yb} ] \; \theta^{(b)} \; w^{\xa i, \yb i}
\cdots 
\\
   & = N_\ep^{(b,d)} \; w_{\xa || \yb}^{\xa i, \yb i} \; \eta_{i \xa}^{-\ep} \; (1 - \eta_{i \xa } )^{\ep} \;
   \left [ {\rm d} \Omega_{\xa}^{(d-1)} \right ] \;  \left [ \rho_{\xa \yb} \frac{{\rm d} x_4}{x_4^{1+2 \ep} }
     \frac{[{\rm d} \Omega_a^{(d-3)}] }{[ \Omega^{(d-3),a} ]} {\rm d} \Lambda \right ]\; C_{\xa \yb} \cdots, 
   \end{split} 
   \label{eq10.14}
   \ee
   where
   \be
N_\ep^{(b,d)} = \frac{\Gamma(1-\ep) \Gamma(1+2\ep) }{\Gamma(1+\ep) }.
\label{eq10.15}
   \ee

As the next step, we  need to integrate over azimuthal angles 
parameterized by  ${\rm d} \Omega_a^{(d-3)}$
and ${\rm d} \Lambda$ in Eq.~(\ref{eq10.14}).
 To facilitate this, 
   we write
   \be
\left ( \frac{ ( n_j \kappa_\yb )}{\rho_{j \xa}} - \frac{ (n_i \kappa_{\yb})}{\rho_{i \xa} } 
   \right )^2 = v_\mu v_\nu \kappa_\yb^\mu \kappa_\yb^\nu,
   \label{eq10.16}
   \ee
   where $v^\mu=(n_i^\mu/\rho_{i \xa}-n_j^\mu/\rho_{j \xa})$, and the four-vector $\kappa$ satisfies
   \be
\kappa_\mu t^\mu = 0, \;\;\;\; \kappa_\mu e_\xa^\mu = 0,
\ee
where $t^\mu = (1,\vec 0)$
and $e_\xa^\mu = (0, \vec n_\xa)$.
The averaging of $\kappa^\mu \kappa^\nu$ over
$ [{\rm d} \Omega_a^{(d-3)}]/[{\rm d} \Omega^{(d-3),a} ] \; {\rm d} \Lambda $ gives 
\be
\langle \kappa^\mu \kappa^\nu \rangle = -\frac{1}{2} g_{\perp, d-2}^{\mu \nu}+\ep r_i^\mu r_i^\nu.
\ee
The  derivation of this formula as well as the definition of vector $r_i^\mu$ are given in Appendix~F in Ref.~\cite{Devoto:2023rpv}.

To further simplify this expression, we write the  metric tensor of the transverse space as follows  
\be
g_{\perp, d-2}^{\mu \nu} = g^{\mu \nu} - t^\mu t^\nu + e_\xa^\mu e_{\xa}^\nu.
\ee
Since
\be
n_\xa^\mu v_\mu = 0, 
\ee
we find
\be
t_\mu v^\mu = -e_{\xa,\mu} v^\mu. 
\ee
This allows us to write 
\be
v_\mu v_\nu \;
\langle \kappa^\mu \kappa^\nu \rangle = 
v_\mu v_\nu \left ( -\frac{1}{2} g^{\mu \nu}_{\perp, d-2} +\ep r_i^\mu r_i^\nu \right ) =
-\frac{1}{2} v_\mu v^\mu+ \ep (r_i \cdot  v)^2. 
\ee
A straightforward computation gives
\be
v_\mu v^\mu = -\frac{2\rho_{ij}}{\rho_{i \xa} \rho_{j \xa}}, 
\ee
and
\be
(r_i v)^2 = \frac{( \rho_{ij} -  \rho_{i \xa} + \rho_{j \xa})^2}{(2 - \rho_{i \xa} ) 
\rho_{i \xa} \rho_{j \xa}^2}
 = 2 \left ( \frac{1}{\rho_{i \xa}}
 + \frac{1}{\rho_{j \xa}}
 \right )
 + W_{ij}^{\xa}.
\ee
We note that the function $W_{ij}^{\xa}$ 
develops  a singularity in the limit $\xa || j$
but this singularity is regulated 
by the partition function.

Finally, we note that the contributions of sectors $b$ and $d$ are equal.   Integrating over
$x_4$ and adding the two sectors, we obtain 
\be
\begin{split} 
\Sigma^{(i)}_{\xa||\yb} &= \frac{N_u}{\ep}
  \int \limits_{0}^{1}
  \frac{{\rm d} \omega  }{\omega^{1+2\ep} }\left \langle C_{\xa \yb} [ {\rm d} \Omega_{\xa \yb} ]\theta^{b+d } w^{\xa i, \yb i} \psi_{\xa \yb}^{4\ep}
\left [ w^2 \tilde S_{ij}^{gg}(\xa, \yb) \right ] \right \rangle_{\xa \yb}
\\
& = -\frac{N_u N_\ep^{b,d}}{  \ep^2} 
\Bigg  \langle
\frac{ \eta^{-\ep}_{i \xa}}{ (1-\eta_{i \xa} )^{-\ep} }
\; \psi_{\xa}^{4 \ep} \;
w_{\xa|| \yb}^{\xa i, \yb i}
\left [ \gamma_g \;A^{(i)}
+ \ep  \delta_g \;B^{(i)}
  \right ]
\Bigg  \rangle_{\xa} .
\end{split}
\label{eq10.25a}
\ee
In Eq.~(\ref{eq10.25a}), we introduced 
the following quantities
\be
    A^{(i)}=\; \frac{\rho_{ij}}{\rho_{i \xa} \rho_{j \xa} },\quad \quad
    B^{(i)}= \left ( 2 \frac{\rho_{i \xa} + \rho_{j \xa}}{\rho_{i\xa} \rho_{j \xa}}+ W_{ij}^{\xa}\right ),
\ee
and the analogs of the cusp anomalous dimension
\be
\begin{split}
\gamma_g  & = \int \limits_{0}^{1}
\frac{{\rm d} \omega }{\omega^{1+2\ep} }\frac{4 \omega}{(1+\omega)^{2-4\ep}} 
\left ( -2 +   \frac{(1-\ep) \omega}{(1+\omega)^2} \right )
= 
-\frac{11}{3}-\frac{137}{9}\ep
\\
& 
+ \left(\frac{22 \pi ^2}{9}-\frac{1646}{27}\right) \ep^2
+ \left(\frac{176}{3}\zeta_3 
+\frac{274 }{27}\pi^2 
-\frac{19760}{81}\right) \ep^3 +{\cal O}(\ep^4),
\\
\delta_g & = (1-\ep)\int \limits_{0}^{1}
\frac{{\rm d} \omega }{\omega^{1+2\ep} } \frac{4 \omega^2}{(1+\omega)^{4-4\ep}}
= \frac{1}{3} + 
\frac{7}{9} \ep
+ \left ( \frac{82}{27} - \frac{2 \pi^2}{9}
\right )\ep^2+{\cal O}(\ep^3) .
\end{split}
\ee
\\

It is convenient to write the result for the sum of $i$- and 
$j$-partitions.  To do this in an optimal way, we need to rearrange  contributions 
that appear in Eq.~(\ref{eq10.25a}).  The 
key point in this 
rearrangement is to 
make the divergent contribution symmetric 
with respect to  the replacement $i \leftrightarrow j$.  We illustrate 
how this is  done by considering the 
term proportional to $\gamma_g$ in Eq.~(\ref{eq10.25a}). 
We write
\begin{align*}
& \Bigg  \langle
\eta^{-\ep}_{i \xa} (1-\eta_{i \xa} )^{\ep} 
\; \psi_{\xa}^{4 \ep} 
w_{\xa|| \yb}^{\xa i, \yb i}
\; \frac{\rho_{ij}}{\rho_{i \xa} \rho_{j \xa} } 
\Bigg \rangle_{\xa } \\
&=  2^\ep\Bigg  \langle
 \psi_{\xa}^{4 \ep} 
w_{\xa|| \yb}^{\xa i, \yb i}
\; \frac{\rho_{ij}^{1+\ep}}{\rho_{i \xa}^{1+\ep} \rho_{j \xa}^{1+\ep} }\; 
(1-\eta_{i \xa} )^{\ep} \left ( 
\frac{\eta_{j \xa} }{\eta_{ij}}
\right )^\ep
\Bigg \rangle_{\xa }
= 2^\ep\Bigg  \langle
 \psi_{\xa}^{4 \ep} 
w_{\xa|| \yb}^{\xa i, \yb i}
\; \frac{\rho_{ij}^{1+\ep}}{\rho_{i \xa}^{1+\ep} \rho_{j \xa}^{1+\ep} }
\Bigg \rangle_{\xa } \numberthis \label{eq10.28} \\ 
& + 2^\ep\Bigg  \langle
 \psi_{\xa}^{4 \ep} 
w_{\xa|| \yb}^{\xa i, \yb i}
\; \frac{\rho_{ij}^{1+\ep}}{\rho_{i \xa}^{1+\ep} \rho_{j \xa}^{1+\ep} }\; 
\left [
(1-\eta_{i \xa} )^{\ep} \left ( 
\frac{\eta_{j \xa} }{\eta_{ij}}
\right )^\ep - 1
\right ]
\Bigg \rangle_{\xa}
,
\end{align*}
and observe that the first term 
in the right-hand side of Eq.~(\ref{eq10.28}) 
 can be easily combined with the  
$\xa || j, \yb || j$ partition and the 
last term is finite. Therefore, it  can be computed by expanding  in $\ep$ before integration.
Note that in this term the 
$\psi_\xa$-dependent contribution appears 
at order ${\cal O}(\ep^2)$ which, given 
the $1/\ep^2$ prefactor in Eq.~(\ref{eq10.25a}) is the last (finite)  order in the  $\ep$-expansion that we care about.

Using these considerations, 
we  combine 
contributions of the $\xa || i, \yb || i$
and $\xa || j, \yb || j$ partitions  and obtain 
\be
\Sigma^{(i+j)}_{\xa || \yb} = 
-\frac{N_u N_\ep^{b,d}}{\ep^2}
\left [ 
\gamma_g A^{(ij)} + \epsilon \delta_g B^{(ij)}
\right ].
\label{eq10.29}
\ee
We note that we have used 
\be
w^{\xa i, \yb i}_{\xa || \yb}
+
w^{\xa j, \yb j}_{\xa || \yb}
=1.
\ee
The term $A^{ij}$ reads
\be
\begin{split}
A^{(ij)}
&= 2^\ep 
\left \langle 
\left ( 1+ 4\ep g_{ij,\xa}^{(4)}
\right )
\frac{\rho_{ij}^{1-3\ep}}{\rho_{i \xa}^{1-3\ep} \rho_{j \xa}^{1-3\ep}}
\right \rangle 
\\
& +
2^\ep
\left \langle 
\frac{\rho_{ij}^{1+\ep}}{\rho_{i \xa}^{1+\ep} \rho_{j \xa}^{1+\ep}}
w^{\xa i, \yb i}_{\xa || \yb}
\psi_{\xa}^{4 \ep}
\left [ 
\left ( \frac{\eta_{j \xa}}{\eta_{ij}} 
\right )^\ep ( 1- \eta_{i \xa})^{\ep}-1
\right ]
\right \rangle 
\\
& +
2^\ep
\left \langle 
\frac{\rho_{ij}^{1+\ep}}{\rho_{i \xa}^{1+\ep} \rho_{j \xa}^{1+\ep}}
w^{\xa j, \yb j}_{\xa || \yb}
\psi_{\xa}^{4 \ep}
\left [ 
\left ( \frac{\eta_{i \xa}}{\eta_{ij}} 
\right )^\ep ( 1- \eta_{j \xa})^{\ep}-1
\right ]
\right \rangle. 
\end{split}
\label{eq10.30}
\ee
As we already mentioned, the 
second and third terms 
on the right-hand side 
of Eq.~(\ref{eq10.30})
can be expanded in $\ep$  because   expressions 
in square brackets protect  the integrands from  developing 
 collinear singularities. 
 Furthermore, since the 
 expressions in square brackets start contributing 
 at 
${\cal O}(\ep)$,  the 
$N$-jettiness dependence that arises from  factors 
$\psi_\xa^{4 \ep}$, 
only appears in  ${\cal O}(\ep^0)$  
contributions to  
the  soft function.

Next, we  discuss the 
$B^{(ij)}$ 
term in Eq.~(\ref{eq10.29}). In this case terms with $W^m_{ij}$
provide integrable contributions. 
Taking this into account and repeating the steps
we employed to rewrite $A^{ij}$, we find  
\be
\begin{split}
B^{(ij)}
&= 2^{1+\ep} 
\left \langle 
\left ( 1+ 4\ep g_{ij,\xa}^{(4)}
\right )
\frac{\rho_{ij}^{-3\ep}
(\rho_{i \xa} + \rho_{j \xa})}{\rho_{i \xa}^{1-3\ep} \rho_{j \xa}^{1-3\ep}}
\right \rangle 
\\
& +
2^{1+\ep}
\left \langle 
\frac{\rho_{ij}^{\ep}(\rho_{i \xa} + \rho_{j \xa}) }{\rho_{i \xa}^{1+\ep} \rho_{j \xa}^{1+\ep}}
\omega^{\xa i, \yb i}_{\xa || \yb}
\psi_{\xa}^{4 \ep}
\left [ 
\left ( \frac{\eta_{j \xa}}{\eta_{ij}} 
\right )^\ep ( 1- \eta_{i \xa})^{\ep}-1
\right ]
\right \rangle 
\\
& +
2^{1+\ep}
\left \langle 
\frac{\rho_{ij}^{\ep}( 
\rho_{i \xa} + \rho_{j \xa} )}{\rho_{i \xa}^{1+\ep} \rho_{j \xa}^{1+\ep}}
\omega^{\xa j, \yb j}_{\xa || \yb}
\psi_{\xa}^{4 \ep}
\left [ 
\left ( \frac{\eta_{i \xa}}{\eta_{ij}} 
\right )^\ep ( 1- \eta_{j \xa})^{\ep}-1
\right ]
\right \rangle
\\
& + 
\left \langle \eta_{i \xa}^{-\ep}
(1-\eta_{i \xa})^{\ep} 
\psi_\xa^{4 \ep} \omega_{\xa || \yb}^{\xa i, \yb i} W_{ij}^{\xa}
\right \rangle 
+
\left \langle \eta_{j \xa}^{-\ep}
(1-\eta_{j \xa})^{\ep} 
\psi_\xa^{4 \ep} \omega_{\xa || \yb}^{\xa j, \yb j} W_{ji}^{\xa}
\right \rangle. 
\label{eq10.31}
\end{split}
\ee
It follows from Eq.~(\ref{eq10.29}) that 
$B^{ij}$  is needed 
through ${\cal O}(\ep)$
to compute the relevant 
contributions to 
the soft function. Therefore, for 
the analysis of the \emph{divergent}
contributions to the soft function,
in Eq.~(\ref{eq10.31}) we require the first term without $g^{(4)}$ and the last term 
at $\ep = 0$.

\subsection{The triple-collinear subtraction term
with \texorpdfstring{$N$}{N}-jettiness constraint}

The second term 
we require is the  triple-collinear limit of the eikonal function subject to the $N$-jettiness constraint. 
The corresponding equation 
reads 
\be
 \frac{N_u}{\ep}
 \sum \limits_{x \in \{i,j\}}
\int \limits_{0}^{1}\frac{{\rm d} \omega}{\omega^{1+2\ep}}  \left \langle (1 - \theta^{b+d}  C_{\xa \yb} )
     \left [ {\rm d} \Omega_{\xa \yb} \right ] C_{x \xa \yb} \; w^{tc} \;
     \psi_{\xa \yb}^{4 \ep}
     \;
     {\bar S}_\omega \;
\left [ w^2 \tilde S_{ij}^{gg}(\xa, \yb) \right ] \right \rangle_{\xa \yb}.
\ee
There are four contributions
to the above equation -- 
the $i$- and $j$-triple-collinear 
partitions and  
the $\xa || \yb$ subtractions 
terms  applied to  the 
triple-collinear limit to 
remove sub-divergences. 
The calculation of all these terms is similar, so that we take 
the $i$-triple-collinear partition and discuss first the 
triple-collinear limit. 

For further reference,  we define 
\be
\Sigma^{(i, tc)}
= 
\frac{N_u}{\ep}
 \int \limits_{0}^{1}\frac{{\rm d} \omega}{\omega^{1+2\ep}}  
 \left \langle 
      (\rho_{i \xa} + \omega \rho_{i \yb})^{4 \ep} \;
 C_{i \xa \yb} \; {\bar S}_\omega \;\left [ w^2 \tilde S_{ij}^{gg}(\xa, \yb) \right ] \right \rangle_{\xa \yb},
\ee
where we have used 
\be
C_{i \xa \yb} w^{tc} =1, 
\;\;\; 
C_{i \xa \yb} \; \psi_{\xa \yb}^{4 \ep}
= ( \rho_{i \xa} + \omega \rho_{i \yb}  )^{4 \ep}.
\ee
It follows from this equation that 
in the triple-collinear limits the dependence on the $N$-jettiness function disappears since
we only need its limiting expression.

To proceed, we  require the triple-collinear limit of the soft-subtracted eikonal function.  It is straightforward to obtain it; we find 
   \be
   \begin{split}
          C_{i \xa \yb} \oS_\omega \; \left [ \omega^2 {\tilde S}_{ij} \right ] & = 
     \frac{2 \omega^2 (1-\ep)  (\rho_{i\xa}-\rho_{i\yb})^2}{(\omega+1)^2 \rho_{\xa\yb}^2 (\rho_{i\xa} + \omega \rho_{i\yb})^2}
     -\frac{\omega \left(\rho_{i\xa}^2+6 \rho_{i\yb} \rho_{i\xa}+\rho_{i\yb}^2\right)}{(\omega+1) \rho_{\xa\yb} \rho_{i\xa}
     \rho_{i\yb} ( \rho_{i\xa} + \omega \rho_{i\yb})}
\\
&
+\frac{\omega (\rho_{i\xa}+\rho_{i\yb} )}{(\omega+1) \rho_{i\xa} \rho_{i\yb} (\rho_{i\xa}  +  \omega \rho_{i\yb} )}.
   \end{split} 
   \label{eq10.35}
   \ee
We then write 
\be
\Sigma^{(i,tc)} = \frac{N_u}{\ep} 
B^{(i,tc)}_{gg},
\ee
where 
\be
\begin{split}
B^{(i, tc)}_{gg} & = 
\int \limits_{0}^{1} \frac{{\rm d} \omega }{ \omega^{1+2\ep}}
\Bigg  \langle \frac{1}{
(\rho_{i \xa }+ \omega \rho_{i \yb})^{1-4\ep}}
\Bigg [
 \frac{2 \omega^2 (1-\ep)  (\rho_{i\xa}-\rho_{i\yb})^2}{(\omega+1)^2 \rho_{\xa\yb}^2 (\rho_{i\xa} + \omega \rho_{i\yb})}
    \\
 & -\frac{\omega \left(\rho_{i\xa}^2+6 \rho_{i\yb} \rho_{i\xa}+\rho_{i\yb}^2\right)}{(\omega+1) \rho_{\xa\yb} \rho_{i\xa}
     \rho_{i\yb}}
    +\frac{\omega (\rho_{i\xa}+\rho_{i\yb} )}{(\omega+1) \rho_{i\xa} \rho_{i\yb} }
    \Bigg ]
\Bigg  \rangle_{\xa \yb}.
\label{eq10.38}
\end{split} 
\ee
To simplify the calculation of this quantity 
we use the (partial) symmetry of the 
integrand under $\omega \to 1/\omega$ transformation as well as the invariance of the integration measure under  
$\xa \leftrightarrow \yb $, 
to extend the integration over $\omega$ in 
Eq.~(\ref{eq10.38}) to $\omega = \infty$ and 
to simplify the integrand. We find 
\be
\begin{split}
B^{(i, tc)}_{gg} & = 
\int \limits_{0}^{\infty} \frac{{\rm d} \omega }{ \omega^{1+2\ep}}
\Bigg  \langle \frac{1}{
(\rho_{i \xa }+ \omega \rho_{i \yb})^{1-4\ep}}
\Bigg [
 \frac{2 \omega^2 (1-\ep)  \rho_{i \xa} (\rho_{i\xa}-\rho_{i\yb})}{(\omega+1)^2 \rho_{\xa\yb}^2 (\rho_{i\xa} + \omega \rho_{i\yb})}
    \\
    & -\frac{\omega \left(\rho_{i\xa}+3 \rho_{i\yb} \right)}{(\omega+1) \rho_{\xa\yb} 
     \rho_{i\yb}}
    +\frac{\omega }{(\omega+1)  \rho_{i\yb} }
    \Bigg ]
\Bigg  \rangle_{\xa \yb}.
\label{eq10.39}
\end{split} 
\ee
\\

It turns out that it is relatively straightforward to compute $B^{(i,tc)}_{gg}$. To do that, it is convenient to  employ  a Mellin-Barnes representation \cite{Dubovyk:2022obc} for 
$1/(\rho_{i \xa} + \omega \rho_{i \yb})^{1-4\ep}$  and then integrate over $\omega$ and the 
directions of $\xa$ and $\yb$.
We note  that this 
requires the  introduction of a 
second Mellin-Barnes representation for integrals that involve $\rho_{\xa \yb}$.  This is typically done using a particular  Mellin-Barnes representation for the hypergeometric function
\be
\begin{split}
{}_2 F_{1}(a,b,c,x) 
&= \frac{\Gamma(c)}{\Gamma (c-b) \Gamma (a) \Gamma (b)\Gamma (c-a)}
\\
& \times \int 
\frac{{\rm d} z }{2 \pi i}
\; \Gamma (-z) \Gamma (a+z) \Gamma (b+z) \Gamma (c-a-b-z)\; (1-x)^z.
\label{eq10.40}
\end{split}
\ee

The two-dimensional Mellin-Barnes integrals that one obtains are  relatively simple as one integration can always be performed using the second Barnes lemma. The remaining integration is either performed 
using the first Barnes lemma or by mapping appropriate products of 
Gamma-functions onto 
parametric integrals using   Euler's representation of a  Beta function~\cite{Dubovyk:2022obc}.
To perform the parametric integrals 
we use {\sf Hyperint}~\cite{Panzer:2014caa}.
We find 
\be
\begin{split} 
\Sigma^{i,tc} = 
\frac{N_u}{\ep }B^{i,tc}_{gg} & = 
\frac{N_u}{\ep} 
\Bigg [ \frac{11}{6 \ep^2} + \frac{1}{\ep} \left ( \frac{73}{18} + \frac{\pi^2}{6} \right )
+3 \zeta_3 +\frac{11 \pi ^2}{18}+\frac{217}{27} 
\\
& 
  +\ep \left( \frac{73 \pi ^2}{54} -\frac{22\zeta_3}{3}+\frac{7 \pi ^4}{45}+\frac{1298}{81}\right)
  +{\cal O}(\ep^2)
  \Bigg ].
\end{split} 
  \ee
\\

We continue with the discussion 
of the double-collinear subtraction of the triple-collinear contribution.  Focusing on the $i$-collinear partition, 
the corresponding expression reads
\be
\Sigma^{i,tc}_{dc} = 
\frac{N_u}{\ep}
 \int \limits_{0}^{1}\frac{{\rm d} \omega}{\omega^{1+2\ep}}   \left \langle \theta^{b+d} \; C_{\xa \yb} 
     \left [ {\rm d} \Omega_{\xa \yb} \right ] C_{i \xa \yb} \;
\psi_{\xa \yb}^{4 \ep}
\; {\bar S}_\omega \;
\left [ w^2 \tilde S_{ij}^{gg}(\xa, \yb) \right ] \right \rangle_{\xa \yb}.
\ee

We now discuss how to compute the required limits. Our starting point is 
Eq.~(\ref{eq10.35}) where 
we are supposed to take 
$\xa || \yb $ limit; this limit 
is made specific through  
the parametrization of scalar 
products in $b$ and $d$ sectors.  The challenge in taking  the limit is related to the fact that $1/\rho_{\xa \yb}^2$ appears in the first term  in Eq.~(\ref{eq10.35}) which is too strong a singularity and so an expansion 
around the limit is needed. Fortunately, this can be avoided if we recall that in $b$
and $d$ sectors 
the following equation holds 
(c.f. Refs.~\cite{Devoto:2023rpv,Caola:2017dug})
\be
\frac{(\rho_{i \xa} - \rho_{i \yb})^2}{\rho_{\xa \yb}}
= 2 x_3 \; N(x_3,1-x_4/2,\lambda).
\label{eq10.43}
\ee
Here   $N$ is the function 
that appears in the phase-space parametrization of 
$b$ and $d$ sectors
\cite{Caola:2017dug,Devoto:2023rpv}.
For the purposes of computing  the double-collinear limit of a triple-collinear limit, we need to take  
$x_4 \to 0$ in Eq.~(\ref{eq10.43}). Upon doing that, we find~\cite{Devoto:2023rpv,Caola:2017dug}
\be
\lim_{x_4 \to 0} \frac{(\rho_{i \xa} - \rho_{i \yb})^2}{\rho_{\xa \yb}} =
8 x_3 (1-x_3) \lambda 
 = 2 \rho_{i \xa} (2 - \rho_{i \xa} ) \; \lambda. 
\ee
If we combined the above expression 
together with Eq.~(\ref{eq10.14}),  we easily obtain 
\be
\begin{split} 
& \theta^{b} \; C_{\xa \yb} 
     \left [ {\rm d} \Omega_{\xa \yb} \right ] C_{i \xa \yb} \;
\psi_{\xa \yb}^{4 \ep}
\;
{\bar S}_\omega \;
\left [ w^2 \tilde S_{ij}^{gg}(\xa, \yb) \right ] 
=  N_\ep^{(b,d)} [{\rm d} \Omega_{\xa}^{(d-1)}]
\frac{{\rm d} x_4}{x_4^{1+2\ep}}
\\
& \times 
\frac{ \eta_{i \xa}^{-\ep}(1-\eta_{i \xa})^{\ep} }{\rho_{i \xa}^{1-4\ep}} 
\frac{4 \omega}{(1+\omega)^{2-4\ep}}
\left ( -2
+ \frac{\omega (1-\ep) (1+2\ep)}{(1+\omega)^2} \left (1 - \eta_{i \xa} \right ) 
\right ).
\end{split}
\ee
Note that we have used 
$\int {\rm d} \Lambda \; \lambda =(1+2\ep)/2$, see  Refs.~\cite{Devoto:2023rpv,Caola:2017dug} for details. 
We also note that the contribution of the sector $d$  is identical. 

The rest of the calculation is 
straightforward.  Integrating 
over directions of $\xa$ 
and over the energy parameter $\omega$, and combining contributions 
of sectors $b$ and $d$ we find 
\be
\begin{split}
\Sigma^{i,tc}_{dc}
&= -\frac{N_u N_\ep^{(b,d)}}{\ep^2}
\Bigg  [ 
-\frac{11}{6 \ep}
-\frac{137}{18}-\frac{11 \ln 2}{3}
+\ep \Big (\frac{11 \pi ^2}{9}-\frac{823}{27}
\\
& -\frac{11}{3} \ln^2 2-\frac{137 \ln 2}{9}\Big )
+ \ep^2 \Big (\frac{88 \zeta_3}{3}+\frac{137 \pi ^2}{27}-\frac{9880}{81}
\\
& -\frac{22}{9} \ln^3 2-\frac{137 \ln^2 2}{9}+\frac{22}{9} \pi ^2 \ln 2-\frac{1646 \ln 2}{27}\Big )
 +{\cal O}(\ep^3) \Bigg  ].
\end{split}
\ee
The quantity $N_\ep^{(b,d)}$
is defined in Eq.~(\ref{eq10.15}).

Combining the different results, we obtain
\be
\begin{split}
 & \frac{N_u}{\ep}
 \sum \limits_{x \in \{i,j\}}
\int \limits_{0}^{1}\frac{{\rm d} \omega}{\omega^{1+2\ep}}  \left \langle (1 - \theta^{b+d}  C_{\xa \yb} )
     \left [ {\rm d} \Omega_{\xa \yb} \right ] C_{x \xa \yb} \; w^{tc} \;
\psi_{\xa \yb}^{4 \ep} \;
{\bar S}_\omega \; 
\left [ w^2 \tilde S_{ij}^{gg}(\xa, \yb) \right ] \right \rangle_{\xa \yb}
\\
& = N_u \Bigg[\frac{1}{\ep^2} 
\left ( \frac{\pi ^2}{6}-\frac{32}{9}-\frac{11}{3} \ln 2
\right )
+\frac{1}{\ep} 
\left ( 3 \zeta_3+\frac{11 \pi ^2}{9}-\frac{202}{9}-\frac{11}{3} \ln^2 2
-\frac{137 \ln 2}{9}
\right )
\\
& +\frac{77}{3} \zeta_3 +\frac{182 \pi ^2}{45}
-\frac{8582}{81}-\frac{22}{9}  \ln^3 2
-\frac{137 \ln ^2 2}{9}
+\left(\frac{11}{9} \pi ^2 -\frac{1646}{27}\right)\ln 2 +{\cal O}(\ep) \Bigg].
\end{split}
\ee
As  expected, we observe the cancellation 
of the leading ${\cal O}(1/\ep^3)$
singularities between the triple-collinear contribution and its  
double-collinear limit, indicating that the subtraction of $\xa || \yb$ subdivergences 
was properly performed.

\section{Correlated emissions without 
\texorpdfstring{$N$}{N}-jettiness constraint}
\label{sect:11}

As we mentioned earlier, we would like 
to use the result of Ref.~\cite{Caola:2018pxp} 
to deduce certain contributions to the 
$N$-jettiness soft function. The 
corresponding expression is given 
in Eq.~(\ref{eq10.10}); our goal is  
to determine the sum of the fourth and the fifth terms in 
the right-hand side 
of that equation   
from the known result for $J_{ij}$.
It is clear that this is only possible if all other terms on the right-hand side of Eq.~(\ref{eq10.10}) are known.  
Below we describe the calculation of the strongly-ordered, 
double-collinear $\xa || \yb$ and the triple-collinear limits 
of the correlated emissions without the $N$-jettiness constraint. 

\subsection{Strongly-ordered limit without \texorpdfstring{$N$}{N}-jettiness constraint}

We begin by considering the following integral
\be
J_{ij}^{\rm so} = 
\frac{N_E}{\ep} \int \limits_{0}^{1} \frac{{\rm d} \omega}{\omega^{1+2\ep}} \left \langle S_\omega [ \omega^2 {\tilde S}_{ij}^{gg}(\xa,\yb) ] \right \rangle_{\xa \yb}. 
\ee
Using the result for the soft limit of the eikonal 
function ${\tilde S}_{ij}^{gg}$, we find 
\be
J_{ij}^{\rm so} = 
-\frac{N_E}{\ep^2}
\left \langle 
\frac{2 \rho_{ij} }{\rho_{\xa \yb} \rho_{i \xa} \rho_{j \yb} }
- \frac{\rho_{ij}^2}{\rho_{i\xa} \rho_{j \xa} \rho_{i \yb} \rho_{j \yb} }
\right \rangle_{\xa \yb}.
\label{eq11.2}
\ee

The two terms that need to be integrated in Eq.~(\ref{eq11.2})
are quite different. The second one is easy to compute because 
integrations over directions of $\xa$ and $\yb$ factorize. 
For each of the integrals we use  
 \be
\left  \langle 
 \frac{\rho_{ij} }{\rho_{i \yb} \rho_{j \yb}}
 \right \rangle_{\yb}
 = -\frac{2^{1-2\ep} \;  \eta_{ij}^{-\ep}}{\ep} K_{ij},
  \ee
  where
  \be
K_{ij} = \frac{\Gamma(1-\ep)^2}{\Gamma(1-2\ep)} {}_2 F_{1}(-\ep,-\ep,1-\ep,1-\eta_{ij}).
\label{eq11.4}
  \ee

Integration of the first term 
on the right-hand side in 
Eq.~(\ref{eq11.2}) requires more effort. 
We first use the 
result in Eq.~(\ref{eqb.5}) to integrate 
over $\yb$, then write  the 
obtained hypergeometric function
using  the representation  in Eq.~(\ref{eq10.40}) and integrate 
over directions of $\xa$. This gives 
another hypergeometric function for which we introduce Mellin-Barnes 
representation again. Finally, we find 
\be
\begin{split}
\left \langle 
\frac{\rho_{ij}}{\rho_{\xa \yb} \rho_{i \xa} \rho_{j \yb}}
\right \rangle_{\xa \yb}
& = \frac{\Gamma(1-\ep)^2}{\Gamma(-2\ep)} 
\int  \frac{{\rm d} z_2 } {2 \pi i} \; 
\frac{ \rho_{ij}^{1+z_2} }{ 2^{1+z_2+4 \ep}} \; 
\Gamma (-z_2) \Gamma (1+z_2)
\\
& 
\times \int \frac{ {\rm d} z_1} {2 \pi i} \; 
\frac{  \Gamma (-z_1)  \Gamma (z_1-\ep) \Gamma^2(1+z_1+z_2) \Gamma( -1-\ep-z_1-z_2) }{\Gamma (1-2 \ep+z_1+z_2)}.
\end{split}
\ee
The integration over $z_1$ can be performed with the help 
of the  Barnes' second  lemma. Once this is done, the resulting 
integral can be transformed to a parametric integral using 
Euler's integral representation of Beta-functions.  This gives 
\be
\begin{split}
& 
\left \langle 
\frac{\rho_{ij}}{\rho_{\xa \yb} \rho_{i \xa} \rho_{j \yb}}
\right \rangle_{\xa \yb}
=  -\frac{6 \; \eta_{ij} \;  2^{-4 \ep} 
\Gamma^5(1-\ep)}{\ep \ \Gamma (1-3 \ep) \Gamma (1-2 \ep)}
\int \limits_{0}^{1} {\rm d} x \; 
\frac{x^{-1-2 \ep} {}_2 F_{1}(-2\ep,-\ep,-3\ep,x) }{
x + (1-x)\eta_{ij} },
\end{split}
\ee
where $\eta_{ij} = \rho_{ij}/2$.
The resulting integral over $x$ is straightforward to compute 
using the program  {\sf Hyperint} \cite{Panzer:2014caa}.

Performing the integration,  combining the results for the two integrals 
that appear in Eq.~(\ref{eq11.2}) and expanding 
 in powers of $\ep$, we find 
\be
\begin{split} 
 J_{ij}^{\rm so}  & = [\alpha_s]^2 \; \frac{E_{\rm max}^{-4\ep} \; 2^{-4 \ep}  }{2 \ep^4 }
  \Bigg [
1 -2 \ep \ln (\eta_{ij}) +\ep^2 \left(4 \text{Li}_2(1-\eta_{ij})+2 \ln ^2(\eta_{ij})-\frac{2 \pi ^2}{3}\right)
\\
& + \ep^3 \Big ( 4 \text{Li}_3(1-\eta_{ij})+12 \text{Li}_3(\eta_{ij})+4 \text{Li}_2(1-\eta_{ij}) \ln (\eta_{ij})
-\frac{4}{3}  \ln^3(\eta_{ij})
\\
& +6 \ln (1-\eta_{ij}) \ln ^2(\eta_{ij})-\frac{2}{3} \pi ^2 \ln (\eta_{ij})-26 \zeta_3 \Big ) 
+\ep^4 \Big ( -\frac{19}{45}\pi^4 + \frac{2}{3} \ln^4 (\eta_{ij}) 
\\
& - \frac{2}{3} \ln^3 (\eta_{ij}) \ln (1-\eta_{ij}) + 10 \ln^2 (\eta_{ij}) \ln^2 (1-\eta_{ij}) - \frac{10}{3} \pi^2 \ln (\eta_{ij}) \ln (1-\eta_{ij})
\\
& + 2 [\pi^2+2 \ln (\eta_{ij}) \ln (1-\eta_{ij}) + \ln^2 (\eta_{ij})] \text{Li}_2(\eta_{ij}) + 2 \text{Li}_2^2(\eta_{ij})-28 \text{Li}_4(\eta_{ij})
\\
& + 4 \text{Li}_4(1-\eta_{ij})+ 32 S_{2,2} (\eta_{ij}) + 4 S_{2,2} (1-\eta_{ij})+ 32 \ln (1-\eta_{ij}) [\text{Li}_3(\eta_{ij})-\zeta_3]
\\
&+ 4 \ln (\eta_{ij}) [\text{Li}_3(\eta_{ij}) + 6 \text{Li}_3(1-\eta_{ij}) + 5 \zeta_3] \Big)+{\cal O}(\ep^5)
\Bigg ].
\end{split}
\ee
In addition to standard polylogarithms, also the 
Nielsen polylogarithm 
$S_{2,2}(x)$ appears in the above formula.

\subsection{Double-collinear limit without \texorpdfstring{$N$}{N}-jettiness constraint}
The easiest contribution to compute is the 
double-collinear $\xa || \yb$ subtraction term since we can borrow the calculation from Section~\ref{subsect:10.1}  nearly verbatim.  We then find 
\be
\begin{split}
 \frac{N_E}{\ep} &
  \int \limits_{0}^{1}\frac{{\rm d} \omega}{\omega^{1+2\ep}} \left \langle C_{\xa \yb} \left [
    {\rm d} \Omega_{\xa \yb} \right ]  \theta^{b+d } \omega^{tc} 
 {\bar S}_\omega \left [ w^2 \tilde S_{ij}^{gg}(\xa, \yb) \right ] \right \rangle_{\xa \yb}
 \\
& = 
-\frac{N_E N_\ep^{b,d}}{\ep^2}
\left [ 
\gamma_{gJ} A_J^{(ij)} + \epsilon \delta_{gJ} B_J^{(ij)}
\right ].
\end{split}
\ee
The various quantities in this expression differ from  similar 
quantities when the  $N$-jettiness constraint is imposed. In particular, 
\be
\begin{split}
A_J^{(ij)}
&= 2^\ep 
\left \langle 
\frac{\rho_{ij}^{1+\ep}}{\rho_{i \xa}^{1+\ep} \rho_{j \xa}^{1+\ep}}
\right \rangle_{\xa \yb}
+
2^\ep
\left \langle 
\frac{\rho_{ij}^{1+\ep}}{\rho_{i \xa}^{1+\ep} \rho_{j \xa}^{1+\ep}}
\omega^{\xa i, \yb i}_{\xa || \yb}
\left [ 
\left ( \frac{\eta_{j \xa}}{\eta_{ij}} 
\right )^\ep ( 1- \eta_{i \xa})^{\ep}-1
\right ]
\right \rangle 
\\
& +
2^\ep
\left \langle 
\frac{\rho_{ij}^{1+\ep}}{\rho_{i \xa}^{1+\ep} \rho_{j \xa}^{1+\ep}}
\omega^{\xa j, \yb j}_{\xa || \yb}
\left [ 
\left ( \frac{\eta_{i \xa}}{\eta_{ij}} 
\right )^\ep ( 1- \eta_{j \xa})^{\ep}-1
\right ]
\right \rangle,
\end{split}
\ee
and 
\be
\begin{split}
B_J^{(ij)}
&= 2^{1+\ep} 
\left \langle 
\frac{\rho_{ij}^{\ep}
(\rho_{i \xa} + \rho_{j \xa})}{\rho_{i \xa}^{1+\ep} \rho_{j \xa}^{1+\ep}}
\right \rangle_{\xa \yb}
\\
& +
2^{1+\ep}
\left \langle 
\frac{\rho_{ij}^{\ep}(\rho_{i \xa} + \rho_{j \xa}) }{\rho_{i \xa}^{1+\ep} \rho_{j \xa}^{1+\ep}}
\omega^{\xa i, \yb i}_{\xa || \yb}
\left [ 
\left ( \frac{\eta_{j \xa}}{\eta_{ij}} 
\right )^\ep ( 1- \eta_{i \xa})^{\ep}-1
\right ]
\right \rangle 
\\
& +
2^{1+\ep}
\left \langle 
\frac{\rho_{ij}^{\ep}( 
\rho_{i \xa} + \rho_{j \xa} )}{\rho_{i \xa}^{1+\ep} \rho_{j \xa}^{1+\ep}}
\omega^{\xa j, \yb j}_{\xa || \yb}
\left [ 
\left ( \frac{\eta_{i \xa}}{\eta_{ij}} 
\right )^\ep ( 1- \eta_{j \xa})^{\ep}-1
\right ]
\right \rangle
\\
& + 
\left \langle \eta_{i \xa}^{-\ep}
(1-\eta_{i \xa})^{\ep} 
\omega_{\xa || \yb}^{\xa i, \yb i} W_{ij}^{\xa}
\right \rangle 
+
\left \langle \eta_{j \xa}^{-\ep}
(1-\eta_{j \xa})^{\ep} 
 \omega_{\xa || \yb}^{\xa j, \yb j} W_{ji}^{\xa}
\right \rangle. 
\end{split}
\ee
Furthermore, the anomalous dimensions read in this case 
\be
\begin{split} 
& \gamma_{gJ} = -\frac{11}{3}
+\ep \left(-\frac{2}{3}-\frac{44}{3} \ln 2\right)
+\ep^2 \left(-\frac{22 \pi ^2}{9}-1-\frac{4}{3} \ln 2\right)
\\
& \qquad \quad +\ep^3 \left( -44 \zeta_3 - \frac{2 \pi^2}{9}+\frac{4}{3}-\frac{16}{3} \ln 2\right) +{\cal O}(\ep^4),
\\
& \delta_{gJ}
= \frac{1}{3}
+\ep \left(\frac{4 \ln 2}{3}-\frac{2}{3}\right)
+\ep^2 \left(\frac{2 \pi ^2}{9}-1-\frac{1}{3} 4 \ln 2\right)
+{\cal O}(\ep^3).
\end{split}
\ee

\subsection{Triple-collinear limit without \texorpdfstring{$N$}{N}-jettiness constraint}

The next contribution that we 
require is the triple-collinear 
one, from which the double-collinear limit is to be subtracted.  Similar to the calculation with the $N$-jettiness constraint, we find it to be convenient to calculate  the two terms separately and then subtract them from each other.  Furthermore, 
the two triple collinear configurations, $\xa || \yb || i$
and $\xa || \yb || j$, 
are the same. Hence, we consider one of them and multiply the result by a factor two to account for the other one. 

 An  important difference
 with respect to a similar computation with the $N$-jettiness constraint is the fact that the $\omega \to 1/\omega$ transformation is not a symmetry anymore.
 For this reason 
 the closed-form integration over $\omega$ becomes impossible.  
Hence, we proceed as follows. 
We first employ  the Mellin-Barnes representations to integrate over 
angles of $\xa$ and $\yb$.  Similar to the $N$-jettiness case, we find that one of the Mellin-Barnes integrations can be performed using the second Barnes lemma.  We then map the 
remaining Mellin-Barnes integrals 
onto  multi-variable integrals
using Euler's integral representation of 
Beta-functions and then integrate
over the new (Euler) variables 
as well as over $\omega$.
We find 
\be
 \begin{split} 
 & \frac{N_E}{\ep}
 \sum \limits_{x \in \{i,j\}}
\int \limits_{0}^{1}\frac{{\rm d} \omega}{\omega^{1+2\ep}}  \left \langle  C_{x \xa \yb}  \; 
w^{tc} \; {\bar S}_\omega \;
\left [ \omega^2 \tilde S_{ij}^{gg}(\xa, \yb) \right ] \right \rangle_{\xa \yb}
= [\alpha_s]^2 E_{\rm max}^{-4\ep}
\Bigg [ 
\frac{11}{12 \ep^3}
\\
&
+\frac{1}{\ep^2} 
\left ( \frac{\pi ^2}{12}-\frac{35}{18} 
\right )
+ \frac{1}{\ep} \left ( \frac{11 \zeta_3 }{4}-\frac{11 \pi ^2}{36}+\frac{205}{54}-11 \ln ^2 2-\frac{1}{3} \pi ^2 \ln 2-\frac{\ln 2}{2} \right )
\\
& +4 \text{Li}_4 \left (\frac{1}{2} \right )-\frac{275 \zeta_3}{12}-\frac{15}{2} \zeta_3 \ln 2+\frac{\pi ^4}{80}+\frac{137 \pi ^2}{216}-\frac{1463}{162} + \frac{\ln^4 2}{6}+\frac{110 \ln^3 2}{3}
\\
& +\frac{1}{2} \pi ^2 \ln ^2 2 +\frac{155 \ln^2 2}{6}-\frac{11}{18} \pi ^2 \ln 2-\ln 2 + {\cal O}(\ep) 
\Bigg ].
\end{split}
\ee
\\
The next step is to compute 
the double-collinear limit 
of the triple-collinear subtraction
term without the $N$-jettiness constraint.  The calculation 
follows the same steps as what is discussed in the previous 
Section where the $N$-jettiness constraint was taken into account. 
We find the following representation of the corresponding 
quantity
\be
\begin{split} 
& \left \langle C_{\xa \yb} [ {\rm d} \Omega_{\xa \yb} ] \theta^{(b+d)} C_{i \xa \yb} \;\omega^{tc} \; {\bar S}_\omega \;
 \left [ \omega^2 \tilde S_{ij}^{gg}(\xa, \yb) \right ] 
 \right \rangle_{\xa \yb} 
\\
   & = -\frac{ N_\ep^{(b,d)}}{2 \ep} 
   \left \langle \eta_{i \xa}^{-\ep-1} (1 - \eta_{i \xa } )^{\ep}
      \frac{4 \omega}{(1+\omega)^2}
         \left [ 
     -2 
     + \frac{(1-\ep)(1+2\ep) \omega}{
     (1+\omega)^2
     }
      \left( 1
     - \eta_{i \xa}
     \right )
         \right ]
         \right \rangle_\xa.   \end{split}
   \ee
The remaining integration over the directions 
of parton $\xa$ is straightforward. 
We find 
 \be
 \begin{split}
&  \frac{N_E}{\ep}
 \sum \limits_{x \in \{i,j\}}
\int \limits_{0}^{1}\frac{{\rm d} \omega}{\omega^{1+2\ep}}  \left \langle \theta^{(b+d)} \; C_{\xa \yb} ) \;
     \left [ {\rm d} \Omega_{\xa \yb} \right ] \; C_{x \xa \yb} \; \omega^{tc}
     {\bar S}_\omega \;
\left [ \omega^2 \tilde S_{ij}^{gg}(\xa, \yb) \right ] \right \rangle_{\xa \yb}
=
\\
& [\alpha_s]^2 E_{\rm max}^{-4\ep} \Bigg [
\frac{11}{12 \ep^3}
+\frac{1}{\ep^2} 
\left ( \frac{11 \ln 2}{6}-\frac{1}{6} 
\right )
+\frac{1}{\ep} 
\left ( 
\frac{11 \pi ^2}{12}+\frac{1}{4}-\frac{1}{2} 11 \ln^2 2-\frac{2 \ln 2}{3}
\right ) 
\\
& +\frac{55 \zeta_3}{6}+\frac{2}{3}-\frac{2 \pi ^2}{9}+\frac{55 \ln^3 2}{9}+\frac{5 \ln^2 2}{3}-\frac{11}{18} \pi ^2 \ln 2-\frac{\ln 2}{2}
+{\cal O}(\ep) \Bigg ].
\end{split}
\ee

\subsection{Extracting the required combinations 
of the various contributions}

As we already mentioned,  $J_{ij}$  defined in Eq.~(\ref{eq10.10}) was 
computed in Ref.~\cite{Caola:2018pxp}. We can use this result,  
supplemented with our computation of the various 
limits without the $N$-jettiness constraint, to derive analytic results 
for the quantity that 
is needed for  computations with the $N$-jettiness constraint.  

Specifically, we use Eq.~(\ref{eq10.10}) to write 
\be
\begin{split} 
 & \frac{N_E}{\ep}  \sum \limits_{x \in \{i,j\}}
  \int \limits_{0}^{1}\frac{{\rm d} \omega}{\omega^{1+2\ep}}  \left \langle (1 - \theta^{b+d}  C_{\xa \yb} ) 
       \left [ {\rm d} \Omega_{\xa \yb} \right ] {\bar  C}_{x \xa \yb} \;w^{x \xa, x \yb}   \;
{\bar S}_\omega\left [ w^2 \tilde S_{ij}^{gg}(\xa, \yb) \right ] \right \rangle_{\xa \yb}
\\
 & +\frac{N_E}{\ep} 
  \int \limits_{0}^{1}\frac{{\rm d} \omega}{\omega^{1+2\ep}} 
  \; \left \langle 
  \left ( 
       w^{\xa i, \yb j} 
       + 
    w^{\xa j, \yb i}  \right )    
       \;
{\bar S}_\omega\left [ w^2 \tilde S_{ij}^{gg}(\xa, \yb) \right ] \right \rangle_{\xa \yb}
\\
 & \;\;\;\; = 
J_{ij}
  -  \frac{N_E}{\ep} 
  \int \limits_{0}^{1} 
  \frac{{\rm d} \omega}{\omega^{1+2\ep}}
  \left \langle S_\omega \left [ \omega^2 {\tilde S}_{ij}(\xa, \yb ) \right ]
\right \rangle_{\xa \yb}
\\
& \;\;\;\; -
 \frac{N_E}{\ep}
  \int \limits_{0}^{1}\frac{{\rm d} \omega}{\omega^{1+2\ep}} \left \langle C_{\xa \yb} \left [
    {\rm d} \Omega_{\xa \yb} \right ]  \theta^{b+d } \omega^{tc} 
 {\bar S}_\omega \left [ w^2 \tilde S_{ij}^{gg}(\xa, \yb) \right ] \right \rangle_{\xa \yb}
\\
& 
 \;\;\;\;  -\frac{N_E}{\ep}
 \sum \limits_{x \in \{i,j\}}
\int \limits_{0}^{1}\frac{{\rm d} \omega}{\omega^{1+2\ep}}  \left \langle (1 - \theta^{b+d}  C_{\xa \yb} )
     \left [ {\rm d} \Omega_{\xa \yb} \right ] C_{x \xa \yb}  \omega^{tc}
{\bar S}_\omega\left [ w^2 \tilde S_{ij}^{gg}(\xa, \yb) \right ] \right \rangle_{\xa \yb}.
\label{eq11.16}
\end{split}
\ee
Using the results of the calculation discussed earlier, we obtain
\be
\begin{split} 
 &\frac{N_E}{\ep}  \sum \limits_{x \in \{i,j\}}
  \int \limits_{0}^{1}\frac{{\rm d} \omega}{\omega^{1+2\ep}}  \left \langle (1 - \theta^{b+d}  C_{\xa \yb} )
       \left [ {\rm d} \Omega_{\xa \yb} \right ] {\bar  C}_{x \xa \yb}   \;
{\bar S}_\omega\left [ w^2 \tilde S_{ij}^{gg}(\xa, \yb) \right ] \right \rangle_{\xa \yb}
\\
 & +\frac{N_E}{\ep} 
  \int \limits_{0}^{1}\frac{{\rm d} \omega}{\omega^{1+2\ep}} 
  \; \left \langle 
  \left ( 
       w^{\xa i, \yb j} 
       + 
    w^{\xa j, \yb i}  \right )    
       \;
{\bar S}_\omega\left [ w^2 \tilde S_{ij}^{gg}(\xa, \yb) \right ] \right \rangle_{\xa \yb}
 \\
& = \frac{1}{\ep} \Bigg ( 
\frac{11 }{12} A_{ij}^{\rm fin}
-\frac{1}{12} B_{ij}^{\rm fin}
-\frac{\pi^2}{6}  \ln (\eta_{ij})+\frac{11}{3} \ln 2 \ln (\eta_{ij})
+\frac{35 \ln (\eta_{ij} )}{9}
\\
& \quad +\frac{11\pi^2}{18}-\zeta_3 
-\frac{1}{9}-\frac{1}{3}\ln 2 
\Bigg ) + {\cal O}(\ep^0),
\end{split}
\label{eq.11.17}
\ee
where 
\be
\begin{split}
& A_{ij}^{\rm fin} 
 = 
 \left \langle 
 \frac{\rho_{ij}}{\rho_{i \xa} 
 \rho_{j \xa}} 
 \left [ 
\omega_{\xa i, \yb i}^{ \xa || \yb } 
\ln \left ( \frac{\eta_{j \xa}(1-\eta_{i \xa}) }{\eta_{ij}}
\right )
+
\omega_{\xa j, \yb j}^{ \xa || \yb } 
\ln \left ( \frac{\eta_{i \xa}(1-\eta_{j \xa}) }{\eta_{ij}}
\right )
\right ]
\right \rangle_{\xa} ,
\\
& B_{ij}^{\rm fin} = 
\left \langle 
\omega_{\xa || \yb}^{\xa i,\yb i} \; W_{ij}^{\xa} 
+
\omega_{\xa || \yb}^{\xa j, \yb j}
\; W_{ji}^{\xa}
\right \rangle_{\xa} .
 \end{split}
\ee
To save space, we do not show 
${\cal O}(\ep^0)$ terms in 
Eq.~(\ref{eq.11.17}) although 
they are needed  for  
the final  result.   It is, however, straightforward to compute them following the preceding discussion. 

We can use the result shown  in Eq.~(\ref{eq.11.17})  
to  compute the remaining contribution to  the $N$-jettiness soft  functions 
that originates from correlated emissions.
Upon doing that, we find that the divergent contributions  cancel and a finite remainder 
is obtained. We present the corresponding 
results in  Section~\ref{sect:13}.

\section{Final-state fermions}
\label{sect:12}

 We continue with the discussion of the contribution of the soft $q \bar q$ pair to the $N$-jettiness 
 soft function. 
 In this case, the entire contribution comes from correlated emissions. 
 It reads
 \be
 S^{q \bar q}_{2,RR,T^2,\tau}
 = \frac{n_f T_R}{2} \sum \limits_{(ij)} {\bf T}_i \cdot {\bf T}_j \;
I^{q \bar q}_{ij,\tau},
 \ee
 where
 \be
    I^{q \bar q}_{ij,\tau} = g_s^4
    \int [{\rm d} p]_\xa [{\rm d} p]_{\yb}
    \delta(\tau - E_\xa \psi_{\xa} - E_{\yb} \psi_{\yb} )\;
          {\tilde S}^{q \bar q}_{ij}(\xa,\yb).
 \ee
 The eikonal function ${\tilde S}^{q \bar q}_{ij}$ is symmetric under the interchange of $q$ and $\bar q$
 momenta; it can be found in Appendix~\ref{app:c}. Hence, we proceed in exactly the same way as with the gluons in that we introduce
 the energy ordering $E_{\yb} < E_{\xa}$, write $E_{\yb} = \omega E_{\xa}$, with $0 < \omega < 1$, and
 integrate over $E_{\xa}$.  After  the Laplace transform, we obtain
 \be
    I^{q \bar q}_{ij} =  \frac{2 N_u}{\ep}
    \int \limits_{0}^{1} \frac{{\rm d} \omega}{\omega^{1+2\ep}}
   \left \langle  \psi_{\xa \yb}^{4 \ep}  
    \omega^2 {\tilde S}^{q \bar q}_{ij}(\xa,\yb)
    \right \rangle_{\xa \yb},
 \ee
 where we have to set $E_\xa$ to  one and $E_{\yb}$ to  $\omega$
 when computing ${\tilde S}^{q \bar q}_{ij}(\xa,\yb)$ in the last equation. 

We continue with the investigation of the  various limits of the above equation.  In principle, these limits
are identical to the ones in the gluon case except that the $q \bar q$
eikonal function does not possess   a strongly-ordered singular limit.  Accounting for this and following what has
been done for the two-gluon final state, we write
\begin{align*}
  I_{ij}^{q \bar q}
  &= \frac{2 N_u}{\ep}
  \int \limits_{0}^{1}\frac{{\rm d} \omega}{\omega^{1+2\ep}} \left \langle C_{\xa \yb} \left [
    {\rm d} \Omega_{\xa \yb} \right ]  \theta^{b+d } w^{tc} \psi_{\xa \yb}^{4 \ep}
   [ \omega^2 {\tilde S}^{q \bar q}_{ij}(\xa,\yb) ] \right \rangle_{\xa \yb}
\\
& 
 +\frac{2 N_u}{\ep}
 \sum \limits_{x \in \{i,j\}}
\int \limits_{0}^{1}\frac{{\rm d} \omega}{\omega^{1+2\ep}}  \left \langle (1 - \theta^{b+d}  C_{\xa \yb} )
     \left [ {\rm d} \Omega_{\xa \yb} \right ] C_{x \xa \yb}  \psi_{\xa \yb}^{4 \ep}
   [ \omega^2 {\tilde S}^{q \bar q}_{ij}(\xa,\yb) ] \right \rangle_{\xa \yb}
  \\
& 
+\frac{2N_u}{\ep}  \sum \limits_{x \in \{i,j\}}
  \int \limits_{0}^{1}\frac{{\rm d} \omega}{\omega^{1+2\ep}}  \left \langle (1 - \theta^{b+d}  C_{\xa \yb} )
  \left [ {\rm d} \Omega_{\xa \yb} \right ] {\bar  C}_{x \xa \yb}  
  \; w^{x \xa, x \yb}\;
     [ \omega^2 {\tilde S}^{q \bar q}_{ij}(\xa,\yb) ] \right \rangle_{\xa \yb}
     \\
     & +\frac{2 N_u}{\ep} \sum \limits_{x \in \{i,j\}}
  \int \limits_{0}^{1}\frac{{\rm d} \omega}{\omega^{1+2\ep}}  \left \langle 
  \; 
  \left ( 
  w^{i \xa, j \yb}
  + 
  w^{j \xa, i \yb}
  \right )
 \;
      [ \omega^2 {\tilde S}^{q \bar q}_{ij}(\xa,\yb) ] \right \rangle_{\xa \yb} \numberthis \label{eq12.4}
\\
& +8 N_u \sum \limits_{x \in \{i,j\}}
  \int \limits_{0}^{1}\frac{{\rm d} \omega}{\omega^{1+2\ep}}  \left \langle (1 - \theta^{b+d}  C_{\xa \yb} )
  \left [ {\rm d} \Omega_{\xa \yb} \right ] {\bar  C}_{x \xa \yb}
  \; w^{x \xa, x \yb}\;
  \ln \psi_{\xa \yb} \;
      [ \omega^2 {\tilde S}^{q \bar q}_{ij}(\xa,\yb) ] \right \rangle_{\xa \yb}
      \\
 & +8 N_u \sum \limits_{x \in \{i,j\}}
  \int \limits_{0}^{1}\frac{{\rm d} \omega}{\omega^{1+2\ep}}  \left \langle 
  \; 
  \left ( 
  w^{i \xa, j \yb}
  + 
  w^{j \xa, i \yb}
  \right )
  \ln \psi_{\xa \yb} \;
      [ \omega^2 {\tilde S}^{q \bar q}_{ij}(\xa,\yb) ] \right \rangle_{\xa \yb}.
\end{align*}
We continue with the discussion of the individual terms. The first one  is the double-collinear $\xa || \yb$
limit. It reads
\be
C_{\xa \yb} [ \omega^2 {\tilde S}^{q \bar q}_{ij}(\xa,\yb) ]
= \frac{2 \omega \rho_{ij} }{(1+\omega)^2 \rho_{\xa \yb} \rho_{i \xa} \rho_{j \xa} }
\left (
-1 + \frac{2 \omega}{(1+\omega)^2} \;  \frac{\rho_{i \xa} \rho_{j \xa}}{\rho_{ij}}\;
v_\mu v_\nu \kappa_\yb^\mu \kappa_\yb^\nu, 
%
%
  \right ).
  \ee
  The vectors $v^\mu$ and 
  $\kappa_\yb^\nu$ are defined 
  in Eq.~(\ref{eq10.16}).
 We compute this contribution following the discussion of the  two-gluon case. Borrowing most of the calculation from there,
  we write the result for the sum of $b$ and $d$ sectors
\be
\begin{split} 
\Sigma^{(i)}_{\xa||\yb} &= \frac{N_u}{\ep}
  \int \limits_{0}^{1}
  \frac{{\rm d} \omega  }{\omega^{1+2\ep} }\left \langle C_{\xa \yb} [ {\rm d} \Omega_{\xa \yb} ]\theta^{b+d } \omega^{\xa i, \yb i} \psi_{\xa \yb}^{4\ep}
\left [ w^2 \tilde S_{ij}^{q \bar q}(\xa, \yb) \right ] \right \rangle_{\xa \yb}
\\
& = -\frac{N_u N_\ep^{b,d}}{  \ep^2} 
\Bigg  \langle
\frac{ \eta^{-\ep}_{i \xa}}{ (1-\eta_{i \xa} )^{-\ep} }
\; \psi_{\xa}^{4 \ep} \;
\omega_{\xa|| \yb}^{\xa i, \yb i}
\left [ \gamma_{q \bar q} \;A^{(i)}_{q \bar q}
+ \ep  \delta_{q \bar q} \;B^{(i)}_{q \bar q}
  \right ]
\Bigg  \rangle_{\xa} ,
\end{split}
\label{eq10.25}
\ee
with
\be
    A^{(i)}_{q \bar q}=\; \frac{\rho_{ij}}{\rho_{i \xa} \rho_{j \xa} },\quad
    B^{(i)}_{q \bar q}= \left ( 2 \frac{\rho_{i \xa} + \rho_{j \xa}}{\rho_{i\xa} \rho_{j \xa}}+ W_{ij}^{\xa}\right ),
\ee
and
\be
\begin{split}
& \gamma_{q \bar q}    = \int \limits_{0}^{1}
\frac{{\rm d} \omega }{\omega^{1+2\ep} }\frac{2 \omega}{(1+\omega)^{2-4\ep}} 
\left ( -1 +   \frac{2 \omega}{(1+\omega)^2} \right )
= 
-\frac{2}{3}-\frac{26}{9}\ep
\\
& 
+ \left(\frac{4\pi ^2}{9}-\frac{320}{27}\right) \ep^2
+ \left(\frac{32}{3}\zeta_3 
+\frac{52 }{27}\pi^2 
-\frac{3872}{81}\right) \ep^3 +{\cal O}(\ep^4),
\\
& 
\delta_{q \bar q} = \int \limits_{0}^{1}
\frac{{\rm d} \omega }{\omega^{1+2\ep} } \frac{4 \omega^2}{(1+\omega)^{4-4\ep}}
= \frac{1}{3} + 
\frac{10}{9} \ep
+ \left ( \frac{112}{27} - \frac{2 \pi^2}{9}
\right )\ep^2+{\cal O}(\ep^3) .
\end{split}
\ee
Structurally, this result is identical to the $\xa || \yb$ contribution in the two-gluon case except
that different collinear anomalous dimensions appear.
The analysis, therefore, proceeds along identical lines and the result can be  written in the same   way as in the
two-gluon case, c.f. 
Eqs.~(\ref{eq10.29}, \ref{eq10.30},
\ref{eq10.31}) up to 
the change in the anomalous 
dimensions and the overall normalization factor. 
\\

As the next step, we need to compute the triple-collinear limit of the $q \bar q$ eikonal function. The calculation
follows what has already been done for the two-gluon case. We obtain 
\be
C_{i \xa \yb} [ \omega^2 \tilde S_{ij}^{q \bar q}(\xa, \yb) ]
= \frac{2 \omega^2}{(1+\omega)^2} \; \frac{(\rho_{i \xa} - \rho_{i \yb})^2}{\rho_{\xa \yb}^2 ( \rho_{i \xa}
  + \omega \rho_{i \yb} )^2}
- \frac{ 2 \omega}{(1+\omega)} \; \frac{1}{\rho_{\xa \yb} (\rho_{i \xa} + \omega \rho_{i \yb})}.
\ee
Since  this expression is also  very similar to the two-gluon case,  we proceed accordingly.
As the first step, we change the integration variable $\omega \to 1/\omega$
and extend the integration over $\omega$ to $\omega = \infty$ using the symmetry of the integrand. We then write 
\be
\Sigma_{q \bar q}^{i,tc} = \frac{N_u}{\ep} 
B^{i}_{q \bar q},
\ee
where 
\be
B^{i}_{q \bar q}
= 
\int \limits_{0}^{\infty} \frac{{\rm d} \omega}{\omega^{1+2\ep}}
\left \langle
\frac{2 \omega}{(1+\omega) (\rho_{i \xa} + \omega \rho_{i \yb} )^{1-4\ep}}
\left (
\frac{\omega}{(1+\omega)} \frac{ (\rho_{i \xa} - \rho_{i \yb} )^2 }{\rho_{\xa \yb}^2 ( \rho_{i \xa} + \omega \rho_{i \yb}) }
- \frac{1}{\rho_{\xa \yb} }
\right )
\right \rangle_{\xa \yb}.
\ee
Using the $\xa \leftrightarrow \yb $ and $\omega \to 1/\omega$ transformations, we rewrite the above expression as
\be
B^{i}_{q \bar q}
=
\int \limits_{0}^{\infty} \frac{{\rm d} \omega}{\omega^{1+2\ep}}\left \langle
\frac{2 \omega}{(1+\omega) (\rho_{i \xa} + \omega \rho_{i \yb} )^{1-4\ep}}
\left (
\frac{2 \omega}{(1+\omega)} \frac{ \rho_{i \xa} (\rho_{i \xa} - \rho_{i \yb} )}{\rho_{\xa \yb}^2 ( \rho_{i \xa} + \omega \rho_{i \yb}) }
-  \frac{1}{\rho_{\xa \yb} }
\right )
\right \rangle_{\xa \yb}.
\ee
The calculation proceeds along the same 
lines as in the two-gluon case. We obtain 
\be
\Sigma_{q \bar q}^{i,tc}=
\frac{N_u}{\ep}
\left [ 
\frac{2}{3 \ep^2}+\frac{16}{9 \ep}+\frac{2 \pi ^2}{9}+\frac{104}{27}
+\ep \left(-\frac{8 \zeta_3}{3}+\frac{16 \pi ^2}{27}+\frac{640}{81}\right)
+{\cal O}(\ep^2) \right ].
\ee
The double-collinear subtraction term of the triple-collinear contribution reads
\be
\Sigma^{i,tc}_{dc, q \bar q} = 
\frac{N_u}{\ep}
 \int \limits_{0}^{1}\frac{{\rm d} \omega}{\omega^{1+2\ep}}   \left \langle \theta^{b+d} \; C_{\xa \yb} 
     \left [ {\rm d} \Omega_{\xa \yb} \right ] C_{i \xa \yb} \;
\psi_{\xa \yb}^{4 \ep}
\;
\left [ w^2 \tilde S^{q \bar q}_{ij}(\xa, \yb) \right ] \right \rangle_{\xa \yb}.
\ee
Following the very same procedure as in the two-gluon case, we arrive at 
\be
\begin{split}
\Sigma_{dc, q \bar q}^{i,tc} &=
\frac{N_u N_{\ep}^{(b,d)}}{\ep^2}
\Bigg [ 
\frac{2}{3 \ep}+\frac{4 \ln 2}{3}+\frac{26}{9}
+\ep \left(\frac{4 \ln^2 2}{3}+\frac{52 \ln 2}{9}-\frac{4 \pi ^2}{9}+\frac{320}{27}\right)
\\
& + \ep^2 \bigg(-\frac{32 \zeta_3}{3} + \frac{8 \ln^3 2}{9} + \frac{52 \ln^2 2}{9} - \left(\frac{8 \pi^2 }{9} - \frac{640}{27} \right)\ln2 
\\
&- \frac{52 \pi^2}{27} + \frac{3872}{81} \bigg)+{\cal O}(\ep^3) \Bigg ].
\end{split}
\ee
Combining the 
above results and accounting for the contribution 
of the second partition function, we obtain the result for  the 
triple-collinear limit 
of the $q \bar q$ contribution 
to the integrated eikonal 
function subject to $N$-jettiness constraint
\be
\begin{split}
& \frac{2 N_u}{\ep}
 \sum \limits_{x \in \{i,j\}}
\int \limits_{0}^{1}\frac{{\rm d} \omega}{\omega^{1+2\ep}}  \left \langle (1 - \theta^{b+d}  C_{\xa \yb} )
     \left [ {\rm d} \Omega_{\xa \yb} \right ] C_{x \xa \yb}  \omega^{tc}
\psi_{\xa \yb}^{4 \ep}
   [ \omega^2 {\tilde S}^{q \bar q}_{ij}(\xa,\yb) ] \right \rangle_{\xa \yb} =
   \\
&  N_{u} 
\Bigg [ 
\frac{1}{\ep^2} 
\left ( -\frac{10}{9}-\frac{4}{3} \ln 2 
\right ) 
+\frac{1}{\ep} 
\left ( 
\frac{4 \pi ^2}{9}-8-\frac{4}{3}  \ln^2 2-\frac{52 \ln 2}{9}
\right ) 
\\
& + \frac{28}{3} \zeta_3 +\frac{14 \pi ^2}{9}-\frac{3232}{81}-\frac{8}{9}  \ln^3 2-\frac{52 \ln^2 2}{9}+\frac{4}{9} \pi ^2 \ln 2-\frac{640 }{27} \ln 2
+{\cal O}(\ep) \Bigg ].
\end{split}
\ee

To determine the collinear-regulated contribution in 
Eq.~(\ref{eq12.4}), we proceed 
in the same way as in the 
two-gluon case. Namely, we write a similar representation for 
the ``energy-regulated''  
integral of the $q \bar q$ 
contribution to the soft function computed in Ref.~\cite{Caola:2018pxp}.
It reads 
\be
\begin{split} 
  J_{ij}^{q \bar q}
  &= \frac{2 N_E}{\ep}
  \int \limits_{0}^{1}\frac{{\rm d} \omega}{\omega^{1+2\ep}} \left \langle C_{\xa \yb} \left [
    {\rm d} \Omega_{\xa \yb} \right ]  \theta^{b+d } \omega^{tc} 
   [ \omega^2 {\tilde S}^{q \bar q}_{ij}(\xa,\yb) ] \right \rangle_{\xa \yb}
\\
& 
 +\frac{2 N_E}{\ep}
 \sum \limits_{x \in \{i,j\}}
\int \limits_{0}^{1}\frac{{\rm d} \omega}{\omega^{1+2\ep}}  \left \langle (1 - \theta^{b+d}  C_{\xa \yb} )
     \left [ {\rm d} \Omega_{\xa \yb} \right ] C_{x \xa \yb}  \omega^{tc}
   [ \omega^2 {\tilde S}^{q \bar q}_{ij}(\xa,\yb) ] \right \rangle_{\xa \yb}
  \\
& 
+\frac{2N_E}{\ep}  \sum \limits_{x \in \{i,j\}}
  \int \limits_{0}^{1}\frac{{\rm d} \omega}{\omega^{1+2\ep}}  \left \langle (1 - \theta^{b+d}  C_{\xa \yb} )
  \left [ {\rm d} \Omega_{\xa \yb} \right ] {\bar  C}_{x \xa \yb}  \;
  w^{x \xa, x \yb} \;
     [ \omega^2 {\tilde S}^{q \bar q}_{ij}(\xa,\yb) ] \right \rangle_{\xa \yb}
          \\
     & +\frac{2 N_E}{\ep} \sum \limits_{x \in \{i,j\}}
  \int \limits_{0}^{1}\frac{{\rm d} \omega}{\omega^{1+2\ep}}  \left \langle 
  \; 
  \left ( 
  w^{i \xa, j \yb}
  + 
  w^{j \xa, i \yb}
  \right )
 \;
      [ \omega^2 {\tilde S}^{q \bar q}_{ij}(\xa,\yb) ] \right \rangle_{\xa \yb}.
\end{split}   
\label{eq12.16}
\ee
We observe that,
up 
to a normalization factor, the last two contributions are the same 
as the missing divergent 
ones  in Eq.~(\ref{eq12.4}). 
Since the result for $J_{ij}^{q \bar q}$ is known 
\cite{Caola:2018pxp}, we can 
extract the contribution 
of the last two terms in Eq.~(\ref{eq12.16}) if the first two terms are known. We note 
that these terms are similar but  not identical to the ones 
that we already computed because the $N$-jettiness function 
$\psi_{\xa \yb}$ does not 
appear in Eq.~(\ref{eq12.16}). However, their computation is nearly identical to what has been discussed in the two-gluon case and for this reason 
we do not repeat it. 
We find 
\be
\begin{split} 
 & \frac{2 N_E}{\ep}  \sum \limits_{x \in \{i,j\}}
  \int \limits_{0}^{1}\frac{{\rm d} \omega}{\omega^{1+2\ep}}  \left \langle (1 - \theta^{b+d}  C_{\xa \yb} )
       \left [ {\rm d} \Omega_{\xa \yb} \right ] {\bar  C}_{x \xa \yb} 
       w^{x \xa, x \yb} \;
\left [ w^2 \tilde S^{q \bar q}_{ij}(\xa, \yb) \right ] \right \rangle_{\xa \yb}
\\
 & +\frac{2 N_E}{\ep} \sum \limits_{x \in \{i,j\}}
  \int \limits_{0}^{1}\frac{{\rm d} \omega}{\omega^{1+2\ep}}  \left \langle 
  \; 
  \left ( 
  w^{i \xa, j \yb}
  + 
  w^{j \xa, i \yb}
  \right )
 \;
      [ \omega^2 {\tilde S}^{q \bar q}_{ij}(\xa,\yb) ] \right \rangle_{\xa \yb}
 \\
 & = \frac{1}{\ep} \Bigg ( 
\frac{A_{ij}^{\rm fin}}{6}-\frac{B_{ij}^{\rm fin}}{12}
+\frac{2}{3} \ln 2 \ln (\eta_{ij})+\frac{ \pi^2}{9}+\frac{8 \ln (\eta_{ij})}{9}-\frac{1}{9}-\frac{\ln 2}{3} +{\cal O}(\ep)
\Bigg ).
\end{split}
\ee
To obtain the above formula, we used the following 
result for the triple-collinear contribution 
in the case without the $N$-jettiness constraint.
\be
\begin{split}
\frac{2 N_E}{\ep} &
 \int \limits_{0}^{1}\frac{{\rm d} \omega}{\omega^{1+2\ep}}  \left \langle C_{x \xa \yb}  \omega^{tc,i}
\left [ w^2 \tilde S^{q \bar q}_{ij}(\xa, \yb) \right ] \right \rangle_{\xa \yb}
=
\frac{2 N_E}{\ep}
\Bigg [ 
-\frac{1}{3 \ep^2}+\frac{19}{18 \ep}
\\
& 
+\frac{\pi ^2}{9}-\frac{83}{54}+4 \ln^2 2+\ln 2
+\ep \Big ( \frac{25 \zeta_3}{3}-\frac{35 \pi ^2}{108}+\frac{1057}{162}
\\
& -\frac{1}{3} 40 \ln^3 2 -\frac{53 \ln^2 2}{3}+\frac{2}{9} \pi ^2 \ln 2+3 \ln 2\Big )
+{\cal O}(\ep^2) \Bigg ]. 
\end{split}
\ee
\\

\noindent The corresponding result for the 
double-collinear subtracted contribution 
without the $N$-jettiness constraint reads 
\be
\begin{split}
 & \frac{2 N_E}{\ep} 
 \int \limits_{0}^{1}\frac{{\rm d} \omega}{\omega^{1+2\ep}}  \left \langle  \theta^{b+d}  C_{\xa \yb} 
     \left [ {\rm d} \Omega_{\xa \yb} \right ] C_{x \xa \yb}  \omega^{tc,i}
\left [ w^2 \tilde S^{q \bar q}_{ij}(\xa, \yb) \right ] \right \rangle_{\xa \yb}
= 
\\
& \frac{2 N_E}{\ep}
\Bigg [ 
 -\frac{1}{3 \ep^2}
+\frac{1}{\ep} 
\left ( \frac{1}{2}-\frac{2 \ln 2}{3} 
\right )
-\frac{\pi ^2}{3}
+2 \ln^2 2+\frac{5 \ln 2}{3}
\\
& +\ep \left(-\frac{10 \zeta_3}{3}+\frac{11 \pi ^2}{18}-\frac{4}{3}-\frac{20}{9} \ln^3 2-\frac{13}{3} \ln^2 2+\left(\frac{8}{3}+\frac{2 \pi^2}{9}\right) \ln 2 \right)
+{\cal O}(\ep^2) \Bigg].
\end{split}
\ee

\section{The renormalized \texorpdfstring{$N$}{N}-jettiness soft function}
\label{sect:13}

We are now in a position 
to present the renormalized 
$N$-jettiness 
soft function through NNLO in the perturbative expansion in QCD. 
We write 
\be
{\tilde S} = 
1 + {\tilde S}_1 
+ {\tilde S}_2 
+{\cal O}(\alpha_s^3).
\ee
It is convenient to introduce the following 
short-hand notations
\be
L_{ij} = \ln( \bar u \sqrt{\eta_{ij}} \mu),
\;\;\; L^\psi_{ij,\xa}
= \ln \left ( 
\frac{\psi_\xa \rho_{ij}}{\rho_{i \xa} \rho_{j \xa}}
\right ).
\ee
The NLO contribution reads 
\be
{\tilde S}_1
= a_s \sum \limits_{(ij)}
{\bf T}_i \cdot {\bf T}_j
\left [ 
2 L_{ij}^2
+ {\rm Li}_{2}(1-\eta_{ij}) + \frac{\pi^2}{12}
+ 
\left \langle  \; 
 L^\psi_{ij,\xa}
\frac{\rho_{ij} }{ \rho_{i \xa} \rho_{j \xa} } \right \rangle_{\xa}
\;
\right ].
\ee
The NNLO contribution 
is constructed from different pieces. 
We write 
\be
{\tilde S}_2 = 
\frac{1}{2} \;
{\tilde S}_1^2 + 
a_s^2 C_A \sum \limits_{(ij)} {\bf T}_i \cdot {\bf T}_j 
\;  G_{ij}
+a_s^2 \; n_f \; T_R 
\sum \limits_{(ij)} {\bf T}_i \cdot {\bf T}_j 
\;  Q_{ij}
+ a_s^2 \pi \sum \limits_{(kij)}
F^{kij} \; \kappa_{kj} G_{kij}^{\text{triple}}.
\ee

The function $G_{ij}$ reads 
{\allowdisplaybreaks
\begin{align*}
 G_{ij} & = \frac{22}{9} L_{ij}^3  
+ \left ( \frac{67}{9} - \frac{\pi^2}{3}
\right ) L_{ij}^2 
+ L_{ij} \Big(
 \frac{11}{3} \left \langle 
 L^\psi_{ij,\xa}
\frac{\rho_{ij} }{ \rho_{i \xa} \rho_{j \xa} } \right \rangle_{\xa}
+\frac{11}{3}{\rm Li}_2(1-\eta_{ij} ) 
\\
& + \frac{202}{27} -
7 \zeta_3 \Big  )  
 + 
\Bigg \langle \frac{\rho_{ij}}{ \rho_{i \xa} \rho_{j \xa} } 
\Bigg ( 
\left ( L^{\psi}_{ij,\xa} \right )^2 \left ( \frac{11}{6}
 - \ln \left( \frac{\eta_{ij} }{\eta_{i \xa} \eta_{j \xa}} \right )
 \right ) 
\\
 & + 
L^{\psi}_{ij,\xa}
\Big ( 
2 \ln^2 \left ( \frac{\eta_{ij} }{\eta_{i \xa} \eta_{j \xa}} \right )
+ \ln \left ( \frac{\eta_{ij} }{\eta_{i \xa} \eta_{j \xa}} \right ) \left (
-\frac{11}{3} + \ln ( \eta_{i \xa} \eta_{j \xa} )
  \right  )
+ \frac{137}{18} - \frac{\pi^2}{2} 
\\
& -\frac{1}{2} \ln^2 \left( \frac{\eta_{i \xa}}{\eta_{j \xa}} \right)+ {\rm Li}_2(1 - \eta_{ij}) + \frac{11}{3} \ln 2
-\frac{11}{6} \ln ( \eta_{i \xa} \eta_{j \xa} )  
- \frac{(\rho_{i\xa} + \rho_{j \xa})}{3 \rho_{ij} }
\Big ) 
\Bigg ) \Bigg \rangle_{\xa}
\\
& + \left \langle A^{\rm fin}_{ij,\xa} 
\left (\frac{11}{6} L_{ij,\xa}^{\psi}  -\frac{11}{6} \ln \frac{\eta_{ij} }{\eta_{i \xa} \eta_{j \xa}}
+\frac{131}{72}
\right) \right \rangle_\xa
\\
&+ \left \langle B^{\rm fin}_{ij,\xa} 
\left (-\frac{1}{6} L_{ij,\xa}^{\psi}  + \frac{1}{6} \ln \frac{\eta_{ij} }{\eta_{i \xa} \eta_{j \xa}}
-\frac{13}{72}
\right ) \right \rangle_\xa -2 G_{-1,0,0,1}(\eta_{ij}) +\frac{7}{2}G_{0,1,0,1}(\eta_{ij})
\\
& 
-\frac{11}{3} {\rm Ci}_3(2 \delta_{ij}) 
-\frac{1}{6 {\rm tan}(\delta_{ij} )}
{\rm Si}_2(2 \delta_{ij}) +17 S_{2,2}(\eta_{ij}) -\frac{1}{2} {\rm Li}_{4} \left( 1-\eta_{ij}^2 \right) 
\\
&
-2 {\rm Li}_{4}\left(\frac{1}{\eta_{ij}+1}\right)
+{\rm Li}_{4}\left(\frac{1-\eta_{ij}}{1+\eta_{ij}}\right)-{\rm Li}_{4}\left(\frac{\eta_{ij}-1}{1+\eta_{ij}}\right)-12 {\rm Li}_{4}(\eta_{ij})
\\
&+{\rm Li}_{3}(\eta_{ij})\left( -\frac{11}{6}+10\ln (1-\eta_{ij}) + 10\ln (\eta_{ij}) - 2\ln (1+\eta_{ij}) \right)
\\
& +{\rm Li}_{3}(1-\eta_{ij}) \Big(10\ln (\eta_{ij}) + 2\ln (1+\eta_{ij}) \Big) - 2 {\rm Li}_{3}(-\eta_{ij}) \ln(1-\eta_{ij}) \numberthis
\\
& +\frac{5{\rm Li}_{2}(\eta_{ij})^2}{4} + {\rm Li}_{2}(\eta_{ij}) \Big(-2{\rm Li}_{2}(-\eta_{ij}) - 4\ln ^2(\eta_{ij})
\\
& +6\ln (1-\eta_{ij}) \ln (\eta_{ij})+\frac{11\ln(\eta_{ij})}{6} +\frac{\pi ^2}{3}-11\ln 2-\frac{131}{12}\Big)
\\
&  - \frac{1}{12} \ln^4 (1+\eta_{ij}) - \ln^3 (\eta_{ij}) \ln (1-\eta_{ij})  + \frac{\pi^2}{12} \ln^2 (1+\eta_{ij})
\\
& + \ln^2 (\eta_{ij}) \left( \frac{11}{2} \ln^2 (1-\eta_{ij}) + \frac{11}{12} \ln (1-\eta_{ij}) - \frac{\pi^2}{6} + \frac{11 \ln 2}{3} + \frac{32}{9} \right)
\\
& + \ln (1-\eta_{ij}) \left( -11 \ln 2 \ln(\eta_{ij}) - \frac{4 \pi^2 \ln(\eta_{ij})}{3} - \frac{131 \ln(\eta_{ij})}{12} - \frac{23 \zeta_3}{2}\right)
\\
& + \ln(\eta_{ij}) \left( - \frac{11 \ln^2 2}{3} - \frac{\pi^2 \ln 2}{3} + \frac{64 \ln 2}{9} - \frac{27 \zeta_3}{4} - \frac{11 \pi^2}{18} + \frac{1631}{108} \right)
\\
& - \frac{7}{4} \zeta_3 \ln (1+\eta_{ij}) + \frac{1}{3} \ln^4 2 + \ln^2 2 \left( \frac{1}{3} - \frac{\pi^2}{3} \right) + \ln 2 \left( 5 \zeta_3 + \frac{11 \pi^2}{18}-\frac{2}{9}\right)
\\
& + 8 {\rm Li}_{4}\left(\frac{1}{2}\right) - \frac{11}{9} \zeta_3 - \frac{11}{80} \pi^4 + \frac{937}{432} \pi^2 + \frac{403}{162}
\\
& -  
\left \langle {\bar C}_{\xa \yb} \ln\left(\frac{\psi_m}{\rho_{im}}\right) 
\ln\left(\frac{\psi_n}{\rho_{jn}}\right) \frac{\rho_{i j}}{\rho_{\xa \yb} \rho_{i \xa} \rho_{j \yb} }
\right \rangle_{\xa \yb}
+ \frac{1}{2} 
\left \langle
L_{ij,\xa}^{\psi} 
\frac{\rho_{ij}} {\rho_{i \xa} \rho_{j \xa}}
\right \rangle_{\xa}^2
\\
& + \frac{1}{2}  \sum \limits_{x \in \{i,j\}}
  \int \limits_{0}^{1}\frac{{\rm d} \omega}{\omega}  \left \langle (1 - \theta^{b+d}  C_{\xa \yb} )
  \left [ {\rm d} \Omega_{\xa \yb} \right ] {\bar  C}_{x \xa \yb} \; 
  w^{x \xa, x \yb} \;
  \ln \psi_{\xa \yb} \; 
  {\bar S}_\omega \; 
      [ \omega^2 {\tilde S}_{ij}(\xa,\yb) ] \right \rangle_{\xa \yb}
      \\
      & + \frac{1}{2} 
  \int \limits_{0}^{1}\frac{{\rm d} \omega}{\omega}  \left \langle ( 
  w^{i \xa, j \yb} 
  +
   w^{j \xa, i \yb} 
) 
  \;
  \ln \psi_{\xa \yb} \; 
  {\bar S}_\omega \; 
      [ \omega^2 {\tilde S}_{ij}(\xa,\yb) ] \right \rangle_{\xa \yb}.
\end{align*}
}
In the above formula $\delta_{ij} = \theta_{ij}/2$
and $\theta_{ij}$ is the relative 
angle between partons $i$ and $j$. The Clausen functions are defined through their relation to complex-valued polylogarithms
\be
\begin{split}
\text{Ci}_n(z) &=\frac{1}{2} \left( \text{Li}_n (e^{iz}) + \text{Li}_n (e^{-iz}) \right), \\
\text{Si}_n(z) &=\frac{1}{2i} \left( \text{Li}_n (e^{iz}) - \text{Li}_n (e^{-iz}) \right). 
\end{split}
\ee
The functions 
$S_{a_1,a_2}$ and $G_{a_1,a_2,\dots, a_m}(x)$ are the Nielsen and Goncharov \cite{Goncharov:1998kja} polylogarithms, respectively.  The auxiliary functions that we 
employed while writing the above  formula read
\be
\begin{split}
& A_{ij,\xa}^{\rm fin} 
 = 
 \frac{\rho_{ij}}{\rho_{i \xa} 
 \rho_{j \xa}} 
 \left [ 
\omega^{\xa i, \yb i}_{ \xa || \yb } 
\ln \left ( \frac{\eta_{j \xa}(1-\eta_{i \xa}) }{\eta_{ij}}
\right )
+
\omega^{\xa j, \yb j}_{ \xa || \yb } 
\ln \left ( \frac{\eta_{i \xa}(1-\eta_{j \xa}) }{\eta_{ij}}
\right )
\right ],
\\
& B_{ij,\xa}^{\rm fin} = 
\omega_{\xa || \yb}^{\xa i,\yb i} \; W_{ij}^{\xa} 
+
\omega_{\xa || \yb}^{\xa j, \yb j}
\; W_{ji}^{\xa}.
 \end{split}
\ee
\\
The function $Q_{ij}$ evaluates to
\be
\begin{split} 
 Q_{ij} & = -\frac{8}{9} L_{ij}^3  
-\frac{20}{9}  L_{ij}^2 
- L_{ij} \Bigg(
 \frac{4}{3} \left \langle 
 L^\psi_{ij,\xa}
\frac{\rho_{ij} }{ \rho_{i \xa} \rho_{j \xa} } \right \rangle_{\xa}
+\frac{4}{3}\ {\rm Li}_2(1-\eta_{ij} ) 
+\frac{56}{27}
\Bigg ) 
\\
& - 
\left \langle 
\frac{\rho_{ij}}{ \rho_{i \xa} \rho_{j \xa} } 
\Bigg ( 
\frac{2}{3} \left ( L^{\psi}_{ij,\xa} \right )^2 
-L^{\psi}_{ij,\xa}
\Big ( \;
\frac{2}{3} \ln \left ( \frac{\eta^2_{ij}}{4\eta_{i \xa} \eta_{j \xa} } \right )
 - \frac{26}{9}
+ \frac{2( \rho_{i \xa} + \rho_{j \xa} ) }{3 \rho_{ij}}
\Big ) 
\Bigg )
\right \rangle_\xa
\\
& + \left \langle A^{\rm fin}_{ij,\xa}
\left (  
-\frac{2}{3} L^{\psi}_{ij,\xa}
+
\frac{2}{3}\ln \left( \frac{\eta_{ij}} {\eta_{i \xa}
\eta_{j \xa} } \right)
-\frac{23}{36} 
\right )
\right \rangle_{\xa}
\\
& + \left \langle B^{\rm fin}_{ij,\xa}
\left ( 
\frac{1}{3} L^{\psi}_{ij,\xa}
-\frac{1}{3} \ln \left(
\frac{\eta_{ij}}{\eta_{i \xa}
\eta_{j \xa} } \right)
+\frac{13}{36} 
\right ) 
\right \rangle_{\xa} 
+\frac{2}{3}  {\rm Li}_3(\eta_{ij}) 
\\
& +{\rm Li}_2(1-\eta_{ij}) \left(\frac{2 \ln (\eta_{ij})}{3}-\frac{7}{2}-4 \ln 2 \right)
+\ln^2 2
\left ( -\frac{2}{3} + \frac{4}{3}
\ln (\eta_{ij}) \right )
\\
& +\ln 2 \left(-\frac{4}{3} \ln ^2(\eta_{ij})-\frac{20}{9} \ln (\eta_{ij})+\frac{4
\pi ^2}{9}+\frac{4}{9}\right)
\\
& - \ln ^2(\eta_{ij})\left(\frac{10}{9}-\frac{1}{3} \ln (1-\eta_{ij})\right) 
+\ln (\eta_{ij}) \left(\frac{\pi ^2}{9}-\frac{335}{54}\right)
\\
& +\frac{4 \zeta_3}{9}+\frac{122}{81}-\frac{47 \pi ^2}{108}
+ \frac{4}{3} {\rm Ci}_3(2 \delta_{ij}) +\frac{1}{3 {\tan} (\delta_{ij} ) } {\rm Si}_2(2 \delta_{ij})
\\
& -  \sum \limits_{x \in \{i,j\}}
  \int \limits_{0}^{1}\frac{{\rm d} \omega}{\omega}  \left \langle (1 - \theta^{b+d}  C_{\xa \yb} )
  \left [ {\rm d} \Omega_{\xa \yb} \right ] {\bar  C}_{x \xa \yb}\;
  w^{x \xa, x \yb}
  \;
  \ln \psi_{\xa \yb} \; 
      [ \omega^2 {\tilde S}^{q \bar q}_{ij}(\xa,\yb) ] \right \rangle_{\xa \yb}
      \\
       & -
  \int \limits_{0}^{1}\frac{{\rm d} \omega}{\omega}  \left \langle ( 
  w^{i \xa, j \yb} 
  +
   w^{j \xa, i \yb} 
) 
  \;
  \ln \psi_{\xa \yb} \; 
      [ \omega^2 {\tilde S}^{q \bar q}_{ij}(\xa,\yb) ] \right \rangle_{\xa \yb},
\end{split} 
\ee
The function $G_{kij}^{\text{triple}}$, defined in Eq.~\eqref{eq7.15}, corresponds to the finite remainder  of the 
triple-color correlation terms.

\section{Numerical implementations and checks}
\label{sec:num}

We have implemented the above formulas into a {\sf Fortran}
code.  The implementation is, in principle, straightforward.  To compute  terms that require integration over  directions of gluons' momenta, we use the phase space parametrization  described in Ref.~\cite{Czakon:2010td, Czakon:2011ve, Caola:2017dug}.  A possible  choice 
of partition functions can be found in 
Ref.~\cite{Asteriadis:2019dte}. 

As we already mentioned, the $N$-jettiness soft function was  recently discussed  in  Ref.~\cite{Bell:2023yso}
where a plethora of numerical results for various kinematic configurations have been  presented.  We have checked many of the presented results for various values of the parameter $N$ and found excellent agreement.
 We present some examples of such a  comparison in Appendix~\ref{app:d}.  

We note that to  obtain the high-precision numbers shown in  Appendix~\ref{app:d}, we have  used a very large number of sample points which  
results in  runtimes that can be as large as a few minutes per dipole per kinematic point.  However, to obtain the same   numbers with a better-than-percent precision, we need a relatively small number of sampling points which results in runtimes of an order of a few  seconds per dipole per phase space point. We  note that runtimes for individual dipoles are largely \emph{independent} of $N$.  Since $(N+2)(N+1)/2$ dipoles  need to be calculated to determine 
the $N$-jettiness soft function for a given phase space point, ${\cal O}(1-2)$ minutes  will be required to do that in case of   three- or four-jet production  at a hadron collider, or  six-jet production at an 
$e^+e^-$ collider. 

The situation with  triple-color correlated terms is quite similar.  Since in this case  integration over the direction of a single gluon is required, 
the integration converges rather fast. 
In general, for an $N$-jettiness case, 
the evaluation of $(N+2)(N+1)N$ independent functions $G_{kij}^{\rm triple}$ is required from which a smaller number of independent color-correlated contributions can be constructed. 
We need about twenty seconds to compute 
all the required  $G_{kij}^{\rm triple}$ functions for $N=2$ and, 
therefore, we will need about a minute to compute them for $N=3$ and about two minutes for $N=4$.

\section{Conclusions}
\label{sect:14}

We have described the computation of  the $N$-jettiness 
soft function through NNLO in QCD. Keeping $N$ as the parameter,  we demonstrated  the cancellation of all $1/\ep$ 
poles \emph{analytically} 
against the soft-function renormalization matrix. Furthermore, 
we derived a  simple representation for 
the finite, jettiness-dependent 
remainder valid for an arbitrary  number of hard partons $N$.  We compared our numerical results for $N=1,2$ and $N=3$ with the ones recently presented in Ref.~\cite{Bell:2023yso}  and found excellent agreement.  We have also found that the representation of  the finite remainder that is derived in this paper  leads to fast and rapidly convergent integration which is important if the results for $N$-jettiness soft function are  to be used for the 
computation of higher-order corrections for multi-jet production at colliders. 

From a more general perspective, this study was motivated by a question of how advances in developing NNLO subtraction schemes for generic processes at the LHC can be used to derive representations for the building blocks 
of modern slicing methods that rely on  choosing  observables to 
define suitable slicing variables. We have  shown that, at least for $N$-jettiness, the benefits of applying subtraction-inspired methods are significant. We note  that the application of the subtraction-inspired methods for computing the $N$-jettiness soft function becomes possible if one departs from the (by now) standard approach \cite{Jouttenus:2011wh,Boughezal:2015eha,Campbell:2017hsw} to such computations where hemisphere soft functions are used as elementary building blocks and, instead,  interprets $N$-jettiness as one of the many infrared safe observables that can be studied using available subtraction schemes.

\section*{Acknowledgments}
We are grateful to G.~Bell for useful exchanges and R. K. Ellis for helpful conversations.  This  research  was supported by the German Research Foundation (DFG, Deutsche Forschungsgemeinschaft) under grant 396021762-TRR 257.

\appendix

\section{Renormalization of the soft function}
\label{app:a}

A suitable  expression for the renormalization of the $N$-jettiness soft function  was recently given in Ref.~\cite{Bell:2023yso}.  It reads 
\be
Z = 1 + Z_1 + \frac{1}{2} Z_1 Z_1 +Z_{2,r},
\ee
where 
\be
Z_1 = a_s \sum \limits_{(ij)} {\bf T}_i \cdot {\bf T}_j \left ( 
\frac{1}{2\ep^2} + \frac{2 L_{ij} + i \pi \lambda_{ij}}{2\ep}
\right ),
\ee
and
\be
Z_{2,r} = a_s^2 
\sum \limits_{(ij)} {\bf T}_i \cdot {\bf T}_j
\left ( 
-\frac{3 \beta_0}{8 \ep^3}
+ \frac{\Gamma_1 - 4\beta_0 (2 L_{ij} + i\pi \lambda_{ij} )}{16 \ep^2}+ \frac{\Gamma_1 (2 L_{ij} + i\pi \lambda_{ij} ) + \gamma^S_1}{8 \ep}
\right ),
\ee
with  
\be
L_{ij} = \ln( \bar u \sqrt{\eta_{ij}} \mu),
\ee

\be
\Gamma_1 = \left ( \frac{67}{9} - \frac{\pi^2}{3} \right )C_A 
- \frac{20}{9} T_R n_f,
\ee
and
\be
\gamma^S_1 = \left ( \frac{202}{27} - \frac{11 \pi^2}{36} - 7 \zeta_3 \right )C_A 
+ \left(- \frac{56}{27} + \frac{\pi^2}{9} \right) T_R n_f.
\ee
where $\Gamma_{1}$ is the cusp anomalous dimension and the non-cusp anomalous dimension $\gamma^S_1$ is known to two-loop order \cite{Stewart:2010qs}.
Furthermore, $\lambda_{ij} = 1$ if both 
$i$ and $j$ are incoming or outgoing and 
$\lambda_{ij} = 0$ otherwise.

\section{Useful phase space integrals}
\label{app:b}

We define measures of angular integration as
\be
   [{\rm d} \Omega^{(d-1)} ] = 
   \frac{{\rm d} \Omega^{(d-1)} }{\Omega^{(d-2)} }.
   \ee
   Then 
   \be
   \frac{{\rm d} \Omega^{(d-1)} }{2 (2 \pi)^{d-1} } 
    = \frac{ \Omega^{(d-2)} }{2 (2\pi)^{d-1}} \; [{\rm d} \Omega^{(d-1)} ],
\ee
where 
\be
\frac{ \Omega^{(d-2)} }{2 (2\pi)^{d-1}} = \left [ \frac{(4 \pi)^{\ep} }{8 \pi^2 \Gamma(1-\ep)}  \right ]. 
\ee

With this integration measure, we find \cite{Somogyi:2011ir}
\begin{equation}
    \left \langle \frac{1}{\rho_{i \xa }^\alpha}
    \right \rangle_\xa
    = 2^{1-\alpha-2\epsilon} \frac{\Gamma(1-\epsilon)\Gamma(1-\alpha-\epsilon)}{\Gamma(2-\alpha-2\epsilon)},
\end{equation}

\begin{equation}
\begin{split}
    \left \langle \frac{1}{\rho_{i \xa}^\alpha \rho_{j \xa}^\beta}
    \right \rangle_\xa
    = 2^{1-\alpha-\beta-2\epsilon} &\frac{\Gamma(1-\beta - \epsilon)\Gamma(1-\alpha-\epsilon)}{\Gamma(2-\alpha-\beta-2\epsilon)}
    \times {}_2 F_1 (\alpha, \beta, 1-\epsilon, 1- \eta_{ij}),
\end{split}
\label{eqb.5}
\end{equation}
where brackets $\langle ... \rangle_\xa$ indicate integration over 
$[{\rm d} \Omega^{(d-1)}_\xa]$.

\section{Eikonal functions}
\label{app:c}

For completeness, we provide the double real-emission eikonal functions that we use to compute the relevant contributions to the 
$N$-jettiness soft function. 
Following Ref.~\cite{Catani:1999ss}, for two soft gluons, we define 
\be
{\tilde S}_{ij}^{gg}(\xa,\yb) = 2 S_{ij}^{gg}(\xa,\yb)  - S_{ii}^{gg}(\xa, \yb) - S_{jj}^{gg}(\xa,\yb).
\ee
The function $S_{ij}^{gg}(\xa,\yb)$ reads
\be
\begin{split}
S_{ij}^{gg}(\xa,\yb)
&= S_{ij}^{gg, \rm so}(\xa,\yb)
- \frac{2p_i \cdot p_j}{
p_\xa \cdot p_\yb [ p_i \cdot (p_\xa + p_\yb)][p_j \cdot (p_\xa + p_\yb) ]
}
\\
&+ \frac{
(p_i \cdot p_\xa)(p_j \cdot p_\yb)+
(p_i \cdot p_\yb)(p_i \cdot p_\xa)
}{[ p_i \cdot (p_\xa + p_\yb) p_j \cdot (p_\xa + p_\yb) ]}
\left [ 
\frac{1-\ep}{(p_\xa \cdot p_\yb)^2} 
- \frac{1}{2} S_{ij}^{gg, \rm so}(\xa,\yb)
\right ],
\end{split}
\ee
where 
\be
\begin{split} 
S_{ij}^{gg, \rm so}(\xa,\yb) 
& = \frac{p_i \cdot p_j}{p_\xa \cdot p_\yb}
\left ( 
\frac{1}{(p_i \cdot p_\xa)(p_j \cdot p_\yb)}
+
\frac{1}{(p_i \cdot p_\yb)(p_j \cdot p_\xa)}
\right )
\\
& -\frac{(p_i \cdot p_j)^2}{(p_i \cdot p_\xa)(p_j \cdot p_\xa) (p_i \cdot p_\yb) (p_j \cdot p_\yb) }.
\end{split}
\ee

In case the soft partons are 
a quark and an anti-quark, we write 
\be
{\tilde S}_{ij}^{q\bar q}(\xa,\yb) = 2 S_{ij}^{q\bar q}(\xa,\yb)  - S_{ii}^{q\bar q}(\xa, \yb) - S_{jj}^{q\bar q}(\xa,\yb),
\ee
where \cite{Catani:1999ss}
\be
\begin{split} 
S_{ij}^{q\bar q}(\xa,\yb) 
= 
\frac{(p_i \cdot p_\xa)(p_j \cdot p_\yb)+ 
(p_i \cdot p_\yb)(p_j \cdot p_\xa) 
- (p_i \cdot p_j) (p_\xa \cdot p_\yb) 
}{(p_\xa \cdot p_\yb)^2 
[ p_i \cdot (p_\xa + p_\yb)][p_j \cdot (p_\xa + p_\yb) ]
}.
\end{split}
\ee

\section{Numerical Checks of N-Jettiness Soft Functions} \label{app:d}
In this section, we compare  results for the $N$-Jettiness soft function obtained in this paper with those in Ref.~\cite{Bell:2023yso}.  For this comparison, 
we define two functions $G_{ij}^{nl} $ and $
Q_{ij}^{nl} $ that are obtained  
from $G_{ij}$ and $Q_{ij}$ by setting 
all terms with $L_{ij}$ to zero
\be
\begin{split}
G_{ij} & = \frac{22}{9} L_{ij}^3  
+ \left ( \frac{67}{9} - \frac{\pi^2}{3}
\right ) L_{ij}^2 
+ L_{ij} \Big(
 \frac{11}{3} \left \langle 
 L^\psi_{ij,\xa}
\frac{\rho_{ij} }{ \rho_{i \xa} \rho_{j \xa} } \right \rangle_{\xa}
+\frac{11}{3}{\rm Li}_2(1-\eta_{ij} ) 
\\
& + \frac{202}{27} -
7 \zeta_3 \Big  ) + G_{ij}^{nl} 
\end{split},
\ee
\be 
 Q_{ij}  = -\frac{8}{9} L_{ij}^3  
-\frac{20}{9}  L_{ij}^2 
- L_{ij} \Bigg(
 \frac{4}{3} \left \langle 
 L^\psi_{ij,\xa}
\frac{\rho_{ij} }{ \rho_{i \xa} \rho_{j \xa} } \right \rangle_{\xa}
+\frac{4}{3}\ {\rm Li}_2(1-\eta_{ij} ) 
+\frac{56}{27}
\Bigg ) + Q_{ij}^{nl}.
\ee

We also note that for the comparison of the 
triple-color correlated contributions, we 
do not set these logarithms to zero but take $\bar u = \mu = 1$ instead. As the result, we should set   $L_{ki} \to 1/2 \ln\eta_{ki}$ in Eq.~(\ref{eq7.15}).

\subsection{1-Jettiness}
In order to compare our numerical results, we parameterize our phase space in a similar way as done in Ref.~\cite{Bell:2023yso}. In the case where there are two back-to-back beams and one jet confined in a plane, we parameterize the scattering by the angle $\theta_{13}$ using the following momenta, 
\be
n_1 =(0, 0, 1), \quad n_2 =(0, 0, -1), \quad n_3 = (\sin \theta_{13}, 0, \cos \theta_{13}).
\ee
In order to compare a phase space point where the jet is separated from the beams, we take
\be
\theta_{13} = \frac{12 \pi}{25}.
\label{eqd4}
\ee
In Table~\ref{tab1} we present the results for the non-logarithmic coefficient of the renormalized $1$-jettiness soft function, which are in agreement with the ones in Ref.~\cite{Bell:2023yso}.

 \begin{table}[ht]
   \centering
    \begin{tabular}{||c|l|l||l|l||}
     \hline \hline
    \multirow{2}{*}{Dipoles} &\multicolumn{2}{c||}{Gluons} &\multicolumn{2}{c||}{Quarks} \\ [0.5ex]
    \cline{2-5} 
    & \multicolumn{1}{c|}{$G_{ij}^{nl}$} & \multicolumn{1}{c||}{Ref.~\cite{Bell:2023yso}} & \multicolumn{1}{c|}{$Q_{ij}^{nl}$} & \multicolumn{1}{c||}{Ref.~\cite{Bell:2023yso}} \\
    \hline \hline
     12 & 48.10 $\pm$ 0.01  & 48.04 $\pm$ 0.07  & -23.504 $\pm$ 0.001 &-23.503 $\pm$ 0.010  \\
    \cline{1-5}    
     13 & 36.81 $\pm$ 0.02 & 36.82 $\pm$ 0.04  & -21.871 $\pm$ 0.003  & -21.875 $\pm$ 0.009  \\
    \cline{1-5} 
     23 & 38.14 $\pm$ 0.01 & 38.13 $\pm$ 0.05  & -22.039 $\pm$ 0.002   & -22.031 $\pm$ 0.009 \\
    \hline
    \end{tabular}
    \caption{Comparison of the selected results for  the 1-jettiness soft function for the kinematic point in Eq.~(\ref{eqd4}). When quoting results for functions $G_{ij}^{nl}$, $Q_{ij}^{nl}$,  we show  {\sf Vegas} integration  errors which, most likely, underestimate the true uncertainties of the result.}
    \label{tab1}
\end{table}
    
\subsection{2-jettiness}

We again consider the configuration where there are two back-to-back beams. We parameterize the scattering of the two jets by $\theta_{13}$, $\theta_{14}$ and $\phi_4$ as
\be
\begin{split}
n_1 =(0, 0, 1), &\quad n_2 =(0, 0, -1), \\
n_3 = (\sin \theta_{13}, 0, \cos \theta_{13}), &\quad n_4 = (\sin \theta_{14}\cos\phi_4, \sin \theta_{14}\sin\phi_4, \cos \theta_{14}).
\end{split}
\ee
Considering a generic phase space point where the beams and jets are separated from each other, we take
\be
\theta_{13}= \frac{6\pi}{25}, \quad \theta_{14}= \frac{13\pi}{25}, \quad \phi_{4}= \frac{\pi}{5}.
\label{eqd.6}
\ee
Results for all possible dipole coefficients are presented on Table~\ref{tab-dip2}. In the case of $2$-jettiness, the tripole contribution can be reduced to a single color structure, $f_{ABC} \ T_1^A T_2^B T_3^C$. The renormalized sum of all tripole contributions in this configuration is $1064.77 \pm 0.08$, which is in agreement with the value ($1064.6 \pm 0.1$) reported in~\cite{Bell:2023yso}.

\begin{table}[ht]
   \centering
    \begin{tabular}{||c|r|r||r|r||}
     \hline \hline
    \multirow{2}{*}{Dipoles} &\multicolumn{2}{c||}{Gluons} &\multicolumn{2}{c||}{Quarks} \\
    \cline{2-5}
    & \multicolumn{1}{c|}{$G_{ij}^{nl}$} & \multicolumn{1}{c||}{Ref.~\cite{Bell:2023yso}} & \multicolumn{1}{c|}{$Q_{ij}^{nl}$} & \multicolumn{1}{c||}{Ref.~\cite{Bell:2023yso}}  \\
    \hline \hline
     12 & 71.15 $\pm$ 0.05 & 71.11 $\pm$ 0.12  & -27.837 $\pm$ 0.001 & -27.841 $\pm$ 0.011  \\
    \cline{1-5}    
     13 & 36.27 $\pm$ 0.02 & 36.17 $\pm$ 0.07 & -21.719 $\pm$ 0.005 & -21.724 $\pm$ 0.009 \\
    \hline 
     23 &75.79 $\pm $ 0.01  & 75.62 $\pm$ 0.09 & -27.804 $\pm$ 0.002  & -27.807 $\pm$ 0.011\\
    \hline
     14 & 65.48 $\pm$ 0.02 &  65.38 $\pm$ 0.09 & -25.660 $\pm$ 0.003  & -25.666 $\pm$ 0.010\\
    \hline
     24 & 46.25 $\pm$ 0.01  & 46.15 $\pm$ 0.06  & -22.933 $\pm$ 0.005 & -22.908 $\pm$ 0.009 \\
    \hline
     34 & 44.86 $\pm$ 0.02 & 44.72 $\pm$ 0.09  & -22.513 $\pm$ 0.004  & -22.518 $\pm$ 0.009 \\
    \hline
    \end{tabular}
    \caption{Comparison of the selected results for  the 2-jettiness soft function for the kinematic point in 
    Eq.~(\ref{eqd.6}).  When quoting results for functions $G_{ij}^{nl}$, $Q_{ij}^{nl}$,  we show {\sf Vegas} integration  errors which, most likely, underestimate the true uncertainties of the result.
    }
\label{tab-dip2}
\end{table}

\subsection{3-jettiness}

We parameterize the scattering of the additional jet by $\theta_{15}$ and $\phi_5$
\be
n_5 = (\sin \theta_{15}\cos\phi_5, \sin \theta_{15}\sin\phi_5, \cos \theta_{15}).
\ee
We compare in Table~\ref{tab3} our 3-jettiness dipole contributions to the benchmark result in \cite{Bell:2023yso} by taking the following phase space point
\be
\theta_{13}= \frac{3\pi}{10}, \quad \theta_{14}= \frac{6\pi}{10},  \quad \theta_{15}= \frac{9\pi}{10}, \quad \phi_{4}= \frac{3\pi}{5}, \quad \phi_{5}= \frac{6\pi}{5}
\label{deq8}.
\ee

\begin{table}[hb]
   \centering
    \begin{tabular}{||c|r|r||r|r||}
     \hline \hline
    \multirow{2}{*}{Dipoles } &\multicolumn{2}{c||}{Gluons} &\multicolumn{2}{c||}{Quarks} \\
    \cline{2-5}
    & \multicolumn{1}{c|}{$G_{ij}^{nl}$} & \multicolumn{1}{c||}{Ref.~\cite{Bell:2023yso}} & \multicolumn{1}{c|}{$Q_{ij}^{nl}$} & \multicolumn{1}{c||}{Ref.~\cite{Bell:2023yso}} \\
    \hline \hline
     12 & 116.20 $\pm$ 0.01 & 116.20 $\pm$ 0.16  & -36.249 $\pm$ 0.001 & -36.244 $\pm$ 0.009  \\
    \cline{1-5}    
     13 & 38.13 $\pm$ 0.03 & 37.63 $\pm$  0.03 & -21.717 $\pm$ 0.007 & -21.732 $\pm$ 0.005\\
    \hline 
     14 & 63.63 $\pm$ 0.01 & 63.66 $\pm $ 0.06 & -25.189 $\pm$ 0.003  & -25.192 $\pm$ 0.006\\
    \hline
     15 & 107.17 $\pm$ 0.01 & 106.99 $\pm$ 0.12 & -35.268 $\pm$ 0.001  & -35.256 $\pm$ 0.009\\
    \hline
     23 & 97.11 $\pm$ 0.01  & 96.97 $\pm$ 0.10  &  -32.875 $\pm$ 0.002  & -32.872 $\pm$ 0.008\\
    \hline
     24 & 67.36 $\pm$ 0.02  & 67.51 $\pm$ 0.08  & -26.821 $\pm$ 0.003 & -26.815 $\pm$ 0.007\\
    \hline
     25 & 30.87 $\pm$ 0.03  & 30.73 $\pm$ 0.04  & -21.561 $\pm$ 0.009 & -21.561 $\pm$ 0.005\\
    \hline
     34 & 69.43 $\pm$  0.01 & 69.24 $\pm$ 0.07   &  -25.854 $\pm$ 0.002  & -25.861 $\pm$ 0.006\\
    \hline
     35 & 106.13 $\pm$ 0.02 & 105.97 $\pm$ 0.13   & -34.799 $\pm$ 0.002  & -34.796 $\pm$ 0.008\\
    \hline
     45 & 74.45 $\pm$ 0.02 & 74.36 $\pm$ 0.09  & -28.247 $\pm$ 0.004  & -28.251 $\pm$ 0.007\\
    \hline
    \end{tabular}
    \caption{Comparison of the selected results for  the 3-jettiness soft function for the kinematic point in Eq.~(\ref{deq8}). 
    When quoting results for functions $G_{ij}^{nl}$, $Q_{ij}^{nl}$,  we show  {\sf Vegas} integration  errors which, most likely, underestimate the true uncertainties of the result.
    }
    \label{tab3}
\end{table}

For the tripole contribution in this configuration, there are four independent color structures. In Table~\ref{tab4} we present our results for each of these configurations, as defined in~\cite{Bell:2023yso}.

\begin{table}[ht]
   \centering
    \begin{tabular}{||c|l|l|l|l||}
     \hline \hline
     & \multicolumn{1}{c|}{$\tilde{c}_{\text{tripoles}}^{(2,124)}$} & \multicolumn{1}{c|}{$\tilde{c}_{\text{tripoles}}^{(2,125)}$} & \multicolumn{1}{c|}{$\tilde{c}_{\text{tripoles}}^{(2,145)}$} & \multicolumn{1}{c||}{$\tilde{c}_{\text{tripoles}}^{(2,245)}$} \\
    \hline \hline
     $\tilde{c}_{\text{tripoles}}$ & -683.25 $\pm$ 0.01 & -2203.3 $\pm$ 0.2  & -6.324 $\pm$ 0.004 & -0.837 $\pm$ 0.008  \\
    \cline{1-5}    
     Ref.~\cite{Bell:2023yso} & -683.23 $\pm$ 0.04 & -2203.5 $\pm$ 0.1 & -6.325 $\pm$ 0.04 & -0.830 $\pm$ 0.039\\
    \hline
    \end{tabular}
    \caption{Same as in Table~\ref{tab3} for the four independent triple-color correlated contributions.}
    \label{tab4}
\end{table}

 \bibliographystyle{JHEP}
 \bibliography{rref}
   
\end{document}